\documentclass[12pt]{article}
\usepackage{a4}
\usepackage{graphicx}
\usepackage{amsfonts}
\usepackage{amscd}
\usepackage{latexsym}
\usepackage{epsf}
\usepackage{amsmath}
\numberwithin{equation}{section}
\newcommand{\uv}{ultraviolet }
\newcommand{\zf}{zeta function }
\newcommand{\e}{\equiv}
\newcommand{\rl}{\right.\nn\\&&\left.}
\newcommand{\rrll}{\right.\right.\nn\\&&\left.\left.}
\newcommand{\G}{{\cal G}}
\newcommand{\I}{{\cal I}}
\newcommand{\J}{{\cal J}}
\newcommand{\tG}{\tilde{{\cal G}}}
\newcommand{\be}{\begin{equation}}
\newcommand{\ee}{\end{equation}}
\newcommand{\bea}{\begin{eqnarray}}
\newcommand{\eea}{\end{eqnarray}}
\newcommand{\ew}{electroweak }
\newcommand{\f}{function }
\newcommand{\fs}{functions }
\newcommand{\eqq}{equation }
\newcommand{\ef}{eigenfunctions }
\newcommand{\ev}{eigenvalues }
\newcommand{\eqqs}{equations }
\newcommand{\rn}{renormalization }
 \newcommand{\rg}{regularization }
\newcommand{\rp}{regularization procedure }
\newcommand{\hke}{heat kernel expansion }
\newcommand{\hkc}{heat kernel coefficients }
\newcommand{\tr}{{\bf tr~} }
\newcommand{\arctg}{{\bf arctg~} }
\newcommand{\hk}{heat kernel }
\newcommand{\gse}{ground state energy }
\newcommand{\jf}{Jost \f}
\newcommand{\lseq}{Lippmann-Schwinger \eqq }

\newcommand{\nn}{\nonumber}
\newcommand{\MVAR}[2]{#1_#2}
\newcommand{\Mvariable}[1]{\MVAR #1}
\newcommand{\fr}{\frac}
\newcommand{\Res}{{\bf Res~\ }}
\newcommand{\Tr}{{\bf Tr\ }}
\newcommand{\pd}{\partial}
\newcommand{\al}{ \alpha }
\newcommand{\la}{ \lambda }

\newcommand{\om}{ \omega }

\newcommand{\vp}{\varphi }

\newcommand{\ra}{\rightarrow}
\newcommand{\half}{\frac{1}{2}}
\newcommand{\E}{{\cal E}}

\begin{document}
\title{\Huge\bf
Vacuum Energy of Quantum Fields
in Classical Background Configurations} 
\author{ {\sc I. Drozdov}\thanks{e-mail: 
Igor.Drosdow@itp.uni-leipzig.de} \\ 
\small University of Leipzig, Institute for Theoretical Physics\\ 
\small Augustusplatz 10/11, 04109 Leipzig, Germany}
\maketitle

{\large
\hspace*{1cm} Der Fakult\"at f\"ur Physik und Geowissenschaften \\ \\
\hspace*{4cm} der Universit\"at Leipzig\\ \\
\hspace*{6cm} eingereichte\\ \\
\hspace*{4cm} D I S S E R T A T I O N \\ \\ 
\hspace*{2.5cm} zur Erlangung des akademischen Grades\\ \\
\hspace*{4cm} Doctor rerum naturalium\\ \\
\hspace*{5.8cm} (Dr.rer.nat)\\ \\
\hspace*{6.1cm} vorgelegt \\ \\
von\hspace*{2cm} Diplom Physiker (UA)\hspace*{2cm}Igor Drozdov \\ \\
geboren am   06.02.1970\ \ \ \ in \ Dnepropetrowsk, Ukraine\\
\\
\\
Leipzig, den 3.11/2003
}

\vspace*{4cm} 

\begin{abstract}
  
  The \gse of a quantum field in the
background of classical field configurations is considered.
 The subject of the \gse in framework of the quantum field theory is explained.

 The short review of calculation methods (generalized \zf and \hke) and its mathematical
 foundations is given. 

 We use the zeta-functional \rg and express the \gse as an
integral involving the Jost function of a two dimensional scattering problem.
We perform the \rn by subtracting the contributions from first
several heat kernel coefficients. The ground state energy is presented as a
convergent expression suited for numerical evaluation.
 The investigation for three models has been carried out: scalar quantum
 field on the background of scalar string with rectangular shape, spinor vacuum polarized by magnetic
 string of the similar shape and spinor vacuum interacting with the Nielsen
 Olesen vortex. 
Using the uniform asymptotic expansion of the special functions entering the
Jost function we are also able to calculate higher order heat kernel
coefficients.

Several features of vacuum energy have been investigated numerically.
We discuss corresponding numerical results.
\end{abstract}


\newpage
\tableofcontents
 \section{Introduction}
\label{introduction}
\subsection{General remarks}

\ \ \ The vacuum energy is a spectacular manifestation of the
  quantum nature of the real objects, in particular the fields of matter.
  In the framework of natural science it is also a powerful tool for investigation
  and of fundamental properties of our world.
  
    It is a really non-trivial fact, that quantum field can be
  observable even in the case that there are no real particles of it present.
   The reason for this phenomena is one of the main
  properties inherent basically to any quantum system, namely the existence
  of the ground state.
    Indeed, it is well known from such a simple example of a quantum
  mechanical system as the linear harmonic oscillator with the eigenfrequency $\om$,
    that the energy spectrum of this system possesses the lowest state and
    the energy of this state $E^0$ is not zero but is proportional to the Planck constant \be E^0=\half\hbar\om.\ee This fact is generally
  also an intrinsic feature of any quantum object. Here we mean first of all quantum fields, namely matter fields (fermionic fields)
  and fields of interaction (bosonic fields).
    It means, one can establish the existence of a quantum field in the nature
  without to observe any real particles (quantums) of this field, but only
  through non zero vacuum values of observables, that are measurable, for example the
vacuum energy.

 The main point of this phenomena is that the energy of the ground
  state of a quantum system is not defined absolutely, that applies by the way in
  general to the energy itself as a physical feature.
  It is defined generally with a relation to some other state. That is why the discussion about the vacuum
  energy of quantum field makes sense in correspondence to some external
  condition, under which it becomes visible (or measurable in physical
  sense). Such an external condition can be some boundary or some classical
  field interacting to the quantum one (background)
  
   The most famous manifestation of the vacuum energy was the Casimir effect
   (H.B.G.Casimir 1948), which consist in an interaction between two parallel
   conducting plates playing the role of boundaries for the
   electromagnetic field in vacuum state. This effect was predicted
   theoretically in \cite{casimir} and measured experimentally 10 years later.

   Sometimes, all the
   problems concerning the calculation of vacuum energy under external
   conditions are being called "Casimir energy", including also the problems of
   vacuum polarization by classical backgrounds. Actually, the vacuum energy of quantum spinor field
   polarized by a special homogeneous magnetic background was calculated by
   W.Heisenberg and H.Euler in 1936, 11 years before the discovery of the Casimir
   force.\\
   
\subsection{Motivation}

   \ \ \  The investigation of vacuum effects in general and vacuum energy in
   particular has a number of applications to wide variety of physical
   problems (see recent reviews \cite{Bordag:2001qi,most,milton}).
   Indeed, in a lot of problems concerning the experiments in particle physics, certain
   effects caused by interaction with vacuum fluctuations must be taken
   into account as radiation corrections.
  For example in the bag model of hadrons in Quantum Chromodynamics the vacuum energy of
   quark and gluon fields makes essential contributions in the total nucleon energy.
   As further examples of such measurements one can refer to the experiments for the exact mass, isotopic charge,
   anomalous magnetic moment and so on.

   Generally speaking, the vacuum  corrections are more or less relevant in all problems of particle and high energy physics.
   
  Moreover, one can imagine a situation that only the sole possibility to establish the existence of a quantum field in
   the nature, (since it is present in the model) is
   the interaction of its vacuum with an observable classical background (and measurable effects reasoned thereby).
Actually, the interaction of the classical field configuration with
   quantum fields being in vacuum state leads to modification of macroscopic
   physical features of this classical object. In particular the energy and its
   dependence on the external parameter of this configuration undergoes an
   alteration. This interplay between the quantum vacuum and classical field is also
   called sometimes "Casimir effect".
   The modification of classical energy by vacuum
   corrections for real objects is usually negligible small. However, one can
   imagine some special field configurations such that the quantum
   corrections become comparable
   with the classical energy of the background.
   
   Therefore it is a reasonable question to ask:
   could the quantum fluctuations change the main
   physical properties of classical object? First of all, the question
   concerns the stability (or instability) of field configurations.

   For example, one can look for a field configuration, that acquires
 stability through an interaction with the
   quantum vacuum, although it would be instable in pure classical approach.

   In particular, the issue of stability of strings is of interest. In
   \ew theories, and in QED in particular, the coupling is small. Hence the
   vacuum energy (being a one loop correction to the classical energy)
   is suppressed by this coupling and indeed in most of the cases is not
   sufficient to change the stability properties. While in QED a magnetic string is 
   intrinsically unstable, in \ew theory there are unstable and stable
   configuration. Here quantum corrections
   may become important for the stability. A question of special interest
   is whether a strong or singular background may have a quantum vacuum
   energy comparable to the classical one. Such an example is considered in
   the present work for two kinds of field.
   
The vacuum correction could provide stability to classical instable
      objects, but vice versa it could also destabilize classical stable
   objects. The good illustration for this speculation can be provided by the
   model with the vortex of the Nielsen-Olesen type. This configuration is known to be
   stable for any integer vorticity $n$, if the parameter $\beta < 1$. If the
    $\beta > 1$, then only the vortex with $n=1$ remains stable, all
   configurations with the vorticity $n>1$ decay into n simple vortices with
   $n=1$. The question arises there: could the quantum vacuum fluctuation
   modify the parameter $\beta$ in such a way that transition from stable configuration
   to instable one takes place? The corresponding model will be also considered
   in the following sections.
   
    The results and techniques gained here can be also applied to the
 problems of cosmology and to Condensed Matter Physics.
 
   In Mathematical Physics the investigation of vacuum polarization has
   stimulated the development of powerful \rg and \rn techniques based on the
   use of  \zf and \hk expansion,  which are also considered below.
  
    The subject of the work is restricted on the investigation of ground state
   energy of vacuum fields in presence of classical static field configurations with
   cylindrical and translational symmetry.\\

\subsection{Historical review}

\ \ \
Some people claim, that
the subject of vacuum energy has no sense since this value is originally
infinite.
  This point of view is in principal true, but the fact of Casimir force
  between two plates suggests the simple thought: if two plates are
  positioned with the distance $d$ from each other, the photon vacuum
  possesses really an infinite energy, and for the distance
   $d'$ the vacuum energy is infinite as well, but not the same, and crucial
   for the problem is that the difference between these two values is finite and
   moreover measurable.

The same concerns the vacuum energy of quantum field, polarized by some
external background. The first known example was the calculation of the
vacuum energy of a spinor field in the background of a magnetic field. This problem was
first considered as early as in 1936 by W. Heisenberg and H. Euler
\cite{euler}. They were interested in the effective action in the
background of a homogeneous magnetic field. It turned out that the global
energy of spinor vacuum in presence of constant homogeneous magnetic field
(the whole space is filled uniform by the field of magnitude $\vec{B}$)
depends on this magnitude non trivialy. In particular
the characteristic behaviour for the vacuum energy in the asymptotics of
strong potential $B\ra \infty$ in the form \be E^0 \sim B^2 \ln B\ee was obtained.
      
Since the classical work \cite{casimir} of H.B.G. Casimir, where the
energy $E^0$ of vacuum polarized by two conducting planes was calculated
analytically with the finite result

 \be E^0= \fr{\pi^2}{720} \fr{\hbar c}{a^3}S ,\ee
a number of similar problems was investigated for various external
conditions (boundary conditions, background potentials etc.).
One may consult, for
example, in the recent review \cite{Bordag:2001qi} or the books
\cite{most,milton}.

The discovery of the Aharonov-Bohm effect \cite{aharonovbohm} gave rise to a series of works
 researching various approaches to the vacuum in presence of configurations similar to the
 Aharonov-Bohm string, so the vacuum in gauge theories claims its non trivial geometrical and
 topological structure, that is closely connected with the structure of gauge
 groups. 

 In this way local properties of photon and electron vacuum have been
 actively investigated. There it has been shown that the structure of vacuum state possesses a more reach
 variety of features as merely the global ground state energy \cite{gornicki}. In particular
 the subjects of interest were various local effects such as distribution of the vacuum energy density \cite{serebryany},
 vacuum currents \cite{fleklein}, \cite{goldhaber} and the backreaction (so called ``feedback effects'') \cite{goldhaber}. 

The other important application of the external field problem is the problem
of non stationary vacuum under external conditions. This activity
includes all the effects of particle creation by time dependent boundaries \cite{Walker:vj},
or non conservative backgrounds, e.g., electric field. The most concerned
problems were investigated in early 70's, like \cite{nikish}, the results were
restricted and confirmed for reduced space dimension \cite{drozdovdip} and
generalized for arbitrary high dimension \cite{gavgit_el}. This interest was
also motivated to explain of Hawking radiation of black holes. 
  

 The class of investigated backgrounds is usually restricted by the
 complication and solvability of a certain problem. In applications of bag
 models \cite{graham}, \cite{cherednik} or in cosmology the spherical symmetric backgrounds are considered
 \cite{BKEL, eliz, borkirhel}, for string like models the cylindrical
 symmetry is present \cite{nieloles, acc_vach, borkir1}. 
 
Pure analytical calculations are only possible for simplest configurations,
  e.g. for the considered classical sample of
  homogeneous field \cite{euler},\cite{savvidy} or a very specific ones such
  as the transparent
  reflectionless domain wall \cite{dunne}. For
  more complicated cases a number of alternative numerical approaches has been
  developed \cite{pasipoularides, gies, gies1}. 
In the present work we restrict ourselves on the problems which
  can be treated analytically, (or analytically to some degree,
  demanding finally numerical evaluations \cite{borkir2,borkir1,scand1,Drozdov:2002um,bednefrkirsant})
  
Although a variety of problems for various backgrounds and in many QFT
 models have been successfully investigated, and a corresponding experience in
 handling these problems have been gained, the
 general understanding of the problem is so far not complete,
  and many features are still unclear. 
 
 The main open question is that no general rule for the dependence of the vacuum
energy on the background properties has been found so far.
In particular, it is unknown how to forecast the sign of the energy. Some
special cases only has been recently investigated.
 An interesting approach is that used e.g. in \cite{fry} where the issue of
the sign and of bounds on fermionic determinants in a magnetic
background had been considered.

\subsection{Examlpes}

In QED, for the background of a flux tube with constant magnetic field
inside the \gse was calculated in \cite{borkir1}. It turned out to be negative remaining of course much smaller than the
classical energy of the background field.

The remarkably simple case of a magnetic background field concentrated
on a cylindrical shell (i.e., a delta function shaped profile) was
considered in \cite{scand2}. Here different signs of the vacuum energy
turned out to be possible in dependence on the parameters (radius and flux).  
 However one has difficulties with a physical interpretation of this result
since the model possesses an infinite classical energy density. So one needs
a consequent interpretation in a suitable limit, as for example was done for
an infinite thin Aharonov-Bohm string \cite{gavgit_magn, bednefrkirsant}.
Here a somewhat intermediate model between the rectangular potential
\cite{borkir1} and $\delta$-shaped potential \cite{scand2} 
is investigated in the present work.

An interesting approach is that used in \cite{fry} where the issue of
the sign and of bounds on fermionic determinants in a magnetic
background had been considered.

Presently, the interest is shifted to string like configurations.  In
\cite{Groves:1999ks} the contribution of the fermionic ground state
energy to the stability of \ew strings was addressed. In
\cite{Diakonov:2002bx} the gluonic ground state energy in the
background of a centrum-of-group vortex in QCD has been considered.

 The ground state energy for vortex type solutions in gauge models with
minimal coupling has been calculated implementing special numerical and
partially analytical approaches to U(1) \cite{gies} and more complicated
SU(2) and SU(3) theories \cite{bordagcolor}. In the case of supersymmetry the
sufficient simplification is established through the high symmetry, that
allowed to calculate recently a non-zero vacuum correction to the mass
\cite{vassilvort} and to the central charge \cite{rebhan} of the SUSY
N=2 vortex.
In the present work the result of \cite{bordagcolor} are extended to the vortex of Nielsen-Olesen type
in the simplest U(1) abelian Higgs model.

\subsection{The aim of the work}

The main goals of the present research are:

Methodologically, 
an improvement of the existent technical tools to calculate the \gse in
classical backgrounds is of interest as well as the applicability of these
technics for more complicated configurations.

Physically, the 
influence of the quantum corrections on the background, in particular on the
stability properties of the field system at all, and the asymptotic behaviour
of the one-loop vacuum contribution to the total energy, are to be
investigated by using the applied method.

\subsection{The method}

 The class of considered problems to which we are restricted in the
 present work excludes basically all the problems with non stationary
 vacuum, since the method used below is not developed for such kind of
 models.
 
  All the considered models contain one quantum field in vacuum state
 polarized by a classical background with special symmetry
 properties that simplify sufficiently the calculation procedure.
 
 The expression for the ground state energy contains the \uv divergences. To make the results of this calculations
 physically meaningful one has to extract the finite value from this infinite
 result, in other words it is necessary to
subtract the \uv divergences using some \rn prescription.

 There are various methods developed to get rid of non-physical
 divergences. In the present paper the \zf \rp has been used. This approach is well known
and we follow in general \cite{borkir1}. Roughly speaking one has to subtract the
contribution of the first few heat kernel coefficients ($a_0$ through
$a_2$ in the given case of (3+1) dimensions). After that one can
remove the intermediate \rg and one is left with finite
expressions. However, in order to obtain these finite expressions in a
form suitable for numerical calculations one has to go one step further.
As described in \cite{borkir1,borkir2}, see also below in Sec.\ref{efineas}, one has to add
and to subtract the certain part of the asymptotic expansion of the
integrand.

In the present work we generalize the analysis done in \cite{borkir1, scand1}
of rectangular shaped background fields (scalar and magnetic) and,
furthermore, for mixed background given in the from of numerical distribution.

 The proceeding has been arranged consequently through the sections 3 - 6
simultaneously for the following three considered models.
 
We consider the vacuum energy of a spinor field in QED for a rectangular
background, the scalar background with the same shape interacting to the
scalar fluctuations, and finally the spinor vacuum polarized by the background configuration in form of the
Nielsen-Olesen vortex in the abelian Higgs model.

The problem of the rectangular magnetic background reproduces in principle the main
features of the one considered in \cite{borkir1}, it is however technically
more involved and allows progress in two directions. First, it allows to refine the
mathematical and numerical tools for such problems and, second, it
allows to address the question how the vacuum energy behaves for an
increasingly singular background (making the rectangle narrower).  So
this model interpolates to some extend between the flux tube with
homogeneous field inside in \cite{borkir1} and the delta shaped one in
\cite{scand2, scand1}.

The basic principles of this analytical calculation procedure are
 described in sections 3-5.
 We start with the well known zeta functional regularization.
 The regularized \gse is represented as
a \zf of a hamiltonian spectrum and treated in terms of the
heat kernel expansion.  A representation of the regularized ground state
energy as an integral of the logarithmic derivative of the Jost
function for wave scattering problem on the external background is obtained.
 The divergent part is identified as that part of the corresponding heat kernel expansion
which does not vanish for large $m$ (mass of the quantum field).
After the subtraction of this divergent part the remaining
analytical expression must be transformed in order to lift the
\rg.  A part of the uniform asymptotic expansion of the
Jost function is used for this procedure.
 The Jost function is obtained from the exact
solutions of the \eqqs  of motion (these are the Klein-Gordon and Dirac \eqqs
  respectively), derived in Sec.\ref{hamiltonians}.
  
 The explicit form of exact and asymptotic Jost functions is considered in
 Sec.\ref{jostfunc} and \ref{lippsw}. In Sec.\ref{spectralsum}, \ref{efineas} the
 representation useful for further numerical evaluations is
derived. Sec.\ref{a52} is devoted to the calculation of the heat kernel
coefficient $a_{5/2}$ for the special magnetic background.

Finally the analytical expressions for the \gse are evaluated
 numerically. Sec.\ref{results} contains some numerical evaluations of the finite \gse for the considered models.
 
  The ground state energy for the Nielsen-Olesen vortex background cannot be
 performed analytically since the background itself is given as the numerical
 solution of Nielsen-Olesen \eqqs. In this case some analytical manipulations
 applied for the rectangular background have been replaced by the
 corresponding numerical procedures.
 
Additionally, the paper contains an introduction to the subject of \gse of
quantum field in Sec.\ref{groundstenergy}. Some technical details
are contained in the Appendix.


\section{The ground state energy of quantized field}
\subsection{The effective action}
\label{effact}

\ \ \ \ In the generalized approach the contribution of the one-loop ground state
energy appears as a first correction to the classical action in the so called
"effective" action. In order to derive this representation we start with the
basic object of the quantum field theory in the functional integration
approach namely the generating functional.

Let the action $S$ of the considered system consist of the classical field $\Phi$
and let the quantum field $\phi$ be defined in the canonical way as

\be
S[\Phi,\phi]= S_{class}[\Phi]+\half\int dx< \phi(x) K[\Phi]\phi(x)>,
\label{action}
\ee
where $\Phi$ is a classical field configuration called up from here and
below "background", and $\phi$ is a quantized field, so called
"fluctuation".
The symbol $K[\Phi]$ relates to the kernel of the corresponding \eqq of
motion, (the so called kernel of free action). It means particulary
$K[\Phi]=\Box+m^2+V[\Phi]$ for scalar field with scalar potential $V$ and
$K[A_\mu]= i\gamma^\mu(\pd_\mu-ieA_\mu)-m$ for spinor field with the gauge
background $A_\mu$.
The edge brackets denote the full scalar product on the space
of $\phi$.
 Proceeding from the defined action $S[\Phi,\phi]$ (\ref{action}), we
 construct the related generating functional for Green \fs of $\phi$ 
\be
Z[\Phi,J]=\int {\cal D} \phi \exp\{ iS[\Phi,\phi]+i\int\phi(x)J(x)dx\},
\label{gen_func}
\ee
 where the explicit dependence on the fluctuation $\phi$ does not appear,
 since it is integrated out. Here we have to notice, that we work in this
 paper with abelian background fields only. For non-abelian models the
 formula (\ref{gen_func}) is in general no more valid \cite{lavrov}.
 
If we restrict the problem only to
connected Green functions, the
corresponding generating functional $W[\Phi,J]$ is related to the $Z[\Phi,J]$
 by
\be
Z[\Phi, J]=\exp\{iW[\Phi,J]\}.
\label{gen_func_connect}
\ee

 
 One can formally restore the function $\phi(x)$ from the $W[\Phi,J]$ by the
functional derivation
  \be
  \phi(x)=\fr{\delta W[\Phi,J]}{i\delta J(x)}
  \label{phi_of_W}
  \ee
  
and express the source parameter $J(x)$ with respect to the field variables
\be
J=J[\Phi(x),\phi(x)].
\ee

This change of variables corresponds to the Legendre transformation $\{\Phi(x),J(x) \}\ra
\{\Phi(x), \phi(x) \}$. The effective action $\Gamma[\Phi(x),\phi(x)]$ is
defined in terms of these new variables as

\be
\Gamma[\Phi,\phi]= W[\Phi, J[\Phi,\phi]]- \int dx <\phi(x)
J[\Phi(x),\phi(x)]>.
\label{effective_action_of_fields}
\ee

 Using the Gaussian integral for the functional integrals of kind

\be
\int \fr{d^nx}{\sqrt{2\pi}^n} e^{-Q(x)}
\label{gauss_int}
\ee

for any quadratic form $Q(x)=\half<x,K x>+<b,x>+\ c$ so that

\be
\int \fr{d^nx}{\sqrt{2\pi}^n} e^{-(\half<x,K x>+<b,x>+c)}=\exp\{
\half<b,K^{-1} b>-c\}(\det K)^{-\half} 
\label{gauss_quadr}
\ee
for the functional integrals (\ref{gen_func}), one can calculate the generating functional $Z[\Phi(x),J(x)]$ explicit:

 \be
 Z[\Phi, J]=e^{iS_{class}[\Phi]} (\det K[\Phi])^{\mp\half}
 e^{-\fr{i}{2}<J K^{-1}J>}.
\label{explicit_gen_func}
 \ee

Here the $\det K[\Phi]$ is the functional determinant of the kernel of the
 free action $K[\Phi]$ (e.g. $\Box -m^2$, more examples for certain models are
 listed below, (Sec.\ref{hamiltonians}).
 
The choice of the sign of the power $(\det K[\Phi])^{\mp\half}$ corresponds
to the bosonic/ fermionic fields $\phi$ respectively. Using the known
representation of the logarithm of determinant through the trace of
logarithms
\be
\ln \det K = \Tr \ln K,
\ee
one can arrive for the connected generating functional $W[\Phi(x),J(x)]$ at
\be
W[\Phi(x),J(x)]=S_{class}[\Phi]\pm \fr{i}{2}\Tr \ln K -\half <J K^{-1} J>
\ee

and for the effective action (\ref{effective_action_of_fields})

\be
 \Gamma[\Phi(x),\phi(x)]=S_{class}[\Phi]\pm \fr{i}{2}\Tr \ln K +\half <\phi K^{-1} \phi>.
\ee

 If the quantum field $\phi$ is considered only in the ground state, the
 corresponding effective action reads

 \be
 \Gamma[\Phi(x),\phi(x)]|_{\phi=0}=\Gamma[\Phi(x)],
 \label{eff_act_vac}
 \ee
whereat it is supposed that there is no spontaneously broken symmetry of the
ground state.
 
The zeta-functional \rg formalism can be applied to the effective
action in the form (\ref{eff_act_vac}). The corresponding \zf- regularized
 effective action is considered in the (Sec. \ref{zetafunc})
 

 \subsection{The ground state energy in the framework\\ of canonical quantization}
\label{groundstenergy}

\ \ \ The explicit expression for calculation of the ground state energy can be
derived also from the canonical quantization formalism, that we even use.
In the following work two kinds of quantum field are discussed-the
Klein-Gordon scalar field and the Dirac spinor field. 

\subsubsection{Quantized scalar field}

Consider a simplest case of the real massive scalar field $\vp$ with the lagrangian
\be {\cal L}=\half  \{ \pd_\mu \vp\pd^\mu\vp - m^2 \vp^2 \}\ee
\label{scalar_lagr}
leading to the Klein-Gordon \eqq
\be
\left\{\Box+m^2 \right\} \vp(\vec{x},t)=0
\label{klein_gordon}
\ee

All the further discussions and calculations are concerned with the Minkowski
space-time $g_{\mu\nu}=diag\{1,-1,-1,-1\}$. Then the (\ref{klein_gordon}) will be rewritten as
\be
\left\{\fr{\pd^2}{\pd {x^0}^2} -\Delta+m^2  \right\} \vp(\vec{x},t)=0.
\label{klein_gordon_mink}
\ee

 Since the \eqq is of second order with respect to time $t=x^0$, we can define
 so called "positive-" and "negative-frequency" solutions $\vp^+ (\vec{x},x^0)$ and $\vp^-(\vec{x},x^0)$.

 The further explanation follows the textbook \cite{bogolubov}.
Let the equation of motion have a continuous spectrum of solutions $\vp^\pm(\vec{x},x^0)$. The $\vp(x)$
 are representable as the Fourier integral:
 \be
 \vp(x)=\fr{1}{(2\pi)^{3/2}}\int\limits_{-\infty}^\infty
 \fr{d^3k}{\sqrt{2k_0}}\left[
   e^{(-ik_0x^0)}e^{(i\vec{k}\vec{x})}\vp^-(k_0,\vec{k})+e^{(ik_0x^0)}e^{-i(\vec{k}\vec{x})}\vp^+(k_0,\vec{k})
   \right].
 \label{phi_fourrier_int}
 \ee

The normalization of the solutions is kept to be the one used in
\cite{bogolubov},
but for our problems it is unimportant.
 The sign switch of the frequency $\theta(k_0)$ is supposed to be contained
in the corresponding Fourier images $\vp^{\pm}(k_0,\vec{k})$.

 We consider the problems with the translational invariance along
 the z-axis. It means, the z-variable is not present in the \eqq of motion
 explicitly. The solution of the \eqq (\ref{klein_gordon_mink}) may contain only the s
 z-dependent factor of kind $e^{\pm (ip_zz)}$ and this factor can be separated for the next step.

  The energy of quantum field will be understood by means of the 00-component
 of the canonical energy momentum tensor $T^{\mu\nu}$, that is defined for a given
 lagrangian ${\cal L}$ as
 \be
 T^{\mu\nu}=\fr{\pd{\cal L}}{\pd(\pd_{\mu}\vp)} \pd^{\nu}\vp -
 g^{\mu\nu}{\cal L}
 \label{energy-momentum}
 \ee

For the zero-zero component of $T^{\mu\nu}$ (the hamiltonian density) of the neutral scalar field we have
\be
{\cal H}(x)=T^{00}(x)=\half \left\{ [\pd_0 \vp(x)]^2+[\nabla
  \vp(x)]^2+m^2\vp(x)^2 \right\}
\label{T_00_scalar}  
\ee

  By (\ref{T_00_scalar}) we have derived the explicit from of the hamiltonian
 density (energy density per unit volume) of the field $\vp$. Then the resulted function of coordinates
$T^{00}(x)$ describes a local spatial distribution of the energy of scalar
 field $\vp$
 
The energy contained in the finite volume V can be obtained through the integration
 over $d^3x$. Suppose we are interesting for the energy of the field
 contained in the layer of finite thickness $L$ between two plates
 transversal to the axis $z$, say $z=0$ and $z=L$. The global field energy
 contained in this layer is the integral

\bea
E_L&=&\int T^{00}=\half\int\limits_0^Ldz \int d^2x \int\fr{d^3p\
 d^3k}{2\sqrt{k_0p_0}}\nn\\&& \left\{
-k_0p_0[
\vp^+(p)\vp^+(k)e^{ix_0 (k_0+p_0)}e^{-i(\vec{k}+\vec{p})\vec{x}}+  
\vp^-(p)\vp^-(k)e^{-ix_0 (k_0+p_0)}e^{i(\vec{k}+\vec{p})\vec{x}}-\right.\nn\\
&&\left.\vp^+(p)\vp^-(k)e^{-ix_0 (k_0-p_0)}e^{i(\vec{k}-\vec{p})\vec{x}}-
\vp^-(p)\vp^+(k)e^{ix_0 (k_0-p_0)}e^{-i(\vec{k}-\vec{p})\vec{x}}]
\right\}\nn
\label{T_00_scalar_global}
\eea

The first and the second summand turn out to be equal zero if the
integration over $d^3x $ is performed, in the remaining two terms the part,
containing the dependence on z-variable can be separated and the integration
over $dz$ builds the Fourier representation of delta-\f $\delta(p_z-k_z)$,
that reads
\be \delta(x)=\int\limits_{-\infty}^\infty \fr{dp}{2\pi} e^{-ipx}\ee

Finally one arrives at the form:

\be
E_L=\int T^{00}=\half L\int \fr{dp_z}{2\pi} \int d^2p\ p_0
[\vp^+(p)\vp^-(p)+ \vp^-(p)\vp^+(p)]
\label{E_L}
\ee
 where the remaining components of the momentum space are denoted as $d^2p$.
 Divided by $L$ this value means the energy density per unit length.
 
 The canonical secondary quantization procedure (for bosonic scalar field)
 means the decomposition of the arbitrary solution $\vp(x,t)$ of (\ref{klein_gordon})
onto the \ef $\vp^\pm_n$

 \be
  \vp(x,t)=\int\fr{d^3k}{(2\pi)^3 2\om_k} [a(k)e^{-ikx} + a^+(k)e^{ikx} ]
  \label{decomposition}
 \ee
with the coefficients $a(k), a^+(k)$ called the annihilation and creation
operators with the momentum $k$. The integration is carried out with the
invariant measure on-shell,\cite{bogolubov,ryder}, $\om_k=\sqrt{\vec{k}^2+m^2}$.

The commutation relations between $a(k), a^+(k)$
implied by the normalization conditions in the Fock space \cite{bogolubov} are

\be
[a(k), a^+(k')]=\delta(k-k');\ \ \ [a^+(k), a^+(k')]=[a(k), a(k')]=0
\ee

Within these relation the hamiltonian ${\cal H}$ takes the form
\be
 H = \int \fr{d^3k}{(2\pi)^3 2k_0} \fr{k_0}{2} \left[ a^+(k)a(k)+ a(k)a^+(k) \right]=\int \fr{d^3k}{(2\pi)^3 2k_0} k_0\left[N(k)+\half \right],
\ee
where $ N(k)=  a^+(k)a^(k)$ 
is the operator of number of particles with the momentum $k$.

The vacuum state is understood as the lowest state of the quantum field in
the Fock space is the vacuum state $|0>$ defined by

\be
a(k)|0>=0, \label{vacuum}
\ee
and the corresponding co-state $<0|$ respectively
\be
<0|a^+(k)=0 \label{co_vacuum}
\ee

Now we define the vacuum energy as a vacuum expectation value of $T^{00}$
in the vacuum state
\be
E_{vac}=\int <0|\ T^{00}\ |0>. \label{E_vac}
\ee

All problems considered below in this paper
concerns with the backgrounds possessing translational symmtery along the
$z$-axis, it is integrated  over the $z$-component of the momentum, $p_z/(2\pi)$
and the transversal component of the momentum, $\vec{p}$ (say, for
cylindrical coordinates $\vec{p}\e\vec{p}_{r\vp}$), can be quantized, so that
it acquires discrete values $\vec{p}_(n)$. Then, by means of the integration
measure of Fourier transformations \cite{vladim}, the
integration over $\fr{dp}{2\pi}$ will be substituted by summation over $p_(n)$  .
Since we mention the vacuum state as a state without real particles, it leads
to the relation
\be
E_{vac}=\half\int \fr{dp_z}{2\pi} \int \fr{d^2p}{(2\pi)^2} \om_p =
\int \fr{dp_z}{2\pi} \sum\limits_{(n)} \om_{(n)}
\ee
 The (\ref{vacuum},\ref{co_vacuum}), substituted into the (\ref{E_L}), provide the final expression for
 vacuum energy density ${\cal E}$ per unit length of the translational
 invariant field configuration
 \be
 {\cal E} =\int T^{00}=\half \int \fr{dp_z}{2\pi} \sum\limits_{(n)} \om_{(n)}
\label{e_dens}
\ee
where $\om_n=\sqrt{\vec{p_n}^2+m^2}$ are the \ev of the hamiltonian (\ref{T_00_scalar}).\\

Here it is suitable to notice that the solutions $ \vp(x)$ of the
Klein-Gordon \eqq should be orthonormalized with respect to the inner
product for scalar field, but the normalization coefficients of the
\ef $ \vp_{(n)}(x)$ does not appear in the final expression
of the \gse because we handle here with the \ev only; thus the
discussion concerning the normalization coefficients can be omitted for
our problem.

Thus we obtained that the vacuum energy defined by \ref{E_vac} is equal to
the half of sum over the spectrum of \ev of hamiltonian, that has been first
defined as the \gse energy. Therefore the both definitions coincide
with. However, it is necessary to notice that this equivalence is true for a
system of not interacting (``free'') fields. It will be mentioned further in
Sec.\ref{zetafunc} in context of the effective action.

\subsubsection{Quantized Dirac field} 

Let the quantum field $\psi(x)$ now be the 4-component spinor field 
with the free lagrangian

\be
{\cal L}=\fr{i}{2}\left\{\bar{\psi}(x)\gamma^\mu\pd_\mu\psi(x)-\pd_\mu\bar{\psi}(x)\gamma^\mu\psi(x))\right\}- m\bar{\psi}(x)\psi(x),
\label{lagr_spinor}
\ee
 and the spinor function $\psi(x)$ obeys the Dirac \eqq

\be
(i\gamma^\mu\pd_\mu-m)\psi(x)=0
\label{dirac_\eqq}
\ee

One can use however the simplified lagrangian that also leads to the \eqq
(\ref{dirac_\eqq}), namely

\be
{\cal L}=\bar{\psi}(x)\left\{i\gamma^\mu\pd_\mu-m  \right\}  \psi(x)
\label{lagr_spinor_simpl}
\ee

 The canonical energy-momentum tensor corresponding to the
 (\ref{lagr_spinor}) reads

 \be
 T^{\mu\nu}=\fr{i}{2}\left\{
   \bar{\psi}(x)\gamma^\mu\pd^\nu\psi(x)-\pd^\nu
 \bar{\psi}(x)\gamma^\mu\psi(x)
 \right\}-g^{\mu\nu}{\cal L}
 \label{T_spinor}
 \ee

 For simplicity we work here with the short form (\ref{lagr_spinor_simpl}). The corresponding energy-momentum tensor $T^{\mu\nu}$ is

 \be
 T^{\mu\nu}(x)=\bar{\psi}(x) \gamma^\mu\pd_\mu \psi(x) 
 \label{T_spinor-simpl}
 \ee
 and the 00-component of them, that we are interested in, is
 the corresponding hamiltonian density ${\cal H}(x)$, that is defined as

 \be
 {\cal H}(x)=T^{00}(x)=\bar{\psi}(x) \gamma^0\pd_0 \psi(x)
 \label{T00_spinor}
 \ee
 
 Let the solutions of the Dirac \eqq be $\psi^+$ and  $\psi^-$, which are
 normalized by

 \be
 \int d^3x\psi^{\dagger}(x)\psi(x) =1,
 \label{norm_cond_spinor}
 \ee
what will be no further discussed since the normalization of wave
functions does not play any role in the resulting \gse (the remark in the end of the previous subsection).

 Now use again the decomposition on the states with positive and negative frequences
 for each of solutions $\bar{\psi}(x),\psi(x) $  as in the scalar case above,  

\bea
\psi^{\pm}(x)=\fr{1}{(2\pi)^{3/2}}\int d^3k e^{\pm ikx} \sum\limits_{\nu=1,2}
a^{\pm}_{\nu}(\vec{k})v^{\nu\pm}(\vec{k})\\
\bar{\psi}^{\pm}(x)=\fr{1}{(2\pi)^{3/2}}\int d^3k e^{\pm ikx} \sum\limits_{\nu=1,2}
\bar{a}^{\pm}_{\nu}(\vec{k})\bar{v}^{\nu\pm}(\vec{k})
\eea
the index $\nu$ enumerates the two states with different spin projections,
$v^{\nu\pm}(\vec{k})$ and $\bar{v}^{\nu\pm}(\vec{k})$ are the 4-component
spinor amplitudes, spinor indices are omitted.

 The resulting hamiltonian density expressed through the amplitudes reads

\be
T^{00}=\int d^3k\ k_0 e^{\pm ikx} \sum\limits_{\nu=1,2}
[\bar{a}^+_{\nu}(\vec{k}) a^-_{\nu}(\vec{k})-\bar{a}^-_{\nu}(\vec{k}) a^+_{\nu}(\vec{k})]
\label{t00_spinor}
\ee

  The secondary quantized spinor field in the canonical formalism will be represented as

\bea
\psi(x)=\psi^+(x)+ \psi^-(x),\ \ \ \ \bar{\psi}(x)=\bar{\psi}^+(x)+ \bar{\psi}^-(x)
\eea
where the creation and annihilation operators $\bar{a}^+_{\nu}, {a}^-_{\nu}$
and $\bar{a}^-_{\nu}, {a}^+_{\nu}$ obey now the anticommutation relations in
order to have a positive definite form for the energy.

\bea
\{ \bar{a}^+_{\mu}(\vec{k}), {a}^-_{\nu}(\vec{k}')\}=\delta_{\mu\nu}\delta(\vec{k}-\vec{k}')\\
\{ \bar{a}^-_{\nu}(\vec{k}), {a}^+_{\nu}(\vec{k}')\}=\delta_{\mu\nu}\delta(\vec{k}-\vec{k}')\\
\{ \bar{a}^+_{\mu}(\vec{k}), {a}^+_{\nu}(\vec{k}')\}=\{
\bar{a}^-_{\mu}(\vec{k}), {a}^-_{\nu}(\vec{k}')\}=\\
\{ \bar{a}^+_{\mu}(\vec{k}), \bar{a}^-_{\nu}(\vec{k}')\}=\{
{a}^+_{\mu}(\vec{k}), {a}^-_{\nu}(\vec{k}')\}=0
  \label{anticomm_spinor}
\eea

 By applying the creation/annihilation operators ${a}^{\pm}$ and complex (but not
 Hermitian) conjugated $\bar{a}^{\pm}$ of them to the vacuum state
\bea
<0| \bar{a}^+ = {a}^-|0>=0\\
<0| {a}^+ = \bar{a}^-|0>=0,
\eea
one arrives at the vacuum expectation value for this hamiltonian

\be H= -2 \sum\limits_{(n)}\om_{(n)}.\ee

 For translational invariant problem the corresponding energy density per
 unit length is

\be {\cal E}= -4 \int \fr{dp_z}{2\pi}\ \half  \sum\limits_{(n)}\om_{(n)}
\label{e_dens_spinor}
\ee


 Thus the hamilonian density, that involves the summation over 2 states with
 different sign of energy
 ( fermoin and anti-fermion energies ) distinguished by the sign of mass,
and over 2 spin projection states, compared with this one for the scalar
 field (\ref{e_dens} of the previous subsection) acquires the resulting
 multiplicative factor -4.  

 In the following sections the formulae (\ref{e_dens_spinor} and
 \ref{e_dens}) will be used as a starting point for calculations of the \gse.

 \section{The \eqqs of motion and solutions}
\label{hamiltonians}
\ \ \  We consider basically the field system which consists of  
 quantum fields $\phi$ (fluctuations) interacting with
 classical fields $\Phi_i$ of fixed geometrical configurations (backgrounds).
 
 The corresponding lagrangian is constructed as
 \be
 {\cal L}={\cal L}_{\Phi_i}+ {\cal L}_{\phi}+ {\cal L}_{int},
\label{total_lagrangian}
 \ee
 where ${\cal L}_{int}$ is an interaction between $\phi$ and $\Phi_i$. The
 background $\Phi_i$ must not necessary satisfy any field
 equations. This simplified approach is somewhat physically incomplete but
 it allows to reduce the problem to solutions for quantum fluctuations only,
 that are supposed to obey the corresponding equations of motion with the
 background field as parameter. Generally speaking, the introduction of an
 external background field in a quantum field model is not trivial and not unique
(see for example \cite{skalozub}).

   A more consistent approach treats any field as consisting of a background
 part and fluctuations \cite{PvN}, \cite{dewitt}.
 
  The total energy density of the system is the sum of the classical and quantum parts
   \be
   {\cal E}={\cal E}^{cl}+{\cal E}^{\phi},
   \ee
where ${\cal E}^{cl}$ is the local energy density as a \f of coordinates of backgrounds ${\Phi_i}$ defined as the 00-component of the canonical energy-momentum tensor
 
  \be
{\cal E}^{cl}(x)=\sum\limits_i {\cal E}^{\Phi_i}(x) =\sum\limits_i T^{00}_{\Phi_i}(x),  \ee
 and the quantum part is the vacuum expectation value of the canonical
 hamiltonian of quantum field ${\phi}$. It is assumed that a considered
 problem is static and the vacuum state in the Fock space is defined in such
 a way that the constructed object is local:

 \be
  {\cal E}^{\phi}(x)={\cal E}^{\phi}_0= <0|{\cal H}(x)|0>.
\label{quant_energy_density}
  \ee

  The object defined in such a way is generally speaking non-local, because
  it depends on the background potential in the whole space. The integral of
  it over the space provides however a well defined global object, which we are
  even interested in, see below.
  The $ {\cal E}^{\phi}$ has been shown in Sec.(\ref{groundstenergy}) to be proportional to the sum over the spectrum of ${\cal H}$

\be
{\cal E}^{\phi}_0= \half\sum\limits_{(n)} \om_{(n)}.
\ee

Here the spectrum is defined as usual
\be
 {\cal H}\Phi_n(x)=\om_n^2 \Phi_n(x),
\ee
and is supposed here to be discrete for technical reasons.

 The energy ${\cal E}$ is treated as the density per unit volume. Since we
 are interested in the global energy $E$ of the field system, it should be
 integrated over the volume V,
 \be
 E=\int\limits_V d^3x\  {\cal E}(x). 
\ee

\subsection{The scalar field with the scalar background}

Consider first a simple model described by the lagrangian density

\bea
 &&{\cal L}
 =\half\{
 \partial_\mu\phi\partial^\mu\phi-m^2\phi^2+\partial_\mu\Phi\partial^\mu\Phi-
 M^2\Phi^2\}-V(x)\phi^2 +[\la \Phi^4]
 \nn\\ 
 &&V(x)= \alpha\Phi(x)^2,
\label{scalar_lagrangian}
 \eea
 where the field  $\phi$ is a quantum fluctuation interacting
 with a scalar background field $\Phi(x)$. The lagrangian can also contain
 the self-interaction term $\la \Phi^4$. It will be shown later that
 such a term is needed for the interpretation of a counter term
 arising from the quantum vacuum fluctuations $\phi$ and leading to a \rn of
 the self-interaction coupling constant $\la$.

 The corresponding classical energy density of the background and the vacuum energy
 can be obtained in a standard way and they are
 
\be
 {\cal E}^{cl}=\half\{ \partial_0\Phi\partial^0\Phi+(\nabla\Phi)^2+ M^2\Phi^2\}
\ee
\be
 {\cal E}^{\phi}=\half<0|\pd_0\pd^0+\nabla^2 +m^2+V(x)|0>.
\ee
The \eqq of motion for the fluctuations $\phi$ corresponding to the
(\ref{scalar_lagrangian}) is the Klein-Gordon \eqq:
\be (\pd_0\pd^0-\Delta+m^2+V(x))\phi(x)=0\ee

From now on we restrict ourselves to cylindrically symmetric potentials $V(x)=V(r)$

For the scalar field $\phi(x)$ in cylindrical coordinates $(\varphi,r,z)$ it is usual to make the ansatz

\be
\phi(x)=\phi(r)e^{-ip_0 x^0}e^{-il\varphi}e^{-ip_z z},
\ee
which leads to an ordinary second order \eqq for the radial part:

\be
\left\{ \fr{\pd^2}{\pd r^2}  + \fr{1}{r} \fr{\pd}{\pd r}
  +\left(k^2-V(r)-\fr{l^2}{r^2} \right)\right\} \phi(r)=0,
\label{scalar_\eqq}
\ee
with $k^2\equiv p_0^2-m^2$. 
This is an ordinary \eqq of the Bessel type, known to have an analytical solution
representable in terms of the Bessel functions.\\
\\
{\bf The rectangular shaped scalar potential}\\
 
Suppose the construction of the background potential looks as follows:
 \be
V(r)=\left\{\begin{array}{cc} V_0=const & R_1 < r < R_2\\ 0 & otherwise  \end{array}\right. 
\label{scalar_potential}
\ee
Then (\ref{scalar_\eqq}) has the following solutions in the domains
restricted by the cylinders $r=R_1$ and $r=R_2$:

\bea
&&I.\ \ \ r<R_1\ \ \phi_{k,l}^I(r)=C^I J_l(kr)\nn\\
&&II.\ \ \ R_1<r<R_2\ \ \phi_{k,l}^{II}(r)=C^{II}_1 H^{(1)}_l(qr) + C^{II}_2  H^{(2)}_l(qr)\nn\\
&&III.\ \ \ r>R_2\ \ \phi_{k,l}^{III}(r)=C^{III}_1 H^{(1)}_l(qr) + C^{III}_2 H^{(2)}_l(qr)
\label{scalar_solutions}
\eea

Here $q=\sqrt{k^2-V_0}$, $C_{i}$ are some normalization coefficients. If the
solution $\phi_{k,l}^I(r)$ is chosen to be regular in $r\ra 0$, then the
coefficients $C^{III}_2$ and $C^{III}_1$ are the corresponding \jf $f_l(k)$
and its complex conjugated, $\bar{ f}_l(k)$, (which will be discussed later,
see Sec.\ref{jostfunc}).

\subsection{The Dirac spinor field with the magnetic background}

The next model to consider is the spinor electrodynamics:
\be
 {\cal L}=-\fr{1}{4}F_{\mu\nu}F^{\mu\nu}+\bar{\psi}[i\gamma^\mu(\pd_\mu-ieA_\mu)-m]\psi,
\ee
where the $A_\mu$ is the classical electromagnetic background,
\be
 F_{\mu\nu}=\pd_\mu A_\nu -\pd_\nu A_\mu, 
\ee as usual. We consider only
a static problem. The background cannot depend on time and must be pure
magnetic, the electric field is excluded, ${\bf E}=0$.

 Like the scalar case considered above we derive the corresponding classical
 energy of the background and the vacuum energy of the spinor quantum fluctuations.
\be
 {\cal E}^{cl}=\fr{1}{4} F_{\mu\nu}F^{\mu\nu}+F^{0\nu}F_\nu^0=\half {\bf B}^2,
\ee
where the ${\bf B}$ is the magnetic field, 

\be
 {\cal E}^{\psi}=<0|\gamma^0\pd_0\{\vec{\gamma}(-i\vec{\nabla}-e\vec{A})+m\}|0>
\ee

Let us chose the potential $ A_{\mu}$ to be of a rectangular shape (like the scalar one discussed above):

\begin{equation} 
\vec{A}= \fr{\Phi}{2 \pi} \fr{a(r)}{r}\vec{e_\vp},
\label{potential_A}
\end{equation}
so that it  contains only the $A_\vp$ component and possesses a cylindrical
 symmetry again. The radial part profile \f $a(r)$ is taken to be\\
\hspace*{1cm}
\begin{equation}
a(r)= \left\{\begin{array}{c} 0, \hfill r< R_1, R_2\\
 \fr{1}{\kappa}(\fr{r^2}{R_1^2}-1)\ R_1\le r \le
 R_2 \\ 1, \hspace{\fill} r > R_1, R_2\end{array}\right. \\
\label{a_r1}
\end{equation}
Then the profile function for the magnetic field ${\bf B}(r)=\fr{\Phi}{2 \pi}
 h(r)\vec{e}_z$ is\\
 \hspace*{2cm}
 \begin{equation}
\ \ h(r)=\fr{1}{r}\fr{\pd}{\pd r}a(r)=\left\{\begin{array}{c} 0, \hfill r< R_1, R_2\\\fr{2}{\kappa R^2_1} \hspace{\fill} R_1\le r \le
 R_2 \\ 0, \hspace{\fill} r > R_1, R_2\end{array}\right.  
\label{h_r1}
\end{equation}
with $\kappa= \fr{R_2^2-R_1^2}{R_1^2}$. The shape of the
background can be interpreted geometrically as an infinitely long flux
tube empty inside.

We proceed now with the Dirac \eqq for spinor field $\psi(x)$ of mass $m$ and charge
$e$
 \begin{equation}
 \left\{i \gamma^{\mu} \fr{\pd}{\pd x^{\mu} }-m-e \gamma^{\mu}
 A_{\mu}\right\}\psi =0 
\label{dirac_eq}
\end{equation}
where $A_{\mu}$ is given by (\ref{potential_A}).
The gamma matrices in our representation are chosen as in \cite{borvor, bensant}: 
\bea
\gamma^0=\left\{\begin{array}{cc} \sigma_3 & 0\\0 & -\sigma_3
 \end{array}\right\} \ \ \ \ \ \gamma^1=\left\{\begin{array}{cc} i\sigma_2 & 0\\0 & -i\sigma_2 \end{array}\right\} \nn\\
\gamma^2=\left\{\begin{array}{cc} -i\sigma_1 & 0\\0 & i\sigma_1
 \end{array}\right\} \ \ \ \ \ \gamma^3=\left\{\begin{array}{cc} 0\  &\ 
 I\\-I\  &\  0 \end{array}\right\},
\label{gamma_matrix}
\eea
$\sigma_j$ are the $2\times 2$ Pauli matrices, $I$ is the $2\times 2$ unit matrix. 

Now we follow the standard procedure and separate the variables.
Using the ansatz
\begin{equation}
\phi(r, \varphi, z) = e^{-ip_0x^0} e^{-ip_z z}\Psi(r, \vp)\\
\end{equation}
 we obtain the equation for the 4- component spinor
$ \Psi(r, \vp)=\left[ \begin{array}{c} \phi(r, \vp) \\ \chi(r, \vp) \end{array}\right]$ 
 \begin{equation}
\left\{\begin{array}{cc} p_0+\hat{L}-m\sigma_3 & p_z I\\ \\p_z I
    & p_0+\hat{L}+m\sigma_3 
 \end{array}\right\}\Psi(r, \vp)=0,
\label{\eqq_Psi}
\end{equation}
where $\hat{L}=i\sum\limits_{i=1}^{2}\sigma_i(\pd_i+ieA_i)$, 
$\phi, \chi $ are the two-component spinors.

Respecting the translational invariance of the system it is convenient
to solve the equations only for $p_z=0$. Then the system (\ref{\eqq_Psi})
decays in two decoupled equations, for  $\phi$ and $\chi$ respectively, which are
distinguished only by sign of mass $m$. 

We consider the first of these two equations. By meaning of the standard ansatz
\begin{equation}
\phi(r, \vp) =\left[ \begin{array}{c} i\psi^u(r) e^{-i(l+1)\vp} \\ \\ \psi^l(r)
 e^{-il\vp} \end{array}\right],
\label{ansatz_phi}
\end{equation}
( $l $ is the orbital quantum number), it reads:

\begin{equation}
\left\{\begin{array}{cc} p_0-m & \fr{\pd}{\pd r}-\fr{l-\Omega
 a(r)}{r}\\ \\-\fr{\pd}{\pd r}-\fr{l+1-\Omega a(r)}{r} & p_0+m 
 \end{array}\right\}\Phi(r)=0
\label{dirac_of_r}
\end{equation}

Here the radial part $\Phi(r)$ denotes
\begin{equation}
\Phi(r)=\left\{\begin{array}{c} \psi^{u}(r)\\ \\ \psi^{l}(r)  
\end{array}\right\} 
\end{equation}
and
\begin{equation}
\Omega=\fr{e\Phi}{2\pi}.
\label{delta}
\end{equation}

In order to use later the symmetry properties of the Jost
function we redefine the parameter $l$ in (\ref{ansatz_phi}) (orbital number)
as $\nu$ according to
\begin{equation}
\nu=\left\{ \begin{array}{c} l+\half\ for\ \ l=0, 1, 2, . . . \\ \\
 -l-\half\ for\ \ l=-1, -2, . . . \end{array}\right.
\label{redef_l_to_nu}
\end{equation}

For the given construction of potential (\ref{a_r1})
we get three equations and three types of solutions for 3 areas of space
respectively:\\
\\

Domain I: $ r<R_1 $. The free wave equation (the magnetic flux is zero), 
the solutions are the Bessel functions $J_{\nu\pm\half}(kr)$ 
\begin{eqnarray}
\Phi_{I}^{-}(r)= \left[\begin{array}{c} \psi^{u-}_{I}(r)\nn\\ \nn\\\psi^{l-}_{I}(r) \end{array} \right]
 ={ \fr{1}{\sqrt{p_0-m}}\left[
\begin{array}{c}-\sqrt{p_0+m}J_{\nu-\half}(k r)\nn\\ \nn\\ \sqrt{p_0-m}J_{\nu+\half}(k r)\end{array} \right]},
\nn\\
\Phi_{I}^{+}(r)= \left[\begin{array}{c} \psi^{u+}_{I}(r)\nn\\ \nn\\\psi^{l+}_{I}(r) \end{array} \right]
={\fr{1}{\sqrt{p_0-m}}\left[
\begin{array}{c}\sqrt{p_0+m}J_{\nu+\half}(k r)\nn\\ \nn\\
 \sqrt{p_0-m}J_{\nu-\half}(k r)\end{array} \right]}
\end{eqnarray}
\label{domain_I}
\hspace*{3cm}$ p_0=\sqrt{m^2+k^2} $\\
 This solution of equation (\ref{dirac_of_r}) in the domain I is chosen
 to be the so called regular solution which is defined as to coincide for
 $r\rightarrow0$ with the free solution.\\ 
\\

 Domain II: $R_1<r<R_2$ The equation with a homogeneous magnetic field $\fr{\Omega}{e}h(r)=\fr{\Omega}{e}\fr{2}{\kappa R^2_1}$
has the solution

\begin{eqnarray}
\Phi_{II}^{-}(r)=
C_r^{-}\left[\begin{array}{c} \psi^{u-}_{II. r}(r)\nn\\
 \nn\\\psi^{l-}_{II. r}(r) \end{array} \right] +
C_i^{-}\left[\begin{array}{c} \psi^{u-}_{II. i}(r)\nn\\
 \nn\\\psi^{l-}_{II. i}(r) \end{array} \right]=\nn\\ \nn\\
\nn\\
C_r^{-} \left[
 \begin{array}{c}
 \fr{(p_0+m)}{2(\tilde{\alpha}+1)}r^{\tilde{\alpha}+1} e^{-\fr{\beta
 r^2}{4} } {_1}F_1(1-\fr{k^2}{2 \beta}, \tilde{\alpha}+2 ;\fr{\beta r^2}{2})
 \nn\\ \nn\\ \hfill r^{\tilde{\alpha}} e^{-\fr{\beta
 r^2}{4}} {_1}F_1(-\fr{k^2}{2 \beta}, \tilde{\alpha}+1 ;\fr{\beta r^2}{2} )
 \end{array} 
 \right]\nn\\
 \nn\\ \\
 + C_i^{-} \left[
 \begin{array}{c}
 \fr{1}{p_0-m}\fr{2 \tilde{\alpha}}{r}(\fr{\beta
 r^2}{2})^{-\tilde{\alpha}} e^{-\fr{\beta r^2}{4}}{_1}F_1(-\fr{k^2}{2
 \beta}-\tilde{\alpha}, -\tilde{\alpha} ;\fr{\beta r^2}{2})\nn\\
 \nn\\ (\fr{\beta
 r^2}{2})^{-\tilde{\alpha}} e^{-\fr{\beta r^2}{4}} {_1}F_1(-\fr{k^2}{2
 \beta}-\tilde{\alpha}, 1-\tilde{\alpha} ;\fr{\beta r^2}{2})
 \end{array} 
 \right].
 \label{domain_II}
\end{eqnarray}

$ \Phi_{II}^{+}(r) $ is the same, but $\tilde{\alpha}$ replaced by $\alpha$\\
here: $\alpha=\nu-\half+\fr{\Omega}{\kappa}$ for $ l\ge 0$\\
\hspace*{2cm}$\tilde{\alpha}=\fr{\Omega}{\kappa}-\nu-\half $ for $ l< 0$\\
\hspace*{2cm}$\beta=\fr{2 \Omega}{\kappa R_1^2}$\\
${_1}F_1$ is a confluent hypergeometric function \cite{abrsteg}, 
\cite{gradst}. 

The coefficients $C_i , C_r$ are some constants which will be unimportant for expressing the Jost function. The indices $u$ and $l$
denote ``upper'' and ``lower'' components of spinor respectively. The
lower index ``i'' or ``r'' corresponds to ``regular'' or ``irregular''
part of solutions dependent on the behaviour of the function if
continued to $r=0$. If the external background vanishes ($ \Omega
\rightarrow 0 $), contributions of irregular parts to the solution
disappear.\\
\\

Domain III: $r>R_2$ The free wave equation outside of the magnetic flux has
the solutions
\begin{eqnarray}
 \Phi_{III}^{-}(r)&=&
\half \bar{f}^{-}_{\nu}(k)  \fr{1}{\sqrt{p_0-m}}\left[\begin{array}{c} \psi^{u-}_{III. r}(r)\nn\\ \nn\\\psi^{l-}_{III. r}(r) \end{array} \right]
+\half f^{-}_{\nu}(k)\fr{1}{\sqrt{p_0-m}}\left[\begin{array}{c} \psi^{u-}_{III. i}(r)\nn\\ \nn\\\psi^{l-}_{III. i}(r) \end{array} \right]
 \nn\\
 &=&\half \bar{f}^{-}_{\nu}(k)  \left[
\begin{array}{r}-\eta H^{(1)}_{\nu-\half+\Omega}(k r)\nn\\ \nn\\ H^{(1)}_{\nu+\half+\Omega}(k
 r)\end{array} \right]+\half f^{-}_{\nu}(k) \left[
\begin{array}{r}-\eta H^{(2)}_{\nu-\half+\Omega}(k r)\nn\\ \nn\\  H^{(2)}_{\nu+\half+\Omega}(k
 r)\end{array} \right],\nn\\
\nn
\end{eqnarray}
\\
\begin{eqnarray}
 \Phi_{III}^{+}(r)&=& \half \bar{f}^{+}_{\nu}(k)\fr{1}{\sqrt{p_0-m}} \left[\begin{array}{c} \psi^{u+}_{III. r}(r)\nn\\
 \nn\\\psi^{l+}_{III. r}(r) \end{array} \right]
+\half f^{+}_{\nu}(k)\fr{1}{\sqrt{p_0-m}} \left[\begin{array}{c} \psi^{u+}_{III. i}(r)\nn\\ \nn\\\psi^{l+}_{III. i}(r) \end{array} \right]
 \nn\\
 &=& \half \bar{f}^{+}_{\nu}(k) \left[
\begin{array}{r}\eta H^{(1)}_{\nu+\half-\Omega}(k r)\nn\\ \nn\\ H^{(1)}_{\nu-\half-\Omega}(k
 r)\end{array} \right]+\half f^{+}_{\nu}(k) \left[
\begin{array}{r}\eta H^{(2)}_{\nu+\half-\Omega}(k r)\nn\\ \nn\\ H^{(2)}_{\nu-\half-\Omega}(k
 r)\end{array} \right],\nn\\ \nn\\
\eta &=&\fr{\sqrt{p_0+m}}{\sqrt{p_0-m}}\nn
\label{domain_III}
\end{eqnarray}
and $f_{\nu}(k)$ is the corresponding Jost function in accordance with the
definition Sec.\ref{jostfunc}, (\ref{jost_func_def}).

\subsection{The Dirac spinor field with the abelian Higgs background}

 Let us consider the abelian Higgs model with the Yukawa type coupling.
 This model is known to be the simplest model with spontaneously
  broken symmetry, where the initially massless fermion field obtains
  a mass through the Higgs mechanism during the U(1) symmetry breaking.
It is interesting since it is also the simplest model possessing
the solution of Nielsen-Olesen (NO) vortex type. This vortex will be treated in the
 considered context as a certain classical background.

  The complete lagrangian of the model reads
\be \label{higgs_lagrangian} 
{\cal L}= {\cal L_{\phi A}}+{\cal L_{\psi}}+{\cal L_{\phi A \psi }}
\ee
where
\bea
{\cal L_{\phi A}}&=&\half (D_\mu \Phi)^\dagger D^\mu \Phi-\fr{1}{4} F_{\mu\nu}
F^{\mu\nu}-\half\lambda ( |\Phi|^2-\fr{\eta^2}{2})^2,
\\
\nn\\
{\cal L_{\psi}}&=&i\ \bar{\Psi}\gamma^\mu \pd_\mu \Psi,
\\
\nn\\
{\cal L_{\phi A \psi }}&=& -\bar{\Psi} (f_e |\Phi|+e \gamma^\mu A_\mu)\Psi
\label{lagr_with_interaction}.
\eea
 

The covariant derivative is defined by $D_\mu =
\pd_\mu-ieA_\mu$, $A_\mu=A_\mu(x)$ -the gauge field (magnetic field),
$\Phi$ is a complex scalar field (Higgs bosonic field ).


 The interaction lagrangian in (\ref{higgs_lagrangian}) is chosen
 in this form to simplify the \eqqs of motion for the spinor quantum
 fluctuations $\Psi$ It will be discussed below in Sec.(\ref{NOinteraction}).

 The classical energy of this scalar-electromagnetic background is in general

\be
  E^{cl}=\int d^2x\left[ |D_0\Phi|^2+|D_i\Phi|^2
  +\la\left(\Phi^\dagger\Phi-\fr{\eta^2}{2}  \right)^2 +\half[{\bf E}^2+{\bf H}^2] \right],
\ee
where the ${\bf E},{\bf H}$ are components of tensor $F_{\mu\nu}$
  corresponding to classical electric and magnetic field respectively. 

 Since we consider this field system on a particular background, namely the NO vortex, it will be useful for further
 calculations to rewrite $E^{cl}$ right now in the terms of the Nielsen-Olesen solution

\be
\Phi=\fr{\eta}{\sqrt{2}} f(r)e^{i n\vp},\ \ qA_{\vp}=nv(r),\ \ A_\rho=A_t=A_z=0,
\ee
that provides:
\bea
\la\left(\Phi^\dagger\Phi-\fr{\eta^2}{2}
\right)^2&=&\fr{\la\eta^4}{4}(f(r)^2-1)^2\nn\\
{\bf E}^2&=&0\nn\\
{\bf H}^2&=&\fr{n^2}{q^2r^2}v'(r)^2,
\label{parts_of_E}
\eea
and finally we arrive at
\be
 {E}^{cl}=\pi\int r\ dr\left\{\fr{1}{q^2r^2}v'(r)^2+\eta^2
 [f'(r)^2+\fr{f(r)^2}{r^2}(1-v(r))^2 ]+ \fr{\la\eta^4}{2}(f(r)^2-1)^2
 \right\}.
 \label{class_energy}
\ee

 The \gse of the spinor quantum fluctuations is derived
 similar to the previous case, as the vacuum expectation of the hamiltonian

\be
 {\cal E}^{\phi}=<0|\gamma^0\{\vec{\gamma}(-i\vec{\nabla}-e\vec{A})+\fr{\eta}{\sqrt{2}}f(r)^2\}|0>,
\ee
and the equation of motion for the spinor field $\Psi(x)$ can be obtained
also in a similar way for the radial component $\Psi(r)$.

\bea
\left\{\begin{array}{cc} p_0-\fr{f_e\eta}{\sqrt{2}}f(r) & \fr{\pd}{\pd r}-\fr{l-n
 v(r)}{r}\\ \\-\fr{\pd}{\pd r}-\fr{l+1-n v(r)}{r} & p_0+\fr{f_e\eta}{\sqrt{2}}f(r)
 \end{array}\right\}\Psi(r)=0.
\label{no_ansatz}
\eea

Comparing with the case (\ref{dirac_of_r}) we see that
\be \fr{f_e\eta}{\sqrt{2}}f(r) \e \mu(r)\label{defofMu}\ee plays the role of space depending mass for the
fermionic field.


 \subsubsection{Nielsen-Olesen vortex}

 The Nielsen-Olesen vortex is a soliton stabilized by a topological charge.
 Here we describe briefly the derivation of the NO \eqqs, following \cite{acc_vach}.
 After that SU(2) invariance of the electroweak model is broken, we have a
 Lagrangian possessing the relic $U(1)$ -symmetry, i.e. the invariance under
 $U(1)$ gauge transformations
 \be
\Phi(x)\ra e^{iq\chi(x)}\Phi(x),\ \ A_{\mu}(x)\ra A_{\mu}(x)+\pd_{\mu}\chi(x). 
 \label{u1_trafo}
 \ee

  We start with the $U(1)$ -invariant Lagrangian
\be
{\cal L}=\int d^4 x\left[
  |D_{\mu}\Phi|^2-\lambda\left(\Phi^{\dagger}\Phi-\fr{\eta^2}{2} \right)^2
-\fr{1}{4} Y_{\mu\nu} Y^{\mu\nu}\right]
\label{origin_ew_lagr}
\ee
 where $D_{\mu}$ is the covarinat derivative $D_{\mu}=\pd_{\mu}-i q A_{\mu}$,
  and the tensor $Y^{\mu\nu}$ is like an electromagnetic field strength
  $Y_{\mu\nu}=\pd_{\mu}A_{\nu}(x)-\pd_{\nu}A_{\mu}(x)$. By applying the
  Euler-Lagrange formalism as usually we obtain the equations of motion for
  both fields $\Phi(x)$ and $A_{\mu}(x)$
  \be
  D_{\mu}D^{\mu}\Phi(x)+2 \lambda\left(\Phi^{\dagger}(x)\Phi(x)-\fr{\eta^2}{2} \right)\Phi(x)=0,
  \label{eom_phi}
  \ee
and
 
\be
\pd^{\mu}Y_{\mu\nu}(x)=-iq \Phi^{\dagger}(x) \stackrel{\leftrightarrow}{D_{\nu}}\Phi(x)
  \label{eom_y}
\ee

The Nielsen-Olesen ansatz \cite{nieloles} is introduced in cylindrical
coordinates $(\tau,r,\varphi,z)$ according to (\ref{no_ansatz}),
where $v(r), f(r)$ have to satisfy the boundary conditions
\be
f(0)=v(0)=0,\ \  f(r),v(r)\ra 1\ \ as\ \ r\ra\infty
\label{bound_cond}
\ee
 Substitution of the ansatz (\ref{no_ansatz}) into the
( \ref{eom_phi},\ref{eom_y}) leads to a system of two second order equation with respect to functions $v(r),f(r)$:

\bea
f''(r)+\fr{f'(r)}{r}-n^2\fr{f(r)}{r^2}[1-v(r)]^2+\lambda\eta^2(1-f(r)^2)f(r)=0,\nn\\
v''(r)-\fr{v'(r)}{r}+q^2\eta^2 f(r)^2[1-v(r)]=0,
\label{no_vortex_eq}
\eea
 which are in general not to be solved analytically, but it is possible to
 obtain several numerical evaluations for $v(r)$ and $f(r)$.

  If we are only interested in the shape of the solutions $v(r), f(r)$, it is
  convenient to introduce a re-scaled variable $r$,
 $r=\rho/(q\eta)$, so that dependence on parameters in \eqqs (\ref{no_vortex_eq})
 can be reduced entirely to one parameter  $\beta=\fr{2\lambda}{q^2}$.
  Now, since the parameter $\beta$ is specified, the corresponding shapes can
 be reconstructed using the boundary conditions (\ref{bound_cond}) and the
 asymptotic behaviour of solutions $v(r), f(r)$ at $r\ra 0$.
 We put
 
 \be
 f(r)=a r +\dots\ \ \ \ v(r)=b r^2 +\dots,
 \label{asymp_r0}
 \ee
 where $a$ and $b$ are some constants, which are fixed by the
 asymptotic condition (\ref{bound_cond}) at $r\ra \infty$.
 
\begin{figure}[t]  \label{NOfig1}
\epsfxsize=14cm
\epsffile{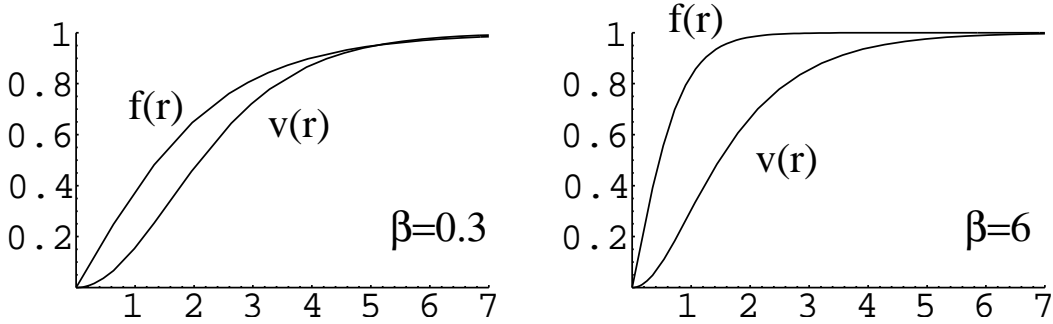}
\caption{The profile functions of the Nielsen-Olesen vortex as
functions of the radius $r$ for $q\eta=1$.}
\end{figure}

 Two examples are shown in Fig.\ref{NOfig1} for $q\eta=1$. The corresponding
values of the constants a,b are $(a,b)=(0.26817,0.17481)$ for $\beta=0.3$ and
$(a,b)=(0.79958,0.42848)$ for $\beta=6$. As the asymptotic values for
$r\to\infty$ are reached with exponential speed it is in fact
sufficient to consider $r$ up to $\approx 15$.
%


\subsubsection{Interaction with the spinor field}
\label{NOinteraction}

  The interaction term ${\cal L_{\phi A \psi }}$ in the form
 (\ref{lagr_with_interaction}) leads to the \eqq of motion for spinor field
 like the ordinary Dirac \eqq with magnetic background, that allows to save a
 lot of analytical calculations by using the results of the solved problem of
 magnetic string, discussed above.
  The interaction (\ref{lagr_with_interaction}) possesses the U(1) gauge
 invariance and the coupling constant $f_e$ is dimensionless, what should
serve to keep the renormalizability of the model.  
 However, the latter is violated through the interaction of the spinor with
 the absolute value of the scalar background $\Phi$, that is originally
 complex, thus the term
\be
   f_e |\Phi| = \sqrt{(\Im \Phi)^2+ (\Re\Phi)^2 }\ee has a non
 polynomial structure.

  As a consequence, the \rn of the one-loop vacuum energy
  requires the introduction of an additional counter-term, as
  considered below, Sec.\ref{renormalize}

  The correct renormalizable form of the interaction lagrangian was supposed,
  e.g. in \cite{vassilvort, rebhan} for N=2 supersymmetric extension of this 
 model.

 Variation of this lagrangian (\ref{lagr_with_interaction}) with respect to the spinor $\Psi$ leads to the
equation
of motion:
\be
\left\{i \gamma^\mu D_\mu -f_e |\Phi(x)|\right\}\Psi(x)=0.
\label{eom_spinor}
\ee

One can see that the scalar
background $\Phi(x)$ plays the role of the variable space-dependent fermionic
mass. It suggests that this problem can be treated in much the same way as
that one for the massive spinor field in magnetic background.


 \section{The zeta functional \rg of the \gse}
\label{zetafunc}
\ \ \ The \gse of the quantum field derived in
Sec.(\ref{groundstenergy}, \ref{hamiltonians}) as the spatial integral over
the infinite volume of the sum over the spectrum of hamiltonian is not well
defined mathematically, since the spectral sum is not necessary convergent
and the spatial integral as well. The one loop vacuum energy is
known to contain \uv divergences.
Even in all further applications the energy
defined as (\ref{e_dens}, \ref{e_dens_spinor}) is divergent.

Quantum field theory operates with a
variety of \rg procedures to treat infinite objects reasonably. We use the
so called zeta-functional regularization.

This procedure is based on the representation of divergent sums and series
through the \zf
and is described in the present section below.

\subsection{The Riemann \zf}

\ \ \  The Riemann \zf (or simlpe "\zf") is the one of the central objects of
the number theory, defined as

 \be
  \zeta(s)=\sum\limits_{n=1}^{\infty} \fr{1}{n^s}.
  \label{riemann_zeta}
 \ee
 It was introduced by Riemann in 1859. It is known that for real $s$ the series (\ref{riemann_zeta}) converges for
  $s>1$.
  
 The next important step to proceed with the \zf is to gain an integral representation of the
 infinite sum (\ref{riemann_zeta}). We perform the standard representation of
 the Euler Gamma-\f in the way
 
 \be
 \Gamma(s)=\int\limits_0^\infty dt\ t^{s-1} e^{-t} = n^s \int\limits_0^\infty dt
\ t^{s-1} e^{-nt}.
 \label{gamma_performed}
 \ee

Then in the product

\be
\zeta(s)\Gamma(s)=\sum\limits_{n=1}^{\infty}\int\limits_0^\infty\ dt
 t^{s-1} e^{-nt},
 \label{zeta_gamma}
\ee
the summation over $n$ and integration over $t$ can be interchanged under
assumption that $s>1$, that admits the resummation of $e^{-nt}$ as a simple
geometrical series, and finally we obtain the desired integral representation

\be
\zeta(s)=\fr{1}{\Gamma(s)} \int\limits_0^\infty\ dt \fr{t^{s-1}}{e^t-1}.
\label{zeta_int_rep}
\ee

 We can proceed from this expression with the analytical continuation of
$\zeta(s)$ in the complex plane. It allows i.e. to establish a number of important
properties of $\zeta(s)$ as a \f of complex variable $s$. Here are
referred to some useful ones.

\begin{itemize}
\item 
The \zf is an analytical one-valued everywhere in the complex plane excepted
the point $s=1$ where it has a simple pole with residue 1.

The corresponding Laurent expansion around the  $s=1$ reads :

\be
\zeta(s)=\fr{1}{s-1}+\gamma_0+\gamma_1(s-1)+\gamma_2(s-1)^2+...\ ,
\label{laurent_series}
\ee
where the irrational coefficients $\gamma_k$ are to calculate as 

\be
\gamma_k=\stackrel{lim}{n\ra\infty}\left[ \sum\limits_{\nu=1}^\infty \fr{(\ln
    \nu)^k}{\nu} - \fr{1}{k+1}(\ln n)^{k+1} \right],
\ee
and the $\gamma_0=\gamma$ is the Euler-Mascheroni constant;

  
\item
 For the $s=0$ we have
  \be
  \zeta(0)=-\half;
  \ee
  
\item
  The real zeros of the \zf are located at negative even integers

  \be
  \zeta(-2n)=0;
  \ee
\item
  Furthermore the \zf has infinitely many complex zeros in the strip $0< \Re s
  <1 $ but  no one outside of it.
  
\end{itemize} 

 The one of the most important generalizations of the Riemann \zf is the
 spectral \zf of some operator ${\cal P}$. This kind of spectral \f (the \f
 defined on the spectrum of operator) is known in mathematics since 1949
 \cite{minakshi}.

 It is defined as
 \be
 \zeta_{\cal{P}}(s)=\Tr {\cal P}^{-s}
 \label{zeta_of_P}
 \ee

 In particular if the spectrum ${\la_n}$ of ${\cal{P}}$ is discrete, the
 (\ref{zeta_trace}) means a sum over the spectrum ${\la_n}$.

 \be
 \zeta_{\cal{P}}(s)= \Tr {\cal P}^{-s}=\sum\limits_n (\la_n)^{-s}
 \label{zeta_trace_sum}
 \ee

  For example, the problem of a linear oscillator or a charged particle in
  homogeneous magnetic field in quantum mechanics \cite{landau} leads to the
  spectral problem of the operator
  \be 
  {\cal V}_z= -\half  \fr{\pd^2}{\pd z^2}+z \fr{\pd}{\pd z} 
  \ee
 on the space of bounded \fs $H(z)$. The solutions of this problem are the
  Hermite polynomials $H_n(z)$, and the corresponding spectrum $\la_n$ is
  $n=0,1,2,3\dots$. If we redefine ${\cal V}_z$ as follows:
  \be {\cal V}_z^+={\cal V}_z+1 \ee to exclude $n=0$ from the spectrum,
   the corresponding \zf
$\zeta_{V_z^+}(s)$ will be even the Riemann \zf
  (\ref{riemann_zeta}). Obviously there are more different spectral problems
  possessing the same spectrum and the same spectral functions, as e.g. \zf
  respectively (so called isospectral problems).
  
 Analogous to the Riemann \zf, the generalized \zf  (considered below, \ref{zeta_trace}) can
be analytically extended on the complex plane $s$ as a meromorphic function
possessing simple poles \cite{seeley}. In particular $\zeta_{\cal{P}}(s)$ is
regular at $s=0$ if ${\cal{P}}$ is of Laplace type.

 One can define also a determinant of the operator formally as

 \be
 \det {\cal P} = \prod\limits_n (\la_n).
 \label{det_of_P}
 \ee
 
 It can be represented through the \zf of the operator ${\cal P}$ in the
 following way

 \be
 \exp\left\{ -\fr{d}{ds}\zeta_{\cal P}(s) \right\}= \prod\limits_n (\la_n)^{(\la_n)^{-s}}.
 \label{exp_zeta}
 \ee
 
For $s=0$ it gives

\be
\left. \det {\cal P} =  \prod\limits_n (\la_n)=
\exp\left\{ -\fr{d}{ds}\zeta_{\cal P }(s) \right\}\right|_{s=0}.
\label{det_P}
\ee

\subsection{The generalized local \zf }

\ \ \ One can also extend the definition of \zf corresponding to the spectrum of
 operator $ {\bf\hat{ P}}$ and define a bilocal \f depending on coordinates
 of two points $({\bf x}, {\bf y})$.

 If we have a typical spectral problem for eigenvalues $\la_{(n)}$ and
 \ef $\phi_{(n)}({\bf x})$ of the positive defined elliptic
 differential operator  $ {\bf\hat{ P}}$:
 \be
 {\bf\hat{ P}}\phi_{(n)}({\bf x}) = \la_{(n)} \phi_{(n)}({\bf x}),
\label{spectral_problem}
 \ee
 where the \ef $\phi_{(n)}({\bf x})$ are the \fs of
 coordinates $({\bf x})$ on the manifold {\cal M} with boundary and
 obey the boundary conditions
\be
{\cal B}\phi_{(n)}({\bf x})|_{\pd{\cal M}}=0
 \ee
 ( ${\cal B}$ is some boundary operator),
 then we can define

%
%
%
%
%
%

\be \zeta_{\bf\hat{ P}}({\bf x,y}|s)=\sum\limits_{(n)} \fr{ \phi_{(n)}({\bf x})  \phi_{(n)}^*({\bf y}) } {\la_{(n)}^s}.\label{local_zeta}\ee

 Thereby the original \zf is naturally constructed as a derived global object being the trace of the local two-point \f (\ref{local_zeta})
\be
\zeta_{\bf\hat{ P}}(s)=\Tr \zeta_{\bf\hat{ P}}(x,y|s),
\label{zeta_trace}
\ee
 that coincides with the definition (\ref{zeta_trace_sum}) (the trace means summation over all discrete indices and integration over
 continual variables $(x,x)$). It is assumed, that for the operator ${\bf\hat{
 P}}$ such a trace exists. For simple examples, where the spectrum of
 ${\bf\hat{P}}$ is known explicitely, we suppose the sum in
 (\ref{zeta_trace_sum}) to be convergent for $\Re s$ being large enough.
More precisely, this assumption is based on the Weyl's theorem, which
 presumes, that for a second-order elliptic differential operator on the
 $D$-dimensional flat euclidean manifold ${\cal M}$ the
 eingenvalues $\la_n$ behave asymptotically for $n\ra \infty$ as
 \be
 \la_n^{D/2}\sim \fr{2^{D-1}\pi^{D/2}D\ \Gamma(D/2) }{\mbox{vol} {\cal M}} n.
\label{weyl_theorem}
 \ee
(We work basically with this class of operators only). 
 In particular, this formula provides the convergence of the sum
 (\ref{zeta_trace_sum}) for $\Re s > D/2 $. \cite{kirsten}.
 
  Here it is also suitable to mention, that the local \zf possesses the
  following useful feature
 
\be
{\bf\hat{ P}} \zeta_{\bf\hat{ P}}({\bf x,y}|s)= \zeta_{\bf\hat{ P}}( {\bf x,y}|s-1),
\label{P_zeta_s-1}
\ee
that can be checked immediately and is useful when applied to problems concerned
with the \rn of Green \f. For $s=0$ one obtains from (\ref{P_zeta_s-1}) the
initial conditions

\be
\zeta_{\bf\hat{ P}}({\bf x,y}|s=0)=\delta({\bf x}-{\bf y}),
\label{P_zeta_s=0_delta}
\ee
where it can be seen explicit, that the local \zf for $s=0$ coincides with
the corresponding Green \f.


 \subsection{The heat kernel}
\label{heatkernel}

\ \ \ The \hk provides a number of tools to use within the \zf \rg
method. Mathematically this object is closely related to the generalized \zf
which is defined on the spectrum of some elliptic self-adjoined
operator.

One defines the simplest (global) \hk as a sum over the spectrum $\la_{(n)}$ of some operator ${\bf\hat{P}}$:

\be
K_{\bf\hat{P}}(t)=\sum\limits_{(n)}e^{-t\la_{(n)}}.
\label{global_heat_kernel}
\ee



We can also define the local heat kernel $K(x,y|t)$:

\be
K_{\bf\hat{P}}({\bf x}, {\bf y}|t)=\sum\limits_{(n)}\phi_{(n)}({\bf x})\phi_{(n)}^*({\bf y})e^{-t\la_{(n)}},
\label{local_heat_kernel}
\ee
where $\phi_{(n)}$ are the normalized \ef of the operator ${\bf\hat{P}}$.

The trace of $K_{\bf\hat{P}}({\bf x}, {\bf y}|t)$ (here trace means summation
over all discrete indices and integration over continuous variables $x$),

\be
K_{\bf\hat{P}}(t)=\Tr K_{\bf\hat{P}}({\bf x}, {\bf y}|t)
\label{heat_trace}
\ee
obviously gives the corresponding global object (\ref{global_heat_kernel}).
The kernel (\ref{local_heat_kernel}) satisfies the equation

\be
\left\{ \fr{\pd}{\pd t} +{\bf\hat{ P}}\right\}K_{\bf\hat{P}}({\bf x}, {\bf y}|t)=0
\label{dt_P_K=0}
\ee
with the initial condition at $t=0$:

\be
K({\bf x}, {\bf y}|t=0)=\delta({\bf x}- {\bf y}).
\label{K_t=0_delta}
\ee

 A natural example of the operator ${\bf\hat{ P}}$ is the minus Laplacian
 $-\Delta$. For this case the (\ref{dt_P_K=0}) becomes the usual heat \eqq

\be
\left\{ \fr{\pd}{\pd t}-\Delta \right\}K_{-\Delta}({\bf x}, {\bf y}|t)=0,
\label{heat_\eqq}
\ee\\
where the name "\hk" comes from. In the applications of mathematical physics ${\bf\hat{ P}}$ means usually the hamiltonian of field
configuration. Some suitable examples are considered in further subsections.

 Transition from the \hk $K_{\bf\hat{P}}({\bf x}, {\bf y}|t)$ to the
local two-point dependent \zf

\be
\zeta_{\bf\hat{ P}}({\bf x,y}|s)=\sum\limits_{(n)} \fr{\phi_{(n)}({\bf x})\phi_{(n)}({\bf y})}{\la_{(n)}^s}
\label{zeta_P_of phi}
\ee\\
can be made by using the standard representation of $\Gamma$-\f 

\be \Gamma(s)=\fr{1}{\la^s}\int\limits_0^\infty dt\ t^{s-1} e^{-t\la}. \ee

This \eqq together with (\ref{zeta_P_of phi}) yields:
\be
\zeta_{\bf\hat{ P}}({\bf x}, {\bf y}|s)=\int\limits_0^\infty dt \fr{t^{s-1}}{\Gamma(s)} K_{\bf\hat{ P}}({\bf x}, {\bf y}|t).
\label{zeta_of_K_Gamma}
\ee

 Taking the trace of both sides of this identity we obtain the connection between
global objects.
\be
\zeta_{\bf\hat{ P}}(s)=\int\limits_0^\infty dt \fr{t^{s-1}}{\Gamma(s)} K_{\bf\hat{ P}}(t).
\label{zeta_of_K_global}
\ee

 Another important spectral function also representable through the \hk is
the Green \f $G(x,y)$ of the operator ${\bf\hat{ P}}$,
that is defined to obey the \eqq

\be
{\bf\hat{ P}}G({\bf x,y})=\delta({\bf x-y})
\label{green_func_def}
\ee
(it will be considered in Sec.\ref{jostfunc} in context of the \jf
).
%

The \f $G(x,y)$ is usually constructed as
 
\be
G({\bf x,y})=\sum\limits_{(n)} \fr{\phi_{(n)}({\bf x}) \phi^*_{(n)}({\bf y})}{\la_{(n)}},
\ee
(that obviously satisfies the (\ref{green_func_def})).
Then representing $1/\la_{(n)}$ as the integral
\be \fr{1}{\la_{(n)}}= \int\limits_0^\infty dt\ e^{-t\la_{(n)}}, \ee we get

\be
G({\bf x}, {\bf y})= \int\limits_0^\infty dt\  K ({\bf x}, {\bf y}|t).
\label{greenf_by_K}
\ee

 In the historical overview it was the first physical application of the
 local \hk \cite{fock}, where the Green \f was proposed to be represented as
 integrals like (\ref{greenf_by_K}) of a kernel satisfying the heat \eqq
 over an auxiliary coordinate (``proper time``), introduced by Fock, 1937.

\ \ In appllications to the \gse it is important that the \hk has the power-law
asymptotics for small $t$. For all strongly ellilptic boundary value there is
an asymptotic expansion 

\be
K(x,y|t)\sim \fr{1}{(4\pi t)^{d/2}} \sum\limits_{n \ge 0} a_n(x,y)t^n,
\label{local_\hke}
\ee
and the asymptotics of the global kernel $K(t)$ (or heat trace) reads respectively

\be
K(t)= \Tr K(x,y|t) \sim \fr{1}{(4\pi t)^{d/2}} \sum\limits_{n \ge 0} a_n t^n
\ee as $t\ra 0$. The $d$ denotes the space dimension and the coefficients
$a_n$ are called the "\hkc". It will be shown below
that the zeta-regularized \gse for massive fields in the limit of large mass
$m\ra\infty$ (\ref{e_reg_zeta}, Sec.\ref{zetafunc}) can be expressed in terms of these coefficients.

\subsection{Heat kernel coefficients}

\ \  There are various methods to calculate the heat kernel coefficients $a_n$
One of this mostly applicable in physics is the consequent iteration of the equation for the local heat
kernel $K(x,y|t)$ (\ref{dt_P_K=0}) with an initial condition (\ref{K_t=0_delta})
using the ansatz with the free solution $K_0(x,y|t)$. This method goes
back to DeWitt \cite{dewitt} and is called in some sources as the ``DeWitt
iterative procedure''.

This procedure was originally developed for manifolds without boundaries
\cite{dewitt} with non zero curvature and can be generalized also for
manifolds with boundaries \cite{McAvity:hy,McAvity:we}.

The short description of it, following \cite{vassilreview} is concluded to
the recursive solution of the chain of \eqqs

\be
(j+ (D^{\mu}\sigma)D_{\mu})a_{2j}+ {\bf\hat{P}} a_{2j-1}=0 
\ee
that are derived by substituting the DeWitt ansatz 

\be K(x,y|t)= K^{(0)}(x,y|t) \sum\limits_{n \ge 0} a_{2n}(x,y| \hat{P} )t^n \ee
into the \eqq (\ref{dt_P_K=0}) with the initial condition
\be
a_0=1,
\ee
and $K^{(0)}(x,y|t)$ obeys the \eqq (\ref{dt_P_K=0}) as $t=0$.

\subsubsection{ Scalar quantum field in scalar background }


Let $\phi(x)$ be a quantum scalar field of mass $m$, interacting with the
 classical background potential $V(x)$. The operator ${\bf\hat{ P}}$ is a hamiltonian ${\bf{\cal H}}$ of this system

\be
  {\bf{\cal H}}=-\Delta+m^2+V(x).
  \label{scalar_hamiltonian}
\ee

The variable $x$ includes now all spatial components, $\Delta \e \pd^i\pd_i$.
 The \eqq (\ref{dt_P_K=0}) reads:

\be
\left\{  \fr{\pd}{\pd t}-\Delta+m^2+V(x)\right\}K(x,y|t) = 0.
\label{eq_heat_kernel_scalar}
\ee

The ansatz for the local heat kernel expansion must be chosen in the form \cite{dewitt}

\be
K(x,y|t)=K^{(0)}(x,y|t) \sum\limits_{n \ge 0} a_n(x,y)t^n;
\label{local_ansatz}
\ee
where $a_n(x,y)$ are the local heat kernel coefficients which depend on
two points $x$ and $y$. Here $K^{(0)}(x,y|t)$ is the zeroth order term
satisying the initial condition at $t=0$ and

for the operator (\ref{scalar_hamiltonian}) the $K^{(0)}(x,y|t)$ is chosen to be

 \be
K^{(0)}(x,y|t)=\fr{1}{(4 \pi t)^{d/2}} exp\left\{ -\fr{(x-y)^2}{4t}-tm^2 \right\}. 
\label{heat_free_sol}
\ee


 Considering the limit of $t\ra 0$ in \ref{heat_free_sol}
we assure that this ansatz

The kernel (\ref{heat_free_sol}) obeys also the initial condition (\ref{K_t=0_delta}). Now substituting the ansatz \ref{local_ansatz} into (\ref{eq_heat_kernel_scalar}) one obtains
 \bea
 \left[\fr{\pd}{\pd t}-\pd_\mu\pd^\mu+V(x)+m^2)\right] \fr{1}{(4 \pi
 t)^{d/2}} exp\left\{ -\fr{(x-y)^2}{4t}-tm^2 \right\}\sum\limits_n a_n t^n
 =0.\nn\\ 
 \eea

%

We equate the coefficient in front of each power of $t$ to zero. This gives
 the following chain of \eqqs

 \bea
&& (x-y)^i \nabla_i a_0(x,y)=0\nn\\
&& \Delta a_0(x,y) + V a_0(x,y) + a_1(x,y)+(x-y)^i \nabla_i a_1(x,y)=0\nn\\
&& \Delta a_1(x,y) + V a_1(x,y) + 2 a_2(x,y)+(x-y)^i \nabla_i a_2(x,y)=0\nn\\
\label{chain_of_\eqqs}
\eea
with the initial condition $a_0(x,x)=1$; $\nabla_i \e \fr{\pd}{\pd x^i}$.


Now let us apply the differentiation $\fr{\pd}{\pd x^k}$ on the first equation
of (\ref{chain_of_\eqqs}) we obtain, that
 \be
2 \pd^k\pd_k a_0(x,y)+(x-y)^i \pd_i \Delta a_0(x,y)=0. 
\label{delta_a0}
\ee


 If the coordinates x and y coincide, this equation gives $\Delta a_0(x,x)=0$

  A similar procedure applied to the second \eqq of (\ref{chain_of_\eqqs}) results in
  \bea
  \Delta^2 a_0(x,y)+2 \nabla a_0(x,y)\nabla V(x)+a_0(x,y)\Delta V(x)+
 \Delta a_0(x,y) V(x)\nn\\+3\Delta a_1(x,y)+(({\bf x}-{\bf y})\nabla) \Delta
  a_1(x,y)=0,
 \label{delta_a1} 
  \eea
where $\Delta a_0(x,y)$ is a solution of (\ref{delta_a0}) and $\Delta^2
  a_0(x,y)$ satisfies a similar \eqq
  
\be
\Delta^2 a_0(x,y)=-\fr{1}{4} (({\bf x}-{\bf y})\nabla) \Delta^2 a_0(x,y).
\label{delta_2_a0}
 \ee

 In the limit  $x=y$ ( \ref{delta_a1} ) simplifies

\be
\Delta a_1(x,y)|_{x=y}=-\fr{1}{3}\Delta V(x)
\label{delta_a1_simple} 
\ee
and
\be
a_1(x,y)|_{x=y}=-V(x)
\label{a1_scalar}
\ee
(the laplacian $\Delta$ acts on the $x$-variable)

Using these results one obtains from the third \eqq (\ref{chain_of_\eqqs}) for $a_2(x,x)$

\be
a_2(x)=a_2(x,x)=-\half\left(\Delta a_1(x,y)|_{x=y}+V(x) a_1(x,y)|_{x=y}\right).
\ee

Finally we get
\be
a_2(x)=a_2(x,x)=\fr{1}{6}\Delta V(x)+\half V(x)^2.
\label{a2_scalar}
\ee

Further coefficients $a_{3}, a_{4}, \dots $ can be calculated in a similar
way \cite{gilkey,avram1,avram2}.

\subsubsection{Scalar field in constant vector  background }

 Following the procedure described in the previous section, we
 construct a chain of recursive \eqqs for $a_i(x,y)$.

 It starts as previously with the same ansatz (\ref{local_ansatz}) into
 the equation (\ref{dt_P_K=0}). The operator ${\bf \hat{P}}$ reads now
 \be
{\bf \hat{P}}=-\Delta+m^2+V(x),
\label{}
 \ee
 where the Laplace operator $\Delta$ is modified now through the gauge field
 $A_\mu(x)$, since it is constructed of gauge covariant derivatives
 $\Delta=g^{\mu\nu}D_\mu D_\nu$, $D_\mu=\pd_\mu-ieA_\mu(x)$

One obtains a similar chain of equations for $a_i(x,y)$ :


\bea
(x-y)^\mu D_\mu a_0(x,y)=0\\
(-D^2+V(x))a_0(x,y)+(1+(x-y)^\mu D_\mu) a_1(x,y)=0\\
(-D^2+V(x))a_1(x,y)+(2+(x-y)^\mu D_\mu) a_2(x,y)=0\\
\label{eqs_a_i_gauge}
\eea
and arrives finally at the following relations:

\be
a_0(x,y)=exp\ \left[-ie \int\limits_x^y A_\mu(z) dz^\mu \right],
\ee
what  yields for the coincidence limit $x=y$ again $a_0(x,x)=1$, 
$F_{\rho \nu}$ is as usually the electromagnetic field strength tensor
$\pd_\rho A_\nu- \pd_\nu A_\rho$;











\be
a_1(x,x) =(D^2-V(x))a_0(x,y)|_{x=y}=-V(x);
\label{a_1_-V}
\ee
 



\be
D^2 a_1(x,y) =\fr{1}{3}(D^2 D^2 -D^2 V(x)) a_0(x,y)|_{x=y}=
-\fr{1}{6} e^2 F(x)^2-\fr{1}{3}\Delta V(x);
\ee

\be
a_2(x,x)= \half (D^2-V(x)) a_1(x,y)|_{x=y} = -\fr{e^2}{12} F(x)^2+\half V(x)^2
-\fr{1}{6}\Delta V(x).
\label{a_2_gauge}
\ee

\subsubsection{Spinor field interacting with gauge vector field }

\ \ \  If we are interesting in the heat kernel expansion for the Dirac field
interacting with external gauge vector field.

We start from the Lagrangian
\be
{\cal L}=-\fr{1}{4} F_{\mu\nu}F^{\mu\nu}+\bar{\psi} [i\gamma^\mu(\pd_\mu-ieA_\mu)-m]\psi.
\label{spinor_lagrangian}
\ee

The \eqq of motion for the spinor field is the Dirac \eqq
\be
[i\gamma^\mu(\pd_\mu-ieA_\mu)-m]\psi(x)= 0.
\ee

The corresponding Hamiltonian is:
\be
{\cal H}=-i\gamma^0\gamma^j D_j +\gamma^0 m.
\label{spinor_hamiltonian}
\ee
(Here the hamiltonian ${\cal H}$ is not to be confused with the definition
(\ref{T00_spinor}, Sec.\ref{groundstenergy}), and is to understand by means
of primary quantization, namely it provides the \ev $p_0$ of energy, acting
on the static states $\psi(x^0,\vec{x})=e^{-ip_0x^0} \psi(\vec{x})$. It can
be obtained immediately from the Dirac \eqq).

The energy spectrum $ \la_n$ for static problems will be defined from the equation:
\be
{\cal H}\psi_n(x)= \la_n \psi_n(x),
\label{eigenvalues_H}
\ee
the \ef $\psi_n(x)$ of the operator ${\cal H}$ depend on spatial coordinates $x$ only.

The global heat kernel $K(t)$ cannot be constructed properly for this operator. Namely since the operator ${\cal H}$ is a fisrt order operator,
it possesses a symmetric spectrum, that broads from $-\infty$ to $\infty$ and
can get some discrete values between $-m$ and $m$ corresponding to bounded
states. Thus $K(t)$ in the form (\ref{global_heat_kernel})
is no more "well defined".

To avoid this difficultness, one can use the operator ${\cal H}^2$ instead of
${\cal H}$. In other words we can deal here with the sum over squared
energies. Indeed, one can apply the operator  ${\cal H}$ to the relation
\ref{eigenvalues_H} with the resulting equation
\be
{\cal H}^2\psi_n(x)= \la_n^2 \psi_n(x)
\ee
with
\be
{\cal H}^2=-\Delta+\half \sigma^{ij}F_{ij}+m^2,
\ee
and by using the recursive procedure to the one described above one obtains

\bea
&&a_1(x) \e a_1(x,y)|_{x=y}= \half \hat{F}\\
&&a_2(x) \e a_2(x,y)|_{x=y}= -\fr{1}{12} F^2_{\mu\nu}+\fr{1}{4} \hat{F}^2+\half \Delta\hat{F}.
\label{dirac_a1_a2}
\eea

Here the notation $\hat{F}$ stays for $\sigma^{ij}F_{ij}$, where
$\sigma^{ij}=\fr{i}{2}[\gamma^i\gamma^j]$ as usual, and the spinor indices
are omitted.

 In applications of global \gse the trace of \hkc (\ref{dirac_a1_a2}) is used:
\bea
a_1=\Tr \int d^3x\  a_1(x)=0\nn\\
a_2=\Tr \int d^3x\  a_2(x)=\fr{2}{3}F^2_{ij}.
\label{a1_a2_spinor}
\eea\\

\subsubsection{Spinor field interacting with an electroweak background}

\ \ \ The electroweak model with spontaneous broken symmetry is represented by
 the Lagrangian Sec.\ref{hamiltonians}, Eq.(\ref{higgs_lagrangian})

The heat kernel coefficients $a_1$ and $a_2$ can be calculated by using
the formulae for the squared Dirac operator, (\ref{dirac_a1_a2}) established above.

\bea\label{Dsquared}
\left(iD\hspace{-8pt}/+\mu(r)\right)\left(iD\hspace{-8pt}/-\mu(r)\right)&=&
-D^2+\frac{q}{2}F_{\mu\nu}\sigma^{\mu\nu}
-if_e \gamma^\mu \frac{\pd}{\pd x^\mu}\mid\Phi\mid
-f_e^2\mid\Phi\mid^2 \nn \\
&\equiv&-D^2+V
\eea
with $\sigma^{\mu\nu}=\frac{i}{2}[\gamma^\mu,\gamma^\nu]$, $\mu(r)$
corresponds to the definition of radial-dependent fermionic mass in
Nielsen-Olesen notations Sec. \ref{hamiltonians} (\ref{defofMu}).
The general expressions for relevant \hkc are

\bea\label{allghkks}a_1&=&\Tr \int d^2x \ (-V),  \nn \\
a_2&=&\Tr \int d^2x \left(
-\frac{1}{12}F_{\mu\nu}^2+\frac12 V^2-\frac16 \Delta V \right),
\eea
where the trace is taken over the spinor indices. Substitution of the expression
for $V$ defined by (\ref{Dsquared}) yields

\bea\label{hkks}a_1 &=& -8\pi\int\limits_0^\infty dr \ r \
\left(\mu(r)^2-m_e^2\right), \nn \\ a_2 &=& 8\pi\int\limits_0^\infty
dr \ r \ \left(\frac13 \frac{v'(r)^2}{r^2} + \frac12 \left(
\mu'(r)^2+(\mu(r)^2-m_e^2)^2\right) \right).
\eea

\subsubsection{Other approaches to the calculation of the \hk\\ coefficients: the expansion of the exponent}

\ \ This approach makes the structure of the global heat kernel coefficients
quite transparent. Its advantage compared to the DeWitt procedure is that it
works with the global objects $a_n$ immediately without need in intermediate
relations for two-point dependent coefficients $a_n(x,y)$.

The approach is based on the generalized representation of $K_{\bf \hat{P}}(t)$
(\ref{heat_trace}) in the symbolic form:

\be
K_{\bf \hat{P}}(t)= \Tr_{L^2} e^{-t{\bf \hat{P}}}.
\ee

It is convenient to replace this object by a more general one:

\be
K_{\bf \hat{P}}(f;t)= \Tr_{L^2}\{f\ e^{-t{\bf \hat{P}}} \},
\label{generalized_heat_trace}
\ee
where a smooth \f $f$ plays the role of smearing \f and keeps $K_{\bf
\hat{P}}(f;t)$ finite even in an infinite space. 

Now consider the calculation of the generalized heat trace
(\ref{generalized_heat_trace}). To this end we have to define the basis in
$L^2$ space where the trace is defined. Following the \cite{vassilreview} we
use the simplest basis of the flat waves in cartesian coordinates.

\be |>=e^{ikx} \ee

The operator ${\bf \hat{P}}$ is some second order operator of the Laplace type

\be
{\bf \hat{P}}=-(g^{\mu\nu} D_\mu D_\nu+E),
\label{laplace_type_operator}
\ee
containing some (possible matrix valued) \f $E$. The metric
$g^{\mu\nu} $ is flat. $D$ is the covariant derivative

\be D= \pd+\om,
\ee
that includes the connection $\om$ (general notation for geometrical or gauge connection).
 Carrying out the calculation for heat trace (\ref{generalized_heat_trace})
of this operator in the flat space we obtain:

\bea
&&\Tr_{L^2}<\ f\ e^{-t{\bf \hat{P}}} \ >=\nn\\
&& \int d^nx\int\fr{d^k}{(2\pi)^n} \ e^{-ikx}\  
\tr f(x)\ e^{-t{\bf \hat{P}}}
\ e^{ikx}\  = \nn\\
&& \int d^nx\int\fr{d^k}{(2\pi)^n} \tr f(x)\ e^{t((D_\mu+ik_\mu)(D^\mu+ik^\mu)+E)}.\eea

The notation $\tr$ denotes now the trace over all remaining indices (spinor,
colour etc.)

The exponent entering here can be represented as a product $e^{-tk^2} e^{t(ik^\mu
D_\mu+ik_\mu D^\mu+D_\mu D^\mu+E)}$ and then the second exponent can be
expanded in the limit $t\ra 0$ with the result

\bea
&& e^{t(ik^\mu
D_\mu+ik_\mu D^\mu +D_\mu D^\mu+E)} = 1+ t(D^2+E)- 2{t^2}(kD)+\nn\\
&& t^2(D^2 D^2+D^2 E+ E D^2+E^2)-\fr{4t^3}{6}[(kD)^2 E+E (kD)^2+(kD)E(kD)]
- \nn\\ &&\fr{4t^3}{6}[(kD)^2 D^2+ D^2 (kD)^2 + (kD) D^2 (kD) ]
+\fr{16t^4}{24} (kD)^4\nn\eea

Keeping in mind that the integrals of terms with odd powers of $k^\mu$
disappear, these terms are already omitted.

Using the integrals of type:
\bea
\int\fr{d^k}{(2\pi)^n}e^{-tk^2}k^{\mu_1} k^{\mu_2}..k^{\mu_{2j}}=
\fr{1}{(4\pi t)^{n/2}}\ \fr {1}{2j(t)^j}[^{\mu_1,\mu_2,\dots \mu_{2j}}],\nn\\
j=0,1,2,\dots,
\eea
where $[\dots]$ denotes the full symmetric combination of indices $\mu_i$,
one can carry out the integration over k and get

\bea
&&\Tr_{L^2}\{f\ e^{-t{\bf \hat{P}}} \}=
\fr{1}{(4\pi t)^{n/2}} \int d^n x \tr f(x) [ 1+ tE+\nn\\
&& \fr{t^2}{2}(D^2 D^2 + D^2 E+ E D^2 +E^2) - \fr{t^2}{3}(D^2 E+ E D^2+D^\mu E
D_\mu +2 D^2 D^2+D^\mu D^2 D_\mu)+\nn\\
&& \fr{t^2}{6} (D^\mu D^\nu D_\mu D_\nu+ D^2 D^2+ D^\mu D^2 D_\mu)+O(t^3)].
\eea

Finally, introducing the corresponding notations for commutators of
connections appearing thourgh commutations of covariant derivatives

\be
\Omega_{\mu\nu}=\pd_\mu \om_\nu- \pd_\nu \om_\mu+\om_\mu\om_\nu-\om_\nu\om_\mu,
\ee
we rewrite the result in the form

\be
K_{\bf \hat{P}}(t)= \fr{1}{(4\pi t)^{n/2}}\int d^n x\tr f(x) [1+ tE+\nn\\
+t^2(\half E^2+\fr{1}{6} E_{;\mu\mu}+\fr{1}{12}\Omega_{\mu\nu}\Omega^{\mu\nu})
+O(t^3)].
\label{Kexpansion}
\ee

 This result is very general comparable with the explicit formulae for certain
 cases obtained above by the DeWitt recursive procedure. In particular if the
Maxwell tensor $F_{\mu\nu}$ corresponding to the electromagnetic potential
 $A_\mu$ is treated as an endomorphism $E$ while the potential itself plays
 at the same time the role of the U(1) gauge connection, we identify the contribution of  order $t^2$ in
 (\ref{Kexpansion}) with the contribution of $a_2$ (\ref{dirac_a1_a2}). The
 $\hat{F}=F_{\mu\nu}\sigma^{\mu\nu}$ corresponds to the $E$ and $F_{\mu\nu}=
\pd_\mu A_\nu-\pd_\nu A_\mu$ to the $\Omega_{\mu\nu}$ respectively, the
 covariant derivative $;_\mu$ in flat space is the ordinary one $\pd_\mu$.

\subsubsection{Some general properties of \hkc}

\ \ We conclude this review of the \hkc with
some general rules inherent to construction of this objects.

 We restrict ourself again on certain class of problems related to the
physical models being handled in this paper.
We consider in general the field configurations in the flat Minkowski
space-time without boundaries. However a boundary can be introduced in intermediate
steps with suitable boundary conditions, that will be
removed later by the transition from the finite volume inside the boundary to
the infinite one. Then in each case it must be respected carefully that the
\hke is transformed properly by this transition.

 The general construction of the \hkc $a_k,\  k=0,1,2,\dots  $ for some
 operator ${\bf \hat{P}}$ of the Laplace type (\ref{laplace_type_operator}) with the smooth
 connection $\Omega$ and smooth endomorphism $E$ acting on the manifold $M$ without
boundaries reads \cite{gilkey}

1.$\ \ 
 a_k(f,{\bf \hat{P}})=0\ \ for\ \ k=2j+1
$

2. $ a_k(f,{\bf \hat{P}})\ \ for\ \ k=2j $ are locally computable in terms of
 geometric invariants
\be
a_{k}(f,{\bf \hat{P}})=\tr \int\limits_M d^nx\sqrt{g}\{f(x)a_k(x,x|{\bf \hat{P}}) \}
=\sum\limits_I \tr \int\limits_V d^nx f(x) u^I{\cal A}_k^I( {\bf \hat{P}}),
\label{struct_a_k_even}
\ee
where ${\cal A}_k^I$ are all possible local invariants of the dimension k
constructed from $E, \Omega$, and their derivatives.
The constants $ u^I $ appearing here contain some rational constants of kind
$p/q, \ p,q \in N $ and powers of $\pi$. These constants define the algebraic
structure of \hkc. In fact since the general structure of the coefficient
$a_k$ is known (\ref{struct_a_k_even}), the problem is reduced to the
calculation of $ u^I $ only. It can be reached by using relations between
$ u^I$. To establish the certain relations one introduces some additional
parameter and applies the known variational properties, described more
detailed in  \cite{vassilreview, gilkey}. 

 The important remark about the smoothness of $E$ must be mentioned. If the
\f $ E(x) $, (for example the scalar potential $V(r)$, Sec.\ref{hamiltonians}) is
not infinite smooth, $E(x)\not\in {\cal C}^\infty $, then the statement 1. is
no more valid. So for the magnetic background with a step \f the coefficient
$a_{5/2}$ appears to be non zero. It will be calculated in Sec. \ref{a52}.

 \subsection{Zeta functional \rg}
\label{zfreg}

 \ \ \ The application of methods using \zf and \hk to the \rg problems of quantum field
theory was introduced independent by  number of authors
\cite{schwinger,dewitt,fock}.
The generalized \zf was used for example in explicit form in problem of \rn of cosmological constant
\cite{hawking}. The usage of \hke for \rg of effective action comes back to
\cite{dowker}.  

The \rp using the \zf can be considered to some degree as a certain extension
of the wide used method of the dimensional \rg in the perturbation theory.

The main advantage of this approach is that it allows to identify and separate
divergences as a pole terms in power expansion for small \rg
argument $s\ra 0$.

  The zeta functional \rg is allied closely with so called
dimensional regularizations while it uses an additional space
dimensionality that will sent to zero and thus detect the divergent terms
by arising poles at $s\ra 0$.
  
   The basic idea of the zeta functional \rg method is the
representation of the regularized object as the generalized zeta function
of an (elliptic) operator $\hat{P}$.
 
The essence of the dimensional \rg consists in the introduction of some
additional (in fact subtractional) dimension $d$ in order to reduce
the dimension of integral measure to make the integrals "conditional"
convergent. After that the performance back to the limit $d \ra 0$ allows
to get the separated divergent contributions. They appears as a terms containing
pole, that will be cut out (renormalized out).

 The zeta functional \rp is also based on the introduction of the additional dummy
 dimension $s$ that will be abrogated afterwards and the pole-containing
 terms will be regularized out, that brings the final results, coinciding
 with the one given by dimensional \rg if the extra dimensions
 are flat \cite{hawking}.
 
\subsubsection{The space dimension and structure of poles}
\ \ \ In the review of this topic we follow \cite{blau, OEBZ} and start with the
definition of \zf in the form Sec.\ref{zetafunc}, (\ref{zeta_trace})
\be
\zeta_{\cal P }(s) = \Tr' \{  (\mu^{-2}{\cal P} )^{-s} \}=
{\sum\limits_{(n)} }' \left( \fr{\la_{(n)} }{\mu^2}\right)^{-s}.
\label{trace_P}
\ee

This definition is generalized compared to the ones (\ref{zeta_trace},\ref{zeta_trace_sum}) by the introduction of the dimension-adjusting constant $\mu$ to keep the \zf dimensionless
for all $s$. The spectrum of ${\cal P }$ can possess some values $\la_{(n)}$
equal to zero, which make the $\zeta_{\cal P }(s)$ ill-defined for positive
$s$. To correct the situation we suppose all zero-modes to be excluded in
(\ref{trace_P}), that is noticed by the prime.  

Let the operator ${\cal P }$ again be a second order elliptic differential operator
 $D_d$ of the Laplace type of dimension $d$  (Sec.\ref{heatkernel}).
(Further we discuss here generally this class of operators only). Second, we
 will work in the Minkowski space-time $d=4$ and assume that only the second
 order time-derivative is present. Thus it can be separated explicit
\be
D_4=-\pd_0^2+D_3.
\ee

The integration over $x^0$ in the global \zf $\zeta_4(s)$ can be carried out
as it was made in Sec.\ref{groundstenergy}:

\be
\zeta_4(s)=\fr{T\mu}{\sqrt{4\pi}} \fr{\Gamma(s-\half)}{\Gamma(s)} \zeta_3(s-\half),
\label{zeta_4-to-zeta_3}
\ee
(here time is intergated oud and $T$ means therefore the age of the system).

Now we use the relation to the global \hk (\ref{zeta_of_K_global}) together
with the expansion of the heat trace obtained from (\ref{local_\hke})  

\be
\tr(e^{-\fr{tD}{\mu^2}})=\left( \fr{\mu^2}{a\pi t}\right)^{d/2}
\left\{ \sum\limits_0^N\left(  \int_\Omega
    a_n(x)(\mu^{-2}t)^n+\int_{\pd\Omega} b_n(y)(\mu^{-2}t)^n \right)+O(t^N)   \right\}.
\label{trace_asymptotic}
\ee

Since the diagonal part of the \hk contains exponentially suppressed terms
of kind $e^{-k(x)/t}$,\cite{blau} it provides a non-zero contribution on manifolds with
boundary as a boundary term present in \ref{trace_asymptotic}.
 The (\ref{zeta_of_K_global}) together with the recently mentioned (\ref{trace_asymptotic})
 yields:
\be
\zeta_d(s)=\fr{1}{\Gamma(s)(4\pi)^{d/2}} \left\{ \sum\limits_0^\infty
  \fr{C_n}{(s-[\half d-n])}+ f(s)\right\},
\label{zeta_d_expansion}
\ee
where the sum runs over half-integers,

\be
C_n=\mu^{d-2n}\left\{ \int_\Omega a_n(x) d^dx+ \int_{\pd\Omega}b_n(y)d^{d-1}y \right\}
\label{C_n}
\ee
and $f(s)$ is an entire analytic \f of $s$.

Thus it can be seen that the location of the poles of \zf depends on the
dimension $d$ of space-time. It is accentuated by the index $d$ at the \zf.
 In particular for $d=4$ the \zf $\zeta_4(s)$ has no pole at $s=0$, that
 allows to define the functional determinant in the form (\ref{det_P}).

\subsubsection{The effective action}
\ \ \ 
 This application of \zf to the \rg of the one loop effective action goes
 back to the paper of Hawking \cite{hawking}, where it was developed for the generalized
 case of curved space-time.

The one loop Feynman graph of perturbation theory is represented as the
 determinant of an operator as follows:
 It starts from the main object of the field theory in functional formulation,
 namely the generating functional for Feynman diagrams

 \be
 Z =\int {\cal D}g {\cal D}\phi e^{iS[g,\phi]},
\label{gen_func_of_g_phi}
 \ee
that is constructed as a functional integral over the product of spaces of
all possible (admissible) metrics $g(x^\mu)$ and matter fields $\phi(x^\mu)$
 with measures ${\cal D}g$ and ${\cal D}\phi$ respectively.

 The $S[g,\phi]$ is  the classical action, that is also a functional of function sets $g(x^\mu)$ and
 $\phi(x^\mu)$.

 We restrict the discussion on the case of flat space and omit the metric
 $g$ in further calculations.
 Now assume that the quantum fields are the perturbations of classical ones (backgrounds) and represent the field variables in the form
\be
\phi=\phi_0+\tilde{\phi}.
\ee
  
In other words we decompose the arguments of the functional $S$ as a sum of
 classical, 'main' fields $\phi_0$ and the quantum additions (fluctuations) $\tilde{\phi}$. The
latter are considered to be definitive small compared to the main classical ones.

 The backgrounds $ \phi_0$ are still supposed to satisfy the classical
equations of motion and boundary or periodic conditions and provide the
 leading contribution to the action $S$. 
 
Respecting the smallness of fluctuations one can represent the complete action $S[\phi]$
as an expansion at  $\phi_0$ with respect to the fluctuations
$\tilde{\phi}$ as

 \be
 S[g,\phi]=S[g_0,\phi_0]+S_\phi[\tilde{\phi}],
 \label{action_expanded}
 \ee

 Where the first corrections to the main action $S[\phi]$ are quadratic in
 the fluctuations; it follows further from (\ref{gen_func}),that

\be
\ln Z=i S[\phi_0]+\ln \int {\cal D}\tilde{\phi}  e^{iS_\phi[\tilde{\phi}]}.
\ee

 The general form of the corrections to the action generated by the
 fluctuations is supposed to be

\be
S_\phi[\tilde{\phi}]=-\half \int d^4 x \ \tilde{\phi}^* {\cal
  P} \tilde{\phi},
\ee
where the operator ${\cal  P}$, is the kernel of free action for the
corresponding field; it is of second order for bosonic fields (e.g. $\Box+m^2$- Klein-Gordon kernel for spin 0), and of the first order for fermionic
fields as well (e.g. Dirac operator $i\gamma^\mu \pd_\mu-m$).

 Suppose, the operator ${\cal P}$ to be elliptic and self-adjoint with the
 spectrum of \ef $\phi_n$ and \ev $\la_n$
\be
{\cal  P} \phi_n= \la_n\phi_n.
\ee

 The \ef $\phi_n$ as a functions of coordinates $x^\mu$ will be orthonormalized as usual

\be
\int d^4 x \phi_n^*\phi_m=\delta_{nm}.
\ee

Then any \f $\tilde{\phi}$ can be decomposed in the basis $\phi_n$

\be
\tilde{\phi}=\sum\limits_n a_n \phi_n
\ee
and the measure of the field space can be expressed through these coefficients
$a_n$

\be
{\cal D}\tilde{\phi} =\prod\limits_n \mu d a_n ,
\ee
fit out with the normalization parameter $\mu$ to keep the correct dimension. Then
one arrives at

\bea
Z[\tilde{\phi}]=\int {\cal D}\tilde{\phi}  e^{iS_\phi[\tilde{\phi}]}=
\prod\limits_n \mu d a_n e^{-\half \la_n a_n^2}=\nn\\
\prod\limits_n \left( \sqrt{ \fr{2\pi}{\la_n} } \right)=
\left[ \det \left( \fr{{\cal  P}}{2\pi \mu^2} \right)\right]^{-\half},
\eea
that confirms the similar fact that the contribution quadratic in fields
reduces to the evaluation of determinant.
 The logarithm of $Z$, treated e.g. in thermodynamical approach as the free energy
reads now in terms of \zf and its derivative

\be
\ln Z[\tilde{\phi}]=\half \zeta_{\cal  P}'(0)+\half \ln \left( \fr{1}{2\pi \mu^2}  \right)\zeta_{\cal  P}(0).
\ee

\subsubsection{The \gse}

\ \ \  The simplest classical example of the zeta \rg is the
  zeta-regularized \gse considered in Sec.(\ref{groundstenergy})

  To consider this application of \zf\rp in details we follow again the
\cite{blau}.  

Here it would be suitable to mention three objects that are in principle
distinguished

 1. The vacuum energy, already defined in Sec.\ref{groundstenergy} as the vacuum
 expectation value of the $T^{00}$ (and therefore called sometimes also vacuum energy):

\be
E_{vac}=\int <0|T^{00}|0>
\label{t00_vev}
\ee

 2. The \gse (or Casimir energy), defined in ( Sec.\ref{hamiltonians},
    Sec.\ref{groundstenergy}) as the half sum over spectral values $\la_{(n)}$ of hamiltonian
 \be   
 E_{0}=\half\sum\limits_{(n)} \la_{(n)}
 \ee
 
 3. The one-loop effective energy, being defined in view of the Sec.(\ref{effact})

\be
E_{eff}=\fr{1}{T} S_{eff}=\fr{1}{2T} \ln \det D_4 = -\fr{1}{2T} \zeta'_4(0)
\ee
with $T=\int dx^0$- age of the system, the index 4 indicates the
4-dimensional \zf and the prime means that the possible zero-modes are not
included. These objects under circumstances of problem are not equal \cite{blau}.
 Following the notations of \cite{blau} we use for further discussion in this
section the dimensionless \zf  ${\bf \zeta}$ of operator $D$ which is related
to the \zf (\ref{zeta_trace_sum}, \ref{zeta_trace}) defined in Sec.4.1, by
\be
{\bf \zeta}(s)= \tr ' \{ (\mu^{-2} D)^{-s}\}={\sum}'  (\mu^{-2}
\la_n)^{-s}=\mu^{2s} {\sum}'(\la_n)^{-s}= \mu^{2s} \zeta_D(s)
\label{zeta_to zetabold},
\ee
 whereat $D$ is supposed to be an usual second order operator. It has
 therefore the dimension $[length]^{-2}$ and the adjusting constant must be of
 dimension $[length]^{-1}$ or $[mass]$.

The 3-dimensional zeta regularized \gse is defined as
\be
{\cal E}_{0}^{reg}(s)= \fr{\mu}{2}
\sum\limits_n(\la_n\mu^{-2})^{1/2-s}=
\half\mu {\bf \zeta}_3(s-\half)
\label{e_reg_3-dim}
\ee
with the dimension-adjusting constant $\mu$ to keep the \zf dimensionless
for all $s$. Then it follows from the expansion (\ref{zeta_d_expansion}) with
$d=3$, that the  ${\cal E}_{0}^{reg}(s)$ is a meromorphic \f of the
complex argument $s$ which has a pole in the \rg limit at $s=0$.
The residuum at this point is
\bea
 \Res_{s=0}\{ {\cal E}_{0}^{reg} \} = -\half C_2(g_3)/(4\pi)^2=\nn\\
 -\half \fr{1}{(4\pi)^2}\{ \int_{\Omega} a_2 + \int_{\pd\Omega} b_2 \}.
\eea

Because of the pole at $s=0$ the regulator $s$ cannot be removed in a unique
way. It means that the vacuum energy contains an ambiguity proportional to
the geometric coefficient $C_2$. The minimal subtraction scheme adopted in
\cite{blau} supposes that

\be
{\cal E}_{0}=\fr{\mu}{2} PP\ {\bf \zeta}_3(s-\half),
\label{minimal_subtraction}
\ee
where $PP$ denotes the "principal part" and means

\be
PP\ {\bf \zeta}_3(s-\half) =\lim_{s\ra 0} \half \left\{ {\bf \zeta}_3(s-\half) + {\bf \zeta}_3(-s-\half)\right\}.
\label{principal_part}
\ee

Another subtraction scheme will be considered below in Sec.\ref{efineas}

Now since we establish the relation between heat kernels in 4 and 3
dimensions respectively
\be   K_4(x,x,t)=\fr{\mu}{4\pi t}K_3(x,x,t), \ee
then under assumption of (\ref{zeta_4-to-zeta_3}) we can apply the expansion
(\ref{zeta_d_expansion}) to both zeta functions with $d=3$ and $d=4$ and
finally we obtain the relation between the \gse and the one-loop
effective one.
\be
E_{eff}=E_{0}+\half [ \psi(1)- \psi(1/2)]\fr{C_2}{(4\pi)^2}.
\label{Eeff_Ecasimir}
\ee
Here $\psi(s)=\fr{d\ln \Gamma(s)}{ds}$ is the polygamma-\f arising from
the expansion of $\Gamma$-\f.

 Thus we can conclude, that the difference between $E_{eff}$ and $E_{0}$
 as well as the ambiguity of the ${\cal E}_{0}$ is caused by the non-zero
 geometric factor $C_2$. If the latter disappears, the \zf regularized
 \gse becomes unique and finite, and  $E_{0}=E_{eff}$.
  However this is not the case for all problems considered
  below, since the \hk coefficient $a_2$ is always non-equal zero.

The vacuum energy defined by (\ref{t00_vev}) in the case that all
interactions are switched off so that $T^{00}\ra T^{00}_{free}$ then
\be { E}_0= \int\ d^3x <0|T^{00}_{free}|0>, \ee  as it
has been assumed in Sec.\ref{groundstenergy}.

  Thus the \gse has been defined to be the sum over the spectrum of eigenvalues $\om_n$ of the hamilton operator ${\cal H}$ (speaking strictly, eigenvalues of  ${\cal H}$ are the squares of eigenfrequences $\om_n^2$ )
  
\be
 {\cal H}\Phi_n(x)=\om_n^2 \Phi_n(x)
 \label{hamiltonian_eigenvalues}
\ee 
and the regularized ground state energy can be just represented as
  \be
 E_0^{reg}(s)=\frac{\mu^{2s}}{2}\zeta_{\cal H}(s-\half)
=\frac{\mu^{2s}}{2}\sum\limits_{(n)}
\om_{(n)}^{1-2s}=\frac{\mu^{2s}}{2}\sum\limits_{(n)} (k_{(n)}^2+m^2)^{\half-s}.
\label{e_reg_zeta}
 \ee

This expression will be used as a starting point in further calculations of
  vacuum energy for different models discussed below.

\subsubsection{The large mass asymptotics for the zeta-regularized\\ \gse of
  massive fields} 


\ \ \ The regularized vacuum energy has been represented above
 Sec.(\ref{groundstenergy}) as the generalized \zf of the corresponding
 hamiltonian ${\cal H}$.
 Here it will be shown how the \uv divergent contributions in the ground state energy can be identified.
  

We start here with the expression for the zeta-regularized ground state
energy of quantum field, derived in Sec.(\ref{groundstenergy}). 

  \be
   {\cal E}_0^{reg}(s) =-\frac{\mu^{2s}}{2}\zeta_{\cal{H}}(s-\half)=-\frac{\mu^{2s}}{2}\int\limits_{0}^{\infty}dt \frac{ t^{s-1}}{\Gamma(s)}K(t)
  \ee
and use the \hke in the form 
  \be
  K_{\cal H}(t)\sim  \fr{e^{-tm^2}}{(4\pi t)^{d/2}} \sum\limits_{n\ge 0} a_n t^n,
  \ee
that corresponds to the global \hk for the operator ${\cal H}$
\be{\cal H} =-\Delta+m^2+V(x),\ee
 $d$ is the space dimension, $n=0,\half,1,\fr{3}{2},2,...$

  Then substituting this into ${\cal E}_0^{reg}(s)$ and re-expanding at $s\ra 0$ one obtains
  the key formula for the \rn of \gse. After collecting of equal powers of
  $m$ together it acquires the form

\bea
{\cal
  E}^{reg}&=&-\frac{m^4}{64\pi^2}\left(
\frac{1}{s}+\ln\frac{4\mu^2}{m^2}-\half \right)a_0
-\frac{m^3}{24\pi^{3/2}}a_{1/2}+\nn\\&\ &\frac{m^2}{32\pi^2}\left( \frac{1}{s}+\ln\frac{4\mu^2}{m^2}-1\right)a_1+\frac {m}{16\pi^{3/2}}a_{3/2}\nn\\
&&-\frac{1}{32\pi^2}\left( \frac{1}{s}+\ln\frac{4\mu^2}{m^2}-2 \right)a_2+ ...\ .
\label{large_mass_expansion}
\eea

  The parameter $t$ has the dimension $[m]^{-2}$, thus the expansion in powers
$t$ at $t\ra  0$  corresponds to the $1/m$ expansion. The first several terms
of this expansion quoted here contain all \uv divergences as poles at $s\ra 0$.

 \section{Analytical treatment of the \\ \gse}
\subsection{The regularized sum over the spectrum of hamiltonian}
\label{spectralsum}

\ \ \  The problem was already formulated in
Secs.\ref{groundstenergy} and \ref{hamiltonians}. We recall briefly the notations
 introduced in Sec.\ref{hamiltonians}.
 The total energy of the system
of fields is considered to consist of two parts
\be
{E}^{tot}={E}_0^{cl}+{ E}_0^{qu},
\label{e_total}
\ee
where
  \be
 E^{cl}=\int\ dx{\cal E}^{cl}(x)=\sum\limits_i\int dx {\cal E}^{\Phi_i}(x)
 =\sum\limits_i \int dx T^{00}_{\Phi_i}(x)
  \ee
denotes the complete global energy of the classical fields ${\Phi_i}$ (backgrounds),
and

\be
E_0^{qu}= \int dx{\cal E}^{\phi}_0 (x) = \int dx <0|{\cal H}(x)|0>=E_0^{\phi}
\ee
is the energy of quantum fluctuations ${\phi}$ in the vacuum state which is defined as
the state without real particles. We work here with the stable vacuum, the
  problem is formulated in the infinite volume without boundaries, thus the
 global quantum vacuum energy $E_0^{\phi}$ has been shown in the
 Sec.(\ref{groundstenergy}) to be equal to the half sum over the spectrum of
 \ev $ \om_{(n)}$ of the hamiltonian  ${\cal H}$ for the
 corresponding quantum field ${\phi}$

\be
{\cal E}^{\phi}_0= C \half\sum\limits_{(n)} \om_{(n)}
\ee
with the symmetry (and statistics) factor $C$ (e.g. for 4-component spinor
 fermionic field $C=-4$). $\om_{(n)}$ obey the \eqq

\be
 {\cal H}\Phi_n(x)=\om_{(n)}^2 \Phi_n(x),
 \label{H_eigenvalues}
\ee
and by using the \zf \rp described in Sec.\ref{zetafunc} ${E}^{\phi}_0$
is represented the operator ${\cal H}$ as a generalized
 \zf the operator ${\cal H}$, (\ref{e_reg_zeta})

\be
{E}_0^{reg}(s)= \frac{1}{2}\mu^{2s}\sum\limits_{(n)}\omega_{(n)}^{1-2s}=
\frac{1}{2}\mu^{2s}\zeta_{\cal H}(s-\half)
\label{e_reg_general}
\ee

The parameter $\mu$ appears here as a usual \rg parameter in all
 procedures and serves to adjust the correct dimension of energy.
 
  It will be shown further, that one does not need to solve the equation
  (\ref{H_eigenvalues}) to obtain the eigenvalues $\om_{(n)}$ explicitly and
  summarize them in sense of (\ref{e_reg_general}). By usage of the \jf
  formalism developed in the scattering theory it is possible to avoid this
  troublesome procedure and express the ${\cal E}_0^{reg}(s)$ in terms of the \jf
  for the corresponding scattering problem.
  
\subsubsection{Scalar field}

 Consider the Klein-Gordon \eqq of motion for the scalar field $\Phi(x)$
  
\be
 \{ \pd_0^2-\Delta+m^2+V(x)\}\Phi(x)=0
\ee
with some cylindrically symmetric background $V(r)$, depending in coordinates
$(\varphi,r,z)$ on the radial coordinate only, and use the ansatz
\be
\Phi(x) = \phi(r)e^{-il\varphi}e^{-ip_z z}.
\label{scalar_ansatz}
\ee

 In accordance with the formula for the zeta-regularized ground state energy,
 derived in Sec. \ref{zetafunc}, we have
\be
{E}_0^{reg}(s)= \frac{1}{2}\mu^{2s}\sum\limits_{(n)}(\vec{k}_{(n)}^2+m^2)^{1-2s},
\ee
${\vec{k}_{(n)}}$ is the spatial momentum of the $(n)$-th wave being an
 eigenfunction of the corresponding hamiltonian.
 
 Because of the translational symmetry of the problem the variables $z$ and
 $k_z$ appear in (\ref{scalar_ansatz}) only in the exponent, then it follows:

\be
  E_0^{reg}(s)=-\frac{1}{2}\mu^{2s}\int\limits_{-\infty}^{\infty} \frac{dk_z}{2\pi}
  \sum\limits_{(n)} (k_z^2+k_{(n)}^2+m^2)^{\half-s}.
  \label{e_reg_p_z}
  \ee

  The symbol $\sum\limits_{(n)}$ still includes the summation over all
  \ev of momenta $k_{(n)}$ transversal to the z-axis and the
  integration over space coordinates, excluding the z-coordinate: this integration has been executed and the resulting global energy will be understood in sense of energy density per unit length (Sec.\ref{groundstenergy}).  
  
  Thus we state that the momentum $k_z$ in (\ref{e_reg_p_z}) can be separated and integrated out explicitly in the analytical form:
 
  \be
  E_0^{reg}(s)=-\frac{\mu^{2s}}{4\sqrt{\pi}} \frac{\Gamma(s-1)}{\Gamma(s-\half)}
\sum\limits_{(n)} (k_{(n)}^2+m^2)^{1-s}.
  \ee

  The sum $\sum\limits_{(n)} $ means now the summation over all remaining
 quantum numbers. In the case of cylindric symmetry these are:
 $k_i$ the radial momentum and $l$ the orbital momentum number,  
then the $\sum\limits_{(n)}$ denotes $\sum\limits_{l=-\infty}^{\infty}\sum\limits_{k_n} $.

 For scalar field $\Phi(x)$ the ansatz (\ref{scalar_ansatz}) leads to the radial wave \eqq
\be
\left\{ \fr{\pd^2}{\pd r^2}  + \fr{1}{r} \fr{\pd}{\pd r} +\left(q^2-\fr{l^2}{r^2} \right)\right\} \phi(r)=0,
\label{scalar_wave_eq}
\ee
where the $q^2$ is in general $r$-dependent and denotes $q^2=k^2-V(r)$.
To proceed with the formalism we have to start with a discrete spectrum of
momenta $k_n$. To this end we suppose that fluctuations $\phi(r)$ are
quantized in the finite cylinder of radius $r=R$ with the suitable boundary
condition. The spectrum $k_n$ is then discrete.

 For example we may use the Dirichlet boundary condition on the boundary $r=R$,
 such that $\phi_k(R)=0$. So we can consider $\phi_k(R)=0$ as a function
 $\phi_R(k):=\phi(k)$ of  $k$, and the set of solutions $k_n$ satisfying this
 condition is the desired spectrum $k_n$. Now since we know the function
 $\phi(k)$ we can obtain the sum over the spectrum $k_n$.
 Note that we don't need to obtain all the values of $k_n$ to summarize it
 over the whole spectrum. Instead of that we can use the approach well known
 from the scattering theory, namely we use the integral properties of special
 construction $\fr{\pd}{\pd k}\ln \phi(k)$, so called "logarithmic
 derivative" of the eigenfunction $\phi(k)$. 
  
Since the eigenvalues $k_n$ correspond to zeros of $\phi(k)$ they correspond
also to poles of the logarithmic derivative $\fr{\pd}{\pd k}\ln \phi(k)$ as
well, because

\be
 \fr{\pd}{\pd k}\ln \phi(k) \e \fr{\pd_k\phi(k)}{\phi(k)}
 \label{logarithm_derivative}
\ee
and the first derivative of $\phi(k)$ is finite for all real $k$ as supposed.
The residuum of (\ref{logarithm_derivative}) in each point $k_n$ is equal to
the order of zero of the ${\phi(k)}$ in $k_n$.
 Then it means that the sum over $k_n$ is equal to the contour integral

 \be
 \sum\limits_{(n)} (k_n^2+m^2)^{1-s} =\fr{1}{2\pi i}\int\limits_\gamma dk (k^2+m^2)^{1-s}  \fr{\pd}{\pd k}\ln \phi(k),
\label{int_k^2+m^2}
 \ee
where the integration contour in the complex k-plane encloses the real
positive axis. The contour integral chosen in such a way contains all the
contributions of residua of the function $\fr{\pd}{\pd k}\ln \phi(k)$, which
are placed on the real half-axis.

 The function $\phi(k)$ can also contain zeros in the k-plane aside from
the boundary conditions on $r=R$  that correspond to the bound states and
build some subset of the discrete spectrum. They can be taken into account
by addition of the sum over $k_i$

\be
 \half \sum\limits_l \sum\limits_{i} (m^2-k_{il}^2)^{1-s}
\ee
to the integral over $k$ (\ref{int_k^2+m^2}).
 For the further transformation of the spectral sum we note that the solution
 $\phi_k(r)$ of the wave \eqq in cylindrical coordinates $(r,\vp)$ may be chosen by the behaviour of this solution at $r\ra 0$.
  For our purposes it is suitable to require this solution to be the so
 called "regular" one \cite{taylor}, that becomes proportional to the
 free solution (of \eqq (\ref{scalar_wave_eq}) with potential $V(r)=0$). For cylindric coordinates it means

 \be
  \phi_{kl}^{reg}(r) \sim J_l(kr),\ \ r\ra 0 ,
 \ee
where $J_l(kr)$ is the (cylindric) Bessel function, having the expansion
\be
 J_{\nu}(z) \sim \fr{1}{\Gamma(\nu+1)}\left(\fr{z}{2}\right)^\nu
\ee
for small arguments $z\ra 0$.
 
  For the solution chosen in such a way the asymptotic behaviour at $r\ra
  \infty$ can be described in terms of the \jf so that:
\be
  \phi_{kl}^{reg}(r) \sim \half \{ \bar{f}_l(k) H_l^{(1)}(kr)+
  f_l(k)H_l^{(2)}(kr) \},
  \label{jost_func_def}
\ee
 where the $ H_l^{(1,2)}(kr)$ are the Hankel functions, that satisfy also the
wave \eqq where the potential becomes constant, (as we even suppose always
for $r\ra\infty$ ). These solution correspond to so called "in-" and
"outcoming waves" respectively. The arising coefficients $f_l(k)$ and
$\bar{f}_l(k)$ are the \jf and its complex conjugated. For the real k the
following useful symmetry property
\be
 f_l(-k)=\bar{f}_l(k)
 \label{schwarz_reflection}
\ee
applies, that is the expression of the Schwarz reflection principle known from the \f
 theory and applied to the \jf in the scattering theory \cite{taylor}. 

 Using the representation (\ref{jost_func_def}) with the property
(\ref{schwarz_reflection}) we can proceed with the transformation of the
 spectral sum of {\cal H} expressed  as a contour integral (\ref{int_k^2+m^2}) for $R\ra \infty$. Rewrite the logarithm of the regular solution $\phi^{reg}$ using the (\ref{jost_func_def}):

 \bea
 \ln \phi^{reg}(k)=\ln \{ \bar{f}_l(k)H_l^{(1)}(kr)+ f_l(k)H_l^{(2)}(kr) \}=\nn\\
=\left\{
 \begin{array}{c}
\ln f_l(k)+\ln H_l^{(2)}(kR)+ \ln \left[1+\fr{\bar{f}_l(k)
    H_l^{(1)}(kR)}{f_l(k)H_l^{(2)}(kR)}  \right]\nn\\ \\
\ln \bar{f}_l(k)+\ln H_l^{(1)}(kR)+\ln \left[1+\fr{f_l(k)H_l^{(2)}(kR)}{\bar{f}_l(k) H_l^{(1)}(kR)} \right]
 \end{array}  \right.
\label{ln_phi}
\eea
(the dropped constant $-\ln 2$ is unimportant for further calculations) and
consider the behaviour of (\ref{ln_phi}) at $R\ra\infty$.

The Hankel functions  $ H_l^{(1,2)}(z)$ are known to behave for a large
 argument $z\ra\infty$ as
 \bea
 H_{\nu}^{(1)}(z)\ \sim\sqrt{\fr{2}{\pi z}} e^{i(z-\fr{\nu\pi}{2}-\fr{\pi}{4})}\\
 H_{\nu}^{(2)}(z)\ \sim \sqrt{\fr{2}{\pi z}}
 e^{-i(z-\fr{\nu\pi}{2}-\fr{\pi}{4})}
 \label{H_asymp_infty}
 \eea

 Keeping in mind these features we use the first expression of (\ref{ln_phi})
for the upper branch of the integration contour $\gamma$ ($\Im k > 0$)
and the second expression of (\ref{ln_phi}) for the lower one $(\Im k < 0)$.
 This approach provides that the contribution of the third summand
 in (\ref{ln_phi})  (the term of kind $\ln(1+...)$) vanishes at
 $R\ra\infty$.
 The construction of the background potential is supposed to be of kind
 (\ref{scalar_potential}) as considered in Sec.\ref{hamiltonians} namely the non zero potential
 is concentrated in some environment of the coordinate origin $r=0$ and
 vanishes at infinity. Then the second term of (\ref{ln_phi}), namely the
 logarithm of Hankel \f does not depend on background properties at
 $R\ra\infty$ and can be treated therefore as a pure (infinite) contribution
 of the empty Minkowski space. It can be dropped on this step under this consideration.
  Thus we are left with the logarithm of the \jf only, that contains all the
 necessary informations in this approach. The following terms remain

 \bea
  E\sim\half \sum\limits_l \left\{ \sum\limits_{i}
  (m^2-k_{il}^2)^{1-s}+
\fr{1}{2\pi i}\int\limits_0^\infty(k^2+m^2)^{1-s} \fr{\pd}{\pd k} \ln \fr{f_l(k)}{\bar{f}_l(k)}\right\}.\label{discret+cont}
 \eea
 
 The sign of the integral is changed because of direction of moving around
 the contour $\gamma$ namely $\infty \ra 0$ above and $0\ra\infty$ below the
 real axis.
 
  This representation can be further simplified. In this order we turn the
  integration contour towards the imaginary axis. This action involves two
  transformations:
  
  1. The upper branch of the contour turns to the positive imaginary axis:
  $k\ra ik$ in the positive direction, so that $ k\ra e^{i\pi/2}k$;
  
  2. The lower branch of the contour turns also to the positive imaginary
  axis in the opposite direction;since some bound states can be present, that corresponds to simple zeros of the \jf on the imaginary axis. Therefore we must turn the upper branch of the contour to the imaginary axis in the negative direction to avoid
  the integration contour crossing the possible poles of $\ln f_l(k)$ in
  order to keep the value of integral unchanged during this transformation.
   Thus the argument $k$ must be transformed as $k\ra e^{-i\pi/2}k$. It
  corresponds to the change of the integral over the lower branch according to
\bea
\int\limits_0^\infty dk (k^2+m^2)^{1-s} \fr{\pd}{\pd k} \ln
  \bar{f}_l(k) \ra \int\limits^0_{-\infty} dk (k^2+m^2)^{1-s} \fr{\pd}{\pd k} \ln
  \bar{f}_l(k)=\nn\\
- \int\limits_0^\infty dk\ (k^2+m^2)^{1-s} \fr{\pd}{\pd k} \ln
  \bar{f}_l(-k)=
- \int\limits_0^\infty dk\ (k^2+m^2)^{1-s} \fr{\pd}{\pd k} \ln
 f_l(k),  
\eea
where the Schwarz reflection principle (\ref{schwarz_reflection}) has been used.      
  Thus we have for the contour integral:
  \bea
 && \left\{ -\int\limits_0^\infty dk (k^2+ m^2)^{1-s}\ln f_l(k) +
  \int\limits_0^\infty dk (k^2+ m^2)^{1-s}\ln \bar{f_l(k)} \right\}=\\
&&{-i}\left\{  \int\limits_0^\infty  dk (k^2- m^2)^{1-s}
  e^{i\pi(1-s)}\ln f_l(ik) - \int\limits_0^\infty  dk (k^2- m^2)^{1-s}
  e^{-i\pi(1-s)}\ln \bar{f_l(ik)}
  \right\}.\nn
  \eea
 Using the symmetry property (\ref{schwarz_reflection}) of the \jf again we get
for the integral
 \be
-\fr{\cos \pi s}{\pi} \int\limits_0^\infty dk (k^2- m^2)^{1-s}\ln  f_l(ik) 
 \ee

 If some bound states are present, that are placed on the imaginary axis
 between $0$  and $im$, then by moving of the contour up to the point $k=im$
we exclude the contribution of these states, that coincides exactly with the
sum over $k_{il}$ once being taken into account in the \ref{discret+cont}. Thus the both
contributions cancel each other and the final formula looks:
\be
E\sim\sum\limits_l \fr{\cos \pi s}{2\pi} \int\limits_m^\infty dk (k^2- m^2)^{1-s}\ln  f_l(ik). 
\ee

This is a very useful representation of the regularized ground state
energy.
 It means, we need only the \jf $f_l(ik)$ (speaking exactly, the logarithm of
 it) of the problem to obtain the regularized spectral sum. Finally we have
 to recall the starting expression of the regularized \gse to get
 
 \be
  E_0^{reg}(s)=-\frac{\mu^{2s}}{4\sqrt{\pi}} \frac{\Gamma(s-1)}{\Gamma(s-\half)}
\sum\limits_{l=-\infty}^\infty\ \fr{\cos \pi s}{2\pi} \int\limits_m^\infty dk
  (k^2- m^2)^{1-s}\ln  f_l(ik).
  \label{e_o_scalar_of_f_l}
  \ee

\subsubsection{Dirac spinor field}

\ \ \  Consider the same problem for the Dirac spinor field with the radial
 dependent potential $A_{\vp}\sim a(r)$. To this end we recall some formulae
 from  the Sec.(\ref{hamiltonians}). There it has been shown
 that the \eqq for 4-component spinor function is reduced to two \eqqs
for two 2-spinor \fs $\Psi$ and $\chi$

 \be
 \Phi(x)=e^{ip_z z}e^{-ip_0 x^0}\left[ \begin{array}{c}\phi\\ \chi \end{array} \right],\
\ee

 It is enough to consider here merely one of these \eqqs, say for $\phi$:
 \be
\left[\begin{array}{cc} p_0-m & L_-\\L_+& p_0+m \end{array}\right]
 \phi(x)=0
\ee

Using the suitable ansatz
\be 
\phi(x)=\left[\begin{array}{c}
    i\psi^u(r)e^{-i(l+1)\varphi}\\ \psi^l(r)e^{-il\varphi}\end{array}
\right]
\ee

we arrive at the radial \eqqs for two-component spinor $\psi^{u,l}(r)$.

\bea
\left\{\begin{array}{cc} p_0-m & \fr{\pd}{\pd r}-\fr{l-
 a(r)}{r}\\ \\-\fr{\pd}{\pd r}-\fr{l+1-a(r)}{r} & p_0+m
 \end{array}\right\}\psi(r)=0
\label{Dirac_of_r}
\eea

 The analytical solutions of this system for the special potential $a(r)$
 (\ref{a_r1}) are obtained in the Section.(\ref{hamiltonians}). Here we use the results of the
 previous subsection to construct the spectral sum via \jf as it has been done
 above with the scalar field problem, namely we proceed as follows:
 
  The component  $p_z$ of the momentum is separated and integrated out again in the
  similar way:
  
\begin{equation}
 E_0^{spin}(s)=-\frac{1}{2}\mu^{2s}\int\limits_{-\infty}^{\infty} \frac{dp_z}{2\pi}
 \sum\limits_{(n)} (p_z^2+k_{(n)}^2+m^2)^{\half-s}.
\end{equation}
 
Here $E_0^{spin}$ has again the meaning of density per unit length.
  Taking the states for $+p_0$ and $-p_0$ energy (upper and lower half of the spectrum
  respectively) and two different projections of spin into account we get the
  factor 4, since all these states possess the same energy. The general sign
  of the energy is minus because of the anticommutation relations for the
  Dirac field operators, as pointed out in Sec.(\ref{groundstenergy}).
  
  The explicit integration over $p_z/(2\pi)$ leads to
  
\begin{equation}
E_0(s)=-\mu^{2s}\frac{1}{4 \sqrt{\pi}} 4\frac{\Gamma(s-1)}{\Gamma(s-\half)}
\sum\limits_{(n)} (k_{(n)}^2+m^2)^{1-s} .
\label{e0sumk}
\end{equation}

Further we proceed like it has been done in \cite{borkir1}. A calculation for
a similar problem has been done in \cite{bednefrkirsant, BKEL}, however
without usage of the \jf formalism. The advance of the latter approach is a
possibility to gain the necessary uniform asymptotic expansion of the
regularized energy in the elegant way by using the \lseq (see
Sec.\ref{lippsw} below).
. 

 For technical reasons we assume again that the system is contained in a
large but finite cylinder of radius $R$ in order to get the discrete spectrum
of the transversal momenta $k_{(n)}$. The suitable boundary on the surface
$r=R$ conditions must be imposed. Though we cannot use the Dirichlet
conditions for all 4 components of the spinor $\Psi(x)$ as in the case of
scalar field because the spinor satisfied the Dirac equation with this
condition is identical zero. One of the possible solution is the so called
"MIT bag conditions'', originally proposed for bag models of hadrons \cite{jaffe} with
the constant homogeneous background field $B$ inside of the ''bag''
\bea
i\ n_{\mu}\gamma^{\mu}\psi =\psi\nn\\
\left( n_{\mu}\fr{\pd}{\pd x_{\mu}}\right)\bar{\psi}\psi=2B
\label{bag_conditions}
\eea

 To proceed with these boundary conditions we redefine the integer orbital
momentum number $l$ of the solutions by the half-integer $\nu$ according to (\ref{redef_l_to_nu}) in
Sec.(\ref{hamiltonians}).

 Choosing the normal unit 4-vector in Minkowski coordinates $(t,x,y,z)$ as
 ${n}^{\mu}=(0,-\cos\vp,-\sin\vp,0)$ and the representation of $\gamma$
 -matrices (\ref{gamma_matrix}) we obtain the spinor matrix $ n_{\mu}\gamma^{\mu}$ as

 \be
 n_{\mu}\gamma^{\mu}=\left[   
\begin{array}{cccc}
0 & -e^{-i\vp} &\  &\ \\  e^{i\vp}& 0 &\ &\ \\ \ &\ & 0 & e^{-i\vp}\\ \ &\ &
-e^{i\vp} & 0\\
 \end{array}
 \right]
 \ee

Then the application of the first condition  (\ref{bag_conditions}) to the
spinor

\be
\psi_{\nu}(R,\vp)=\half\bar{f}_{\nu}(k) \left[\begin{array}{c} i\sqrt{p_0+m}\Psi^{(1)}_{\nu+1/2}\\
    \sqrt{p_0-m}\Psi^{(1)}_{\nu-1/2}\\ i\sqrt{p_0-m}\Psi^{(1)}_{\nu+1/2}\\
    \sqrt{p_0+m}\Psi^{(1)}_{\nu-1/2}\\ \end{array}\right]+
\half f_{\nu}(k) \left[\begin{array}{c} i\sqrt{p_0+m}\Psi^{(2)}_{\nu+1/2}\\
    \sqrt{p_0-m}\Psi^{(2)}_{\nu-1/2}\\ i\sqrt{p_0-m}\Psi^{(2)}_{\nu+1/2}\\
    \sqrt{p_0+m}\Psi^{(2)}_{\nu-1/2}\\ \end{array}\right]
\ee
with
\be
\Psi^{(1,2)}_{\rho}=H^{(1,2)}_{\rho-\delta}(kR)e^{-i\rho\vp}
\ee
for positive $\nu$ on the cylindric boundary $r=R$ gives

\be
\left[
\begin{array}{cc} \sigma_1+I & 0\\0\ &\sigma_1-I \end{array}\right] \ \
\left(
  \begin{array}{c} \Phi^+(R)\\ \ \Xi^+(R)\end{array}
  \right)=0,
  \label{boundary_\eqqs}
\ee
 where
\bea
 \Phi^{+}(R)&=&
  \nn\\
 &=& \half \bar{f}^{+}_{\nu}(k) \left[
\begin{array}{r}\sqrt{p_0+m} H^{(1)}_{\nu+\half-\delta}(k R)\nn\\ \nn\\
  \sqrt{p_0-m} H^{(1)}_{\nu-\half-\delta}(k R)\end{array} \right]+\half f^{+}_{\nu}(k) \left[
\begin{array}{r}\sqrt{p_0+m} H^{(2)}_{\nu+\half-\delta}(k R)\nn\\ \nn\\  \sqrt{p_0-m}H^{(2)}_{\nu-\half-\delta}(k
 R)\end{array} \right],\nn
\eea
and $\Xi^+(R)$ can be obtained from $\Phi^{+}(R)$ through the substitution $m\ra-m$.
The $ \sigma_1$ denotes the Pauli matrix
\be \sigma_1=           
\left[
\begin{array}{cc} 0 & 1\\1\ & 0 \end{array}
\right]
\ee
and $I$ is the identity matrix,
therefore the determinants of $\sigma_1-I$ as well as  $\sigma_1+I$ are
zero. It means that the first couple of 4 \eqqs (\ref{boundary_\eqqs})
are in fact identical to each other. The same concerns the second couple of
(\ref{boundary_\eqqs}).

Finally the boundary condition (\ref{bag_conditions} ) provides only two
independent \eqqs:
\bea
&&{\cal B}^{(+E)}_{\nu}(k)\equiv \bar{f}^+_{\nu}(k)\left(\sqrt{p_0+m} H^{(1)}_{\nu+\half-\delta}(k R)+\sqrt{p_0-m}
H^{(1)}_{\nu-\half-\delta}(k R)\right)+\nn\\
&&\ \ \ \ \ \ \ f^+_{\nu}(k)\left(\sqrt{p_0+m} H^{(2)}_{\nu+\half-\delta}(k R)+\sqrt{p_0-m}
H^{(2)}_{\nu-\half-\delta}(k R)\right)=0\nn\\
&&{\cal B}^{(-E)}_{\nu}(k)\equiv \bar{f}^+_{\nu}(k)\left(\sqrt{p_0+m} H^{(1)}_{\nu-\half-\delta}(k R)-\sqrt{p_0-m}
H^{(1)}_{\nu+\half-\delta}(k R)\right)+\nn\\
&&\ \ \ \ \ \ \ f^+_{\nu}(k)\left( \sqrt{p_0+m} H^{(2)}_{\nu-\half-\delta}(k R)-\sqrt{p_0-m}
H^{(2)}_{\nu+\half-\delta}(k R)\right)=0.\nn\\
\label{boundary_functions}
\eea

Here we introduced the notations for boundary functions ${\cal B}^{(\pm)}$
for positive and negative values of the energy $p_0$ respectively.

To summarize the \ev $(m^2+k_{(n)}^2)$ of this boundary problem that are
implied by the boundary conditions (\ref{bag_conditions}) we proceed in the
similar way as in the previous subsection for scalar field. The \ev
$k_{(n)}$ are being viewed as the zeroes of the boundary functions
(\ref{boundary_functions}). We use again the properties of the logarithmic
derivative and write:

\be
\sum\limits_{(n)} (k_{(n)}^2-m^2)^{1-s} = \int\limits_\gamma dk
(k^2-m^2)^{1-s} \fr{\pd}{\pd k}[ \ln {\cal B}^{(+E)}_{\nu}(k)+ \ln {\cal B}^{(-E)}_{\nu}(k)],
\label{B+B-}
\ee
where the contributions of zeros from $ {\cal B}^{(+)}$ and $ {\cal
  B}^{(-)}$ are summarized independent. Finally we get for the contribution
of positive orbital momenta $\nu$, that

\be
E^{\nu+}_0(s)=-\mu^{2s}\frac{1}{\sqrt{\pi}}
\frac{\Gamma(s-1)}{\Gamma(s-\half)} \sum\limits_{\nu=1/2,3/2,\dots} 
\int\limits_\gamma\frac{dk}{2\pi i}(k^2+m^2)^{1-s} \partial_k \ln [{\cal B}^{(+E)}_{\nu}(k) {\cal B}^{(-E)}_{\nu}(k) ]. 
\label{E0_positiv_nu}
\ee

As it follows from the (\ref{redef_l_to_nu}) of Sec.\ref{hamiltonians}, it can be
easily changed from the positive orbital momenta to the negative ones
through the substitution of $-\nu$ instead of $\nu$. Note that the \jf in this
case is different from that considered above, and will be denoted as
$f^-_{\nu}(k)$.

  The change to the representation via \jf is similar to the one described
  above for the scalar case. This procedure is described in details in the App.(\ref{bag_perform}) 

Now after the subtraction of contribution from empty Minkowski space
(identified as the term divergent at $R\rightarrow\infty$, which is
independent of the background and corresponds to the contribution from the \hk
coefficient $a_0$,(which is simple the integral over empty space), and rotation of the path towards
the imaginary positive axis as described above one obtains \cite{borkir1}

\bea
&&E_0(s) = -\mu^{2s}\frac{1}{\sqrt{\pi}}
\frac{\Gamma(s-1)}{\Gamma(s-\half)}\frac{\cos \pi s}{\pi} 
\sum\limits_{l=-\infty}^{\infty} \int\limits_m^{\infty} dk
(k^2-m^2)^{1-s} \partial_k \ln f_l(ik)\nn\\
=&&\fr{-\mu^{2s}}{\sqrt{\pi}}
\frac{\Gamma(s-1)}{\Gamma(s-\half)}\frac{\cos \pi s}{\pi} 
\sum\limits_{\nu=1/2,3/2,\dots} \int\limits_m^{\infty} dk
(k^2-m^2)^{1-s} \partial_k [\ln f^+_\nu(ik)+\ln f^-_\nu(ik)]\nn,
\label{e0_of_lnf}
\eea

 A merit of this representation through the \jf of imaginary argument
 is the absence of oscillations of the integrand for large arguments (in
 contrary to the real \jf which is known to have an oscillating behaviour)\footnote{Note that bound states
are also taken into account in $E_0$ (\ref{e0_of_lnf}). In the
considered problem bound states appear if the flux is larger than one
flux unit. Strictly speaking, these are zero modes located on the
lower end of the continuous spectrum, i.e. at k=0 (this is known from
\cite{aharonov_casher}).}.



 \subsection{ The exact Jost \fs }
\label{jostfunc}
\ \ \ The \jf for problem with cylindrical symmetry was defined in the
Sec.\ref{spectralsum} as follows: The regular solution

\be
\vp^{reg}(r)\sim J_\nu (qr)\label{reg_sol}
\ee
of radial equation is defined to behave in the coordinate origin $r\ra
0$ as a free one. The solution chosen in such a way is known \cite{taylor}
to have at infinity a representation
\be
 \vp^{reg}(r)\sim\half[ \bar{f}(k) H_\nu^{(1)}(kr)+f(k) H_\nu^{(2)}(kr)].
 \label{reg_sol_infty}
\ee

$J_\nu(qr),H_\nu(kr)$ are the cylindrical Bessel/ Hankel \fs
respectively. Then the coefficients $f(k)$ and $\bar{f}(k)$ are the \jf
and its complex conjugated. It is assumed that the scattering potential $V(r)$
entering in the radial wave equation becomes constant $V(r)\ra V$ for large $r$ fast enough,
so that the relation \ref{reg_sol_infty} is valid. Practical this means
usually that the scattering area is localized around the symmetry axis of the
cylindric background.

\subsubsection{Scalar field with rectangular scalar background}
\ \ \  The scalar background $V(r)$ is considered to have a rectangular shape
  (Sec.\ref{hamiltonians},\ref{scalar_potential})
  
 \be
V(r)=\left\{\begin{array}{cc} V=const & R_1 < r < R_2\\ 0 & otherwise  \end{array}\right. 
 \ee
and the solutions
$\phi_{k,l}^I(r),\phi_{k,l}^{II}(r),\pd_r\phi_{k,l}^{III}(r) $ of the \eqq
\ref{scalar_\eqq} are chosen in the form of (\ref{scalar_solutions})
for three space domains under consideration of the potential shape.

Now we require the wave \f to be continuous and to have a continuous
first derivative. It imposes the matching conditions on the domain boundaries, i.e.:

\bea
&&\phi_{k,l}^I(r)=\phi_{k,l}^{II}(r)|_{r=R_1}\nn\\
&&\phi_{k,l}^{II}(r)=\phi_{k,l}^{III}(r)|_{r=R_2}\nn\\
&&\pd_r\phi_{k,l}^I(r)=\pd_r\phi_{k,l}^{II}(r)|_{r=R_1}\nn\\
&&\pd_r\phi_{k,l}^{II}(r)=\pd_r\phi_{k,l}^{III}(r)|_{r=R_2}.
\eea

 Thus no more undefined coefficients in (\ref{scalar_solutions}) remain and the resolution of this
 system of algebraic equations with respect to $f(k)$ provides us the
 \jf of the real momentum $k$ expressed in terms of the Bessel \fs
 $J$ and $H$. In order to proceed with the spectral sum we need to get
 the \jf $f(ik)$ of the imaginary argument. The transformation $k\ra ik$
 leads to the replacement of the Bessel \fs $H$ and $J$ by the
 corresponding modified Bessel \fs $K$ and $I$. The resulting
 expression allows to be simplified by Wronski determinants of $K$ and $I$
 and finally one arrives at
{\small
 \bea
&&f_l(ik) = R_1 R_2\times\\ &&\left\{
\left[q I_l(k R_1)  K'_l(q R_1)- k  I'_l(k R_1) K_l(q R_1)\right]
\left[k I_l(q R_2)K_l'(k R_2)-q I_l'(q R_2)K_l(k R_2)\right]\right. \nn\\ 
&&-\left.\left[q I_l(k R_1)I_l'(q R_1)-k I_l'(k R_1)I_l(q R_1)\right]
\left[k K_l(q R_2)K_l'(k R_2)-q K_l'(q R_2)K_l(k R_2)\right]\right\}\nn
\label{scalar_jost_exact}
\eea
 }
Here primes denote the derivation with respect to $r$. It is suitable to check
the useful symmetry property of the \jf. Taking in mind that the $l$ is
 integer and using the explicit symmetry relations for the modified Bessel
 \fs $K$ and $I$ \cite{gradst}:
 \bea
 I_l(z)=I_{-l}(z),\nn\\
 K_l(z)=K_{-l}(z)
 \eea
we can see  that
\be f_l(ik)=f_{-l}(ik) \label{symmetry_l}.\ee
This relation allows to consider further the \jf only for $l\ge 0.$

\subsubsection{Spinor field with rectangular magnetic background}

\ \ \  The radial Dirac equation (\ref{dirac_of_r})
corresponding to this problem with the chosen special
 form of background (\ref{a_r1},\ref{h_r1}), has been already solved in the
 Sec.\ref{hamiltonians}. The solutions $\Phi_I,\Phi_{II},\Phi_{III}$ in three
 space domains (\ref{domain_I} - \ref{domain_III}) are obtained in terms of Bessel and confluent hypergeometric \fs.
 
To get a \jf we need to impose certain matching conditions
like in the scalar problem considered above.
It is known that for a continuous potential $a(r)$ as we consider here, the
spinor wave \f must be continuous too. Thus
for the spinor wave \f (\ref{domain_I} - \ref{domain_III}) we demand only to
be continuous at the boundaries between I-II and II-III domains.
It means:
\begin{eqnarray}
\Phi_{II}^{\mp}(r) =\Phi_{I}^{\mp}(r)\Big|_{r=R_1}, \nn\\
\Phi_{III}^{\mp}(r) =\Phi_{II}^{\mp}(r)\Big|_{r=R_2}. 
\end{eqnarray}

Resolving these equations with respect to $f_{\nu}^{\pm}$ we obtain, that \\
\begin{eqnarray}
f_{\nu}^{\pm} (k)&=& 2\fr{ A + B }{ W_{III} W_{II} }\nn\\
 A &=& (\psi_{I}^{u\pm}\psi_{II. i}^{l\pm}-\psi_{I}^{l\pm}\psi_{II. i}^{u\pm})\Big|_{R_1}(\psi_{II. r}^{u\pm}\psi_{III. r}^{l\pm}-\psi_{II. r}^{l\pm}\psi_{III. r}^{u\pm})\Big|_{R_2},\nn\\
B &=& (\psi_{I}^{l\pm}\psi_{II. r}^{u\pm}-
 \psi_{I}^{u\pm}\psi_{II. r}^{u\pm})\Big|_{R_1}(\psi_{II. i}^{u\pm}\psi_{III. r}^{l\pm}-\psi_{II. i}^{l\pm}\psi_{III. r}^{u\pm})\Big|_{R_2}.
\end{eqnarray}

The denominator of this expression can be written using the Wronskians
of hypergeometric and Bessel \fs as follows

\begin{eqnarray}
W_{III}=(\psi_{III. i}^{u}\psi_{III. r}^{l}-\psi_{III. i}^{l}\psi_{III. r}^{u})\Big|_{R_2}=
(\fr{p_0+m}{p_0-m})^{\half} \fr{4i}{\pi k R_2}\nn\\
\\W_{II}=(\psi_{II. r}^{u}\psi_{II. i}^{l}-\psi_{II. r}^{l}\psi_{II. i}^{u})\Big|_{R_1}=
-\fr{\alpha\beta}{p_0-m}(\fr{\beta R_1}{2})^{-\alpha-1}.
\end{eqnarray} 

To calculate $E_0$ we need the \jf of imaginary momentum 
(\ref{e0_of_lnf}), so we need to replace $k$ by $ik$. As a result we obtain
the new expression for $f(ik)$ that contains now modified Bessel \fs $I_{\nu\pm \half}$ instead
of $J_{\nu\pm \half}$ and modified Bessel \fs $K_{\nu\pm \half\mp \delta}$
 (McDonald \fs) instead of $H^{(1, 2)}_{\nu\pm \half\mp \delta}$.
 
\begin{eqnarray}
 f^{+}_\nu(ik)&&= \fr{\pi k R_1 R_2 e^{-\delta /\kappa+\fr{i\pi\delta}{2}}}
 {\half- \tilde{\nu}} \times \nn \\
&& \left\{ \left[ \fr{kR_1}{2\tilde{\nu}+1}\  
{_1}F_1(1+\fr{k^2}{2\beta},\ \tilde{\nu}+\fr{3}{2};\ \fr{\beta R_1^2}{2})
\ I_{\nu-\half}(k
 R_1)-\right.\right.\nn\\&&\left.\hspace*{4cm}{_1}F_1(\fr{k^2}{2\beta},\
 \tilde{\nu}+\half;\ \fr{\beta R_1^2}{2})
\ I_{\nu+\half}(kR_1) \right]\times\nn\\ 
&&\left[\fr{2\tilde{\nu}+1}{R_2}\
 {_1}F_1(\half-\tilde{\nu}+\fr{k^2}{2\beta},\ \half-\tilde{\nu};\ \fr{\beta R_2^2}{2})\ K_{\nu-\half-\delta}
 (kR_2)-\right.\nn\\ 
&&\left.\hspace*{3cm} k\ {_1}F_1(\half-\tilde{\nu}+\fr{k^2}{2\beta},\
 \fr{3}{2}-\tilde{\nu};\ \fr{\beta R_2^2}{2})\ K_{\nu+\half-\delta}(kR_2)\right]- \nn\\
&&\left[k\
 {_1}F_1(\half-\tilde{\nu}+\fr{k^2}{2\beta},\ \fr{3}{2}-\tilde{\nu};\ \fr{\beta
 R_1^2}{2})\ I_{\nu+\half}(kR_1)+\right.\nn\\
&&\left.\hspace*{3cm}\fr{2\tilde{\nu}-1}{R_1}\
 {_1}F_1(\half-\tilde{\nu}+\fr{k^2}{2\beta},\ \half-\tilde{\nu};\ \fr{\beta
 R_1^2}{2})\ I_{\nu-\half}(kR_1)\right]\times\nn\\ 
&&\left[\fr{kR_2}{2\tilde{\nu}+1}\
 {_1}F_1(1+\fr{k^2}{2\beta},\ \tilde{\nu}+\fr{3}{2};\ \fr{\beta R_2^2}{2})\
 K_{\nu-\half-\delta}(kR_2)+\right.\nn\\&&\left.\left.\hspace*{4cm}{_1}F_1(\fr{k^2}{2\beta},\
 \tilde{\nu}+\half;\ \fr{\beta
 R_2^2}{2})\ K_{\nu+\half-\delta}(kR_2)\right]\right\},\nn
\label{f_v_of_ik+}
\end{eqnarray}
where $\tilde{\nu}= \nu + \fr{\delta R_1^2}{R_1^2-R_2^2} $, and

\begin{eqnarray}
 f^{-}_\nu(ik)&&= \fr{\pi k R_1 R_2 e^{-\delta /\kappa-\fr{i\pi\delta}{2}}}
 {\half+ \bar{\nu}} \times \\
&& \left\{ \left[ \fr{kR_1}{2\bar{\nu}-1} 
\ {_1}F_1(\bar{\nu}+\half+\fr{k^2}{2\beta},\ \bar{\nu}+\half;\ \fr{\beta R_1^2}{2})
\ I_{\nu+\half}(rR_1)-\right.\right.\nn\\
&&\left.\hspace*{4cm}{_1}F_1(\bar{\nu}+\half+\fr{k^2}{2\beta},\
 \bar{\nu}+\fr{3}{2};\ \fr{\beta R_1^2}{2})
\ I_{\nu-\half}(kR_1) \right]\times\nn\\ 
&&\left[\fr{2\bar{\nu}-1}{R_2}\  
  {_1}F_1(1+\fr{k^2}{2\beta},\ \fr{3}{2}-\bar{\nu};\ \fr{\beta R_2^2}{2})
  \ K_{\nu+\half+\delta}(kR_2)-\right.\nn\\
&&\left.\hspace{4cm}
 k\ {_1}F_1( \fr{k^2}{2\beta},\ \half-\bar{\nu};\ \fr{\beta R_2^2}{2})
\ K_{\nu-\half+\delta}(kR_2)\right]- \nn\\
&&\left[k\ {_1}F_1(\fr{k^2}{2\beta},\ \half-\bar{\nu};\ \fr{\beta R_1^2}{2})
\ I_{\nu-\half}(kR_1)+\right.\nn\\
&&\left.\hspace*{4cm}\fr{2\bar{\nu}+1}{R_1}
  \ {_1}F_1(1+\fr{k^2}{2\beta},\ \fr{3}{2}-\bar{\nu};\ \fr{\beta R_1^2}{2})
  \ I_{\nu+\half}(kR_1)\right]\times\nn\\ 
&&\left[\fr{kR_2}{2\bar{\nu}-1}
  \ {_1}F_1(\bar{\nu}+\half+\fr{k^2}{2\beta},\ \bar{\nu}+\half;\ \fr{\beta R_2^2}{2})
  \ K_{\nu+\half+\delta}(kR_2)+\right.\nn\\
&&\left.\left.\hspace*{4cm}{_1}F_1(\bar{\nu}+\half+\fr{k^2}{2\beta},\
 \bar{\nu}+\fr{3}{2};\ \fr{\beta R_2^2}{2})
 \ K_{\nu-\half+\delta}(kR_2)\right]\right\},\nn
\label{f_v_of_ik-}
\end{eqnarray}
where $\bar{\nu}= \nu - \fr{\delta R_1^2}{R_1^2-R_2^2} $.

\subsubsection{Spinor field with background of Nielsen-Olesen Vortex}

\ \ \  The technical distinction of this case from those recently considered
 above is the not explicit background \f, given by two \fs $f(r)$ and $v(r)$,
 being the solutions of the system of two second-order equations
 (\ref{no_vortex_eq}) - Nielsen-Olesen (NO) \eqqs. This system has no known solutions representable in analytical
 form. Therefore one cannot apply the experience of two previous models to
 obtain some analytical expression for the \jf. 
  Nevertheless the system of Nielsen-Olesen equations is known to be solvable
at least by numerical methods. Let we have the radial part of 2-component spinor \f $\Phi_{l,k}(r)$ to be a (numerical) solution of the 
system

\be
\left\{\begin{array}{cc} p_0-\fr{f_e\eta}{\sqrt{2}}f(r) & \fr{\pd}{\pd r}-\fr{l-n
 v(r)}{r}\\ \\-\fr{\pd}{\pd r}-\fr{l+1-n v(r)}{r} & p_0+\fr{f_e\eta}{\sqrt{2}}f(r)
 \end{array}\right\}\Phi(r)=0
\label{Dirac_of_r}
\ee
with
\be
\Phi(r)=\left\{\begin{array}{c} \psi^{u}(r)\\ \\ \psi^{l}(r)  
\end{array}\right\}, 
\label{spinor_2comp}
\ee
indices $l$ and $u$ denote ``upper'' and ``lower'' components of spinor respectively.
These \eqqs can be solved numerically for any specified \fs $v(r)$
and $f(r)$.

 In order to define the \jf properly we start as usual with the definition of
 the regular solution for the system (\ref{Dirac_of_r}).
   The behaviour of the potentials $f(r)$ and $v(r)$ is known for $r\ra 0$ and
   $r\ra \infty$, since they define a vortex solution.

   The \f  $\fr{f_e\eta}{\sqrt{2}}f(r)$ entering here plays a role of
   space dependent fermion mass and for further purposes it will be denoted
   as $\mu(r)$.
   
   The usual boundary conditions for (\ref{dirac_of_r}) are generally speaking
  
\be
f(0)=v(0)=0;\ \ \ \ \  f(\infty)= v(\infty)=1;
\label{NO_init_cond}
\ee
 for the Nielsen-Olesen vortex. It corresponds to the situation that the
 fermionic field in the center of the vortex become massless
 ( $\fr {f_e\eta}{\sqrt{2}}f(r)= \mu(r)= 0$), and  outside of the vortex (far
 enough from the vortex core) has the observable mass $m_e$. In order to
 check the symmetry properties of solutions better, we suppose for a moment
 that
\be
\mu(0)=m_0, 
\ee
$m_0$ is some constant, that will be later set to zero.

 The behaviour of the $f(r),v(r)$ for vorticity $n=1$ closed to zero follows
 from the corresponding asymptotics of the NO-\eqqs (\ref{no_vortex_eq}) and
 is known to be

\be
\left. \begin{array}{ccc}f(r)& \sim & J_1(\eta \sqrt{\la}r)\\v(r)&\sim&r^2
  \end {array}\right\},\ \ r\ra 0.
\ee

 Further we must take respect to this fact, that the fermionic mass
 distinguishes for $r\ra 0$ and $r\ra \infty$ and are equal to $m_0$ and $m_e$
 respectively. Note also that $0 < m_0 < m_e$. Thus we have to define two radial momenta
 \bea
 q=\sqrt{p_0^2-m_0^2},\nn\\
 k=\sqrt{p_0^2-m_e^2}.
\label{momenta_k_q}
 \eea

 Now the definition of the \jf demands some generalization. Here we follow
 the approach derived in \cite{borkirhel}. The regular
 solution of  (\ref{Dirac_of_r}) will be chosen to be proportional to

 \be
 \psi^{reg}(r)\sim \left( \fr{k}{q}\right)^{l+1}\left\{\begin{array}{cc} \sqrt{p_0+m_0} &
 J_{l-n+1}(q r)
 \\ \\\sqrt{p_0-m_0}  & J_{l-n}(q r).
 \end{array}\right\}
 \ee
 
 Further we represent the regular solution $\psi^{reg}(r)$ at arbitrary $r\not=0$ in the form:
\be
\psi(k,r)_{\infty}= \half\left\{ \bar{f}_l(k,r) \psi(k,r)_{\infty}^{(1)}+
  f_l(k,r) \psi(k,r)_{\infty}^{(2)}\right\},
\label{generalized_\jf}
\ee
where 

\bea
\psi_l(k,r)_{\infty}^{(1,2)}=\left\{\begin{array}{cc} \sqrt{p_0+m_e} &
 H^{(1,2)}_{l-n+1}(k r)
 \\ \\\sqrt{p_0-m_e}  & H^{(1,2)}_{l-n}(k r)
 \end{array}\right\}.
\eea

 Here the coefficients $f_l,\tilde{f_l}$ depend now in general on the radius
$r$ of the cylindrical shell. The background potential enclosed inside of this
is considered as the scattering potential.

 Then it follows from the asymptotic behaviour $r\ra \infty$ of the profile \fs
 $f(r),v(r)$, that for the radius $r$ large enough the \fs
 $\tilde{f_l(q,r)}$, $f_l(q,r)$ do not depend more on $r$ and coincide
 therefore with the \jf $f_l(k)$ and its complex conjugate $\bar{f}_l(k)$
 respectively for this scattering problem.
  Thus we define the \f $f_l(q,r)$, which for large $r$ approximates the \jf.
  The detailed motivation of this assumption is given in \cite{borkirhel}.

  The $l$ is replaced up from here by $\nu$ as usual by means of
  (\ref{redef_l_to_nu}, Sec.\ref{hamiltonians}). Thus for
  $\nu=1/2,3/2,5/2,\dots$ we have two solutions for positive and negative
  orbital momenta, denoted below as $\nu+$ and $\nu-$ respectively.
\bea
&&\psi_{\nu+}(k,r)^{\infty}=\\
&&\half\left\{
\bar{f}_{\nu}^+(k,r)\left[\begin{array}{cc} \sqrt{p_0+m_e} &
 H^{(1)}_{\nu+1/2-n}(k r)
 \\ \\\sqrt{p_0-m_e}  & H^{(1)}_{\nu-1/2-n}(k r)
 \end{array}\right]+\right. \nn\\&&\left.\hspace*{4cm}
f_{\nu}^+(k,r)\left[\begin{array}{cc} \sqrt{p_0+m_e} &
 H^{(2)}_{\nu+1/2-n}(k r)
 \\ \\\sqrt{p_0-m_e}  & H^{(2)}_{\nu-1/2-n}(k r)
 \end{array}\right]
\right\}\nn\\
&&\psi_{\nu-}(k,r)^{\infty}=\\
&&\half\left\{
\bar{f}_{\nu}^-(k,r)\left[\begin{array}{cc} \sqrt{p_0+m_e} &
 H^{(1)}_{-\nu+1/2-n}(k r)
 \\ \\-\sqrt{p_0-m_e}  & H^{(1)}_{-\nu-1/2-n}(k r)
 \end{array}\right]+\right.\nn\\&&\left.\hspace*{4cm}
f_{\nu}^-(k,r)\left[\begin{array}{cc} \sqrt{p_0+m_e} &
 H^{(2)}_{-\nu+1/2-n}(k r)
 \\ \\-\sqrt{p_0-m_e}  & H^{(2)}_{-\nu-1/2-n}(k r)
 \end{array}\right]
\right\}\nn
\eea

The spinor \fs $\psi(k,r)_{\nu\pm}^{\infty}$ are the (numerical) solutions of (\ref{Dirac_of_r}).
 
 Thereby the generalized (r-dependent) \jf by means of (\ref{generalized_\jf}) can be calculated as follows
 
\bea
&&f^{+}_{\nu}(k,r) =\nn\\&&\fr{-i\pi r}{2}\left\{
 \sqrt{p_0-m_e} H^{(1)}_{\nu-\half-n}(kr)\ \psi^{u+}_{\nu}(k,r)-\sqrt{p_0+m_e} H^{(1)}_{\nu+\half-n}(kr)\ \psi^{l+}_{\nu}(k,r) 
\right\}\nn\\
\\
&&f^{-}_{\nu}(k,r)=\nn\\&& \fr{i\pi r}{2}\left\{
 \sqrt{p_0-m_e} H^{(1)}_{\nu+\half+n}(kr)\ \psi^{u-}_{\nu}(k,r)+\sqrt{p_0+m_e} H^{(1)}_{\nu-\half+n}(kr)\ \psi^{l-}_{\nu}(k,r) 
\right\},\nn\\
\label{f_l_of_k_plus_minus}
\eea
with the initial condition
  \be\left.
 \psi_{0}^\pm(r)=\left( \fr{k}{q}\right)^{\nu+\half}\left\{\begin{array}{cc} \sqrt{p_0+m_0} &
 J_{\nu\pm\half}(q r)
 \\ \\ \pm\sqrt{p_0-m_0}  & J_{\nu\mp\half}(q r)
 \end{array}\right\}\right|_{r=0}.
\label{init_cond_nu}
 \ee
 
The spinor upper and lower components \fs $\psi^{l\pm,u\pm}_{\nu}(k,r) $ obey the \eqqs:
\bea
\left\{\begin{array}{cc} p_0-\mu(r) & \fr{\pd}{\pd r}-\fr{\nu-\half-n
 v(r)}{r}\\ \\-\fr{\pd}{\pd r}-\fr{\nu+\half-n v(r)}{r} & p_0+\mu(r)
 \end{array}\right\} \left\{\begin{array}{c} \psi^{u+}_{\nu}(r)\\ \\
 \psi^{l+}_{\nu}(r)\end{array}\right\}=0\ \ \nn\\
\left\{\begin{array}{cc} p_0-\mu(r) & \fr{\pd}{\pd r}-\fr{-\nu-\half-n
 v(r)}{r}\\ \\-\fr{\pd}{\pd r}-\fr{-\nu+\half-n v(r)}{r} & p_0+\mu(r)
 \end{array}\right\} \left\{\begin{array}{c} \psi^{u-}_{\nu}(r)\\ \\
 \psi^{l-}_{\nu}(r)\end{array}\right\}=0.\ \ \
\label{Dirac_nu_r}
\eea

However we are interesting again for the \jf of the imaginary momentum $ik$, so
we make the transformation $k\ra ik$, like the one considered above for
magnetic background. By means of (\ref{momenta_k_q}) this implies

\bea
p_0\ra p=\sqrt{m_e^2-k^2}=i\sqrt{k^2-m_e^2}=i\tilde{p}\nn\\
q=\sqrt{k^2-m_0^2+m_e^2}\ \ra \sqrt{m_e^2-k^2-m_0^2}=i\sqrt{k^2-(m_e^2-m_0^2)}=
i\tilde{q},
\label{k_to_ik}
\eea
resulting finally in

\bea
&&f^{+}_{\nu}(ik,r)=-(-i)^{\nu-\half-n} r \left\{
 \sqrt{i\tilde{p}-m_e} K_{\nu-\half-n}(kr)\ \psi^{u+}_{\nu}(ik,r)+\right.\\
 &&\left.\hspace*{5cm} i\sqrt{i\tilde{p}+m_e} K_{\nu+\half-n}(kr)\ \psi^{l+}_{\nu}(ik,r) 
\right\}\nn\\
\nn\\
&&f^{-}_{\nu}(ik,r)=(-i)^{\nu+\half+n} r\left\{
 \sqrt{i\tilde{p}-m_e} K_{\nu+\half+n}(kr)\ \psi^{u-}_{\nu}(ik,r)+\right.\\
 &&\left.\hspace*{5cm} i\sqrt{i\tilde{p}+m_e} K_{\nu-\half+n}(kr)\ \psi^{l-}_{\nu}(ik,r) 
\right\}\nn
\label{f_l_of_ik_plus_minus}
\eea

 The \fs $\psi^{l\pm,u\pm}_{\nu}(ik,r) $ entering here obey the modified
 Dirac \eqq which is to obtain from the original one (\ref{Dirac_nu_r})
through the replacement of $p_0$ by $i\tilde{p}$ according to (\ref{k_to_ik}).

 From the (\ref{f_l_of_ik_plus_minus}) together with the \eqqs
 (\ref{Dirac_nu_r}) modified by means of  (\ref{k_to_ik}) follows in
 particular, that for the \jf of imaginary momentum a change of the sign of
 scalar potential $\mu$ corresponds to complex conjugation and a change in
 the sign of the magnetic background $v(r)$ to opposite one $v(r)\ra -v(r)$
 corresponds to an exchange of positive and negative orbital momenta
 $f^{+} \leftrightarrow f^{-}$. It leads to the simplification of the
 resulting formula for the regularized \gse, which includes the sum of
 logarithms of $f^{+}$ and $f^{-}$. The odd powers of $v(r)$ disappear and
 pure imaginary contributions of $\mu$ as well. These features are described
 and argued in details in \cite{Bordag:2003at}.
 
 Using these formulae we could be able to calculate the needed \jf
 numerically. To this end we must have numerical solutions $
 \psi^{u\pm}_{\nu}(ik,r)$ of the Dirac \eqq with the profile \fs $f(r),v(r)$
 which are calculated from the NO-\eqqs (\ref{no_vortex_eq}) numerically as well.\\
\\

{\bf Numerical calculation of $f^{\pm}_{\nu}(ik,r)$}\\

The practical experience shows the need of higher raised precision to get
  correct results for $\psi^{l\pm,u\pm}_{\nu}(ik,r) $ by usage of "NDSolve'' package of "Mathematica" for the radial
Dirac \eqq directly in the form

\be
\left\{\begin{array}{cc}i\tilde{p} -\mu(r) & \fr{\pd}{\pd r}-\fr{\nu-\half-n
 v(r)}{r}\\ \\-\fr{\pd}{\pd r}-\fr{\nu+\half-n v(r)}{r} & i\tilde{p}+\mu(r)
 \end{array}\right\} \left\{\begin{array}{c} \psi^{u+}_{\nu}(r)\\ \\
 \psi^{l+}_{\nu}(r)\end{array}\right\}=0,\label{Dirac_imag}\ee
and starting with the initial condition following from (\ref{init_cond_nu})
   \be
 \psi_0^\pm(r) = \left.i^{\nu+\half}\left( \fr{k}{\tilde{q}}\right)^{\nu+\half}\left\{\begin{array}{cc} \sqrt{\tilde{p}+m_0} &
 I_{\nu\pm\half}(\tilde{q} r)
 \\ \\ \mp i\sqrt{\tilde{p}-m_0}  & I_{\nu\mp\half}(\tilde{q} r)
 \end{array}\right\}\right|_{r=0}.
\label{init_cond_iq_nu}
 \ee

To avoid this trouble we apply the representation of regular solution at the
  point $r=0$ (\ref{init_cond_iq_nu}) in the form

\bea
 \psi^{u\pm}_{\nu}(k,r)= I_0(\nu) \bar{\psi}^{u\pm}_{\nu}(k,r),\nn\\
 \psi^{l\pm}_{\nu}(k,r)= I_0(\nu+1) \bar{\psi}^{l\pm}_{\nu}(k,r),\nn
\eea
where $I_0$ are defined from the asymptotic behaviour of
  \be\left[
\begin{array}{c}
    I_{\nu+\half}(kr)\\
    I_{\nu-\half}(kr)
 \end{array}\right]
\ee
for small arguments, namely:
\be
 I_0(\nu)=I_{\nu-\half}(kr)|_{r\ra 0} \sim \fr{(kr)^{\nu-1/2}}{2^{\nu-1/2}\Gamma(\nu+1/2)}
\ee

\be
I_0(\nu+1)=\fr{kr}{2\nu+1}I_0(\nu).
\ee

Thus we can rewrite the initial condition as
\be
\psi^-(r)|_{r=0}=
\left[\begin{array}{c}
I_0(\nu)\\
-iI_0(\nu+1)    
\end{array}\right];\ \ \ \  
\ \ 
\psi^+(r)|_{r=0}=
\left[\begin{array}{c}
iI_0(\nu+1)\\
I_0(\nu)    
\end{array}\right]. 
\ee

Then for $r\ra 0$ the initial condition for numerical iterations of the
system (\ref{Dirac_imag}) in terms of $ \tilde{\psi^\pm(r)} $ looks as
\bea
\tilde{\psi}^\pm(r)|_{r=0}=
\left[\begin{array}{c}
    1\\1
  \end{array}\right].
\eea

 Now the final formulae to calculation of the \jf in terms of the redefined
 wave \fs looks

\bea
&&f^+_{\nu}(ik,r)=\\&&\ \ \ -I_0(\nu)(-i)^{\nu-\half-n}r
\left\{
\sqrt{i\tilde{p}-m_e} K_{\nu-\half-n}(kr) \fr{kr}{2(\nu+\half)} \tilde{\psi}^{u+}+i\sqrt{i\tilde{p}+m_e}K_{\nu-\half+n}(kr) \tilde{\psi}^{l+}
  \right\}\nn\\
&&f^-_{\nu}(ik,r)=\\&&\ \ \ I_0(\nu)(-i)^{\nu+\half+n}r
\left\{
\sqrt{i\tilde{p}-m_e} K_{\nu+\half+n}(kr)\tilde{\psi}^{u-}+i\sqrt{i\tilde{p}+m_e} K_{\nu-\half+n}(kr)\fr{kr}{2(\nu+\half)} \tilde{\psi}^{l-}
  \right\},
  \nn
\eea
which are to evaluate numerically.


 \subsection{The uniform asymptotics of the \jf from the\\ \lseq}
\label{lippsw}
\ \ \ The present section contains the explicit derivation of the uniform
asymptotics of the \jf that is necessary to construct the regularized ground
state energy $E^{reg}$ has been used in the previous Section (\ref{efineas})

The uniform asymptotics of the \jf $f_{\nu}(ik)$ is the power
expansion in the combined limit $k,\nu \ra \infty$ at $k/\nu\equiv z=const$.
 It can be obtained in various ways. One can make for example the explicit
 substitution of asymptotic expansions for $K_l(kr),\ I_l(kr)$ and other
 \fs that the \jf $f_{\nu}(ik)$ consists of. Another approach is the iteration
 of the \lseq for the corresponding scattering problem.

This procedure is the most laborious analytical topic of the method.

\subsubsection{The \lseq }

\ \ \  The \lseq is one of important tools of the scattering theory.
 This is simple an expression of that fact, that the
 scattered wave function $\phi(x)=\phi^{out}(x)$ observed far enough from the
 scattering area can be represented through the initial (unscattered)
 wave function $\phi^{0}= \phi^{in}(x)$ and integral of interaction with
 the scattering potential.

  It can be shown to follow from the theory of ordinary linear differential
 \eqqs:

   We suppose $\phi^{0}(x)$ to be the general solution for the homogeneous
 \eqq (or "free solution" by notation in scattering theory)

\be
  {\bf\hat{P}} \phi^{0}(x)=0
  \label{hom_eq}
\ee
 and $\phi^{V}(x)$ to be any particular solution for the inhomogeneous equation
of kind
\be
  {\bf\hat{P}} \phi^{V}(x)=V(x)\phi^{V}(x)
  \label{inhom_eq}
\ee
with the V(x)- generally speaking some \f of coordinates $x$, in our
case it plays the role of the potential.
Then we have that the general solution $\phi(x) $ of the inhomogeneous
 equation (\ref{inhom_eq}) reads 
\be
\phi(x)=\phi^{0}(x)+\phi^{V}(x),
\label{general_solution}
\ee
in accordance to the statement about general solutions of
 inhomogeneous equation \ref{inhom_eq}, (e.g. \cite{vladim}).

 On the other hand the  particular solution $\phi^{V}(x)$ of the inhomogeneous
\eqq \ref{inhom_eq} can be expressed by using the Green function $g(x,x')$
of the operator $\bf{\hat{P}}$, satisfying the definition

\be
\bf{\hat{P}} g(x,x')=\delta (x-x').
\label{green_def}
\ee

Suppose such a \f is known. Then the solution of any inhomogeneous \eqq
(\ref{inhom_eq}) can be restored from the inhomogeneous part (right-hand
part) by the integration 
\be
\phi^{V}(z)=\int\limits_0^x g(x,x')V(x')\phi(x') dx'.
 \label{inhom_green}
\ee
Thus we obtain the integral equation for the general solution $\phi(x)$
\be
 \phi(x)=\phi^{0}(x)+\int\limits_0^x g(x,x')V(x')\phi(x') dx',
\label{lipp_schw_gen} 
\ee
that calls in applications to scattering problems ``Lippmann-Schwinger equation'' 

 We consider below the several cases of usage of the equation \ref{lipp_schw_gen} for potentials $V(x)$ with cylindric symmetry.
 
\subsubsection{Scalar field}

\ \ \ The \eqq of motion for the massive scalar field $\vp(x)$ of the
mass $m$ in the scalar background $V(x)$ depending in cylindrical coordinates
on the radial variable $r$ only, has been reduced to the one for radial part
(Sec.\ref{hamiltonians})

\be
\left\{ \fr{\pd^2}{\pd r^2}  + \fr{1}{r} \fr{\pd}{\pd r} +\left(k^2-\fr{l^2}{r^2} \right)\right\} \phi(r)=0.
\label{scalar_motion}
\ee

Then the \lseq (\ref{lipp_schw_gen}) reads:
\be
 \phi_l(z)= \phi_l^0(z)+ \int\limits_0^{\infty}dz'G_l(z,z')V(z')\phi_l(z'),
\label{lippman_schw}
\ee
 where the Green \f $G_l(k,r,r')$ of this \eqq is constructed
 according to the known prescription \cite{taylor, ryder} as

\be
G_l(z,z')= \fr{i\pi}{2} [ J_l(z) H_l^{(1)}(z')-J_l(z') H_l^{(1)}(z) ],
\label{green_scalar}
\ee
and $z$ denotes $z\e kr$.

 It can be easily proved by explicit substitution that (\ref{green_scalar})
 satisfies the definition of the Green \f (\ref{green_def}) applied to
 the \eqq (\ref{scalar_motion}).

 Further, substituting the Green function (\ref{green_scalar}) into the
 \lseq (\ref{lippman_schw}) and choosing the solution
$ \phi_l^0(z)$ of the homogeneous \eqq (i.e. at $V(z)=0$) as
\be
 \phi_l^0(z)= J_l(z),
 \label{homogen_sol_scalar}
\ee
we have for the \lseq

 \be
 \phi_l(k r)=J_l(k r)+\fr{i\pi}{2}\int\limits_0^r r' dr' V( r')[J_l(k
 r)H_l^{(1)}(k r')- J_l(k r') H_l^{(1)}(k r) ]\phi_l(k r').
 \label{lipp_scw_of_j_h}
 \ee

 Now we need to obtain the expression for our required \jf $f_l(k)$
 defined in (\ref{jost_of_h1_h2}). Let the solution of the inhomogeneous equation
 $\phi^{V}(k r)$ \ref{inhom_eq} far from the scattering area be denoted as
 $\phi^{V}_{r\ra \infty}(k r)$. Since the solution at $r\ra 0$ has been
 chosen to be a "regular one'' (Sec.\ref{jostfunc}, Eq.\ref{reg_sol} ), it can
 be represented at $r\ra \infty$ (Sec.\ref{jostfunc}, Eq.\ref{reg_sol_infty}) as 
 
 \be
 \phi^{V}_{r\ra \infty}(k r)=\half [\bar{f}_l(k)H_l^{(1)}(k r)+
 f_l(k)H_l^{(2)}(k r)].
 \label{jost_of_h1_h2}
 \ee

 Substituting this relation into (\ref{lipp_scw_of_j_h}) and using the
 certain relations between Bessel and Hankel functions ( $J_l$ and
 $H_l^{1,2}$), \cite{gradst} 
\be
 J_{\nu}(z)=\half\{H^{(1)}_{\nu}(z)+ H^{(2)}_{\nu}(z) \}
\ee
 we arrive at the relation

\bea
&& \half [\bar{f}_l(k)H_l^{(1)}(k r)+ f_l(k)H_l^{(2)}(k r)] =\nn\\
&&\half H^{(2)}_{l}(kr)\left[ 1+\frac{i\pi}{2}\int\limits_0^\infty r' dr' V( r') H_l^{(1)}(k r') \phi_l(k r') \right]+\nn\\
&&\half H^{(1)}_{l}(kr)\left[ 1-\frac{i\pi}{2}\int\limits_0^\infty r' dr' V( r') H_l^{(2)}(k r') \phi_l(k r') \right]
\label{to_\jf_implicit}  
\eea
whereby the \jf $f_l(k)$ is left to be identified as

 \be
 f_l(k)=1+\fr{i\pi}{2}\int\limits_0^\infty r' dr' V( r') H_l^{(1)}(k r')
 \phi_l(k r')
 \label{jost_func_of_LS}
 \ee
 and for $\bar{f_l(k)}$ one obtains 

\be
 \bar{f}_l(k)=1-\fr{i\pi}{2}\int\limits_0^\infty r' dr' V( r') H_l^{(2)}(k r') \phi_l(k r')
\label{jost_func_bar_of_LS}
\ee
respectively, 
that is really not surprising, and confirms the supposition once more, that
$f_l(k)$ and $\bar{f_l(k)}$ are indeed complex conjugated to each other, what
was not initially adopted.

 Going on with the \rg procedure we have to transform the obtained
 real \jf $f_l(k)$ to the complex one, changing the argument $k$ to
 $i k$ respectively. It provides \cite{gradst, abrsteg}

 \be
 f_l(ik)=1+i^{-l}\int\limits_0^\infty r' dr' V(r') K_l(k r') \phi_l(ik r'),
\label{jost_func_ik_of_LS}
 \ee
where  $K_l^{(1)}(k r)$ is the modified Bessel \f (McDonald \f).
 The wave function $\phi_l(ik r')$ entering here obeys now the \eqq derived
 from (\ref{lipp_scw_of_j_h}) by the substitution $k\ra ik$

 \be
 \phi_l(ik r)=i^l I_l(k r)+\int\limits_0^r r' dr' V( r')[I_l(k
 r)K_l(k r')- I_l(k r') K_l(k r) ]\phi_l(ik r'),
 \label{lipp_scw_of_k_i}
 \ee
 ($I_l(z)$ is the modified Bessel \f, \cite{gradst, abrsteg}). On comparing
 (\ref{jost_func_ik_of_LS}) with (\ref{lipp_scw_of_k_i}) it can be seen that
 the factor $i^{\pm l}$ provides only the redefinition of the wave \f and is
 irrelevant for further calculations. We omit these factors in both equations.
 
 Solving the \eqq (\ref{lipp_scw_of_k_i}) by iterations of $\phi_l(ik r')$ and
 beginning from the $\phi_l^{(0)}(ik r)=I_l(k r)$ as a zero order approximation
 of $\phi_l$ one obtains after two iterations:

 \bea
&&\phi_l^{(1)}(ik r)=I_l(k r)+\int\limits_0^r r'dr'V(r')[I_l(kr)
K_l(kr')-I_l(kr') K_l(kr)] I_l(kr')\nn\\
&&\phi_l^{(2)}(ik r)=I_l(k r)+\int\limits_0^r r'dr'V(r')[I_l(kr)
K_l(kr')-I_l(kr') K_l(kr)] I_l(kr')+\nn\\
&&\int\limits_0^r r'dr'V(r')[I_l(kr)
K_l(kr')-I_l(kr') K_l(kr)]\times \\ &\times&\left(\int_0^{r'}r''dr'' V(r'')I_l(kr')K_l(kr'')I_l(kr'')-
 \int_0^{r'} r''dr''V(r'')K_l(kr')I_l(kr'')^2 \right)\nn
\label{second_iter}
\eea
 
 The considered \rg method doesn't operate with the \jf immediately,
 but with it's logarithm $\ln f_l(ik)$ (See Sec.\ref{spectralsum}). Then by
 using the standard series representation of logarithm:
 \be
 \ln (1+x)=x-\half x^2+ \fr{1}{3}x^3-...
 \label{series_of_ln}
 \ee
 we can derive from (\ref{jost_func_ik_of_LS}) by means of (\ref{second_iter}) that
 \be
 \ln f_l(ik)= \ln f_l^{(1)}(ik)+\ln f_l^{(2)}(ik)+\ln f_l^{(3)}(ik)+\ln f_l^{(4)}(ik)+\dots,
 \label{lnf_of_ik}
 \ee
where the terms $\ln f_l^{(1)}(ik)$ denotes the resulting expressions, that
after substitution of (\ref{second_iter}) into (\ref{jost_func_ik_of_LS}) and
re-expansion and re-ordering of terms contain integrations $n$ times
respectively (or the same, the $n$-th power of $V$). They are represented in
Math.App.(\ref{JOSTASscalar}).

The given approach is though applicable for all models considered in the
present work. However it will be shown later in Sec.\ref{efineas} that for
\rg procedure the only minimal subtraction scheme is used. So for
the scalar field we need only the contributions up to $\ln f_l^{(2)}(ik)$ to
take into account in $\ln f_l^{as}$.

The procedure of simplification of  $\ln f_l^{(2)}(ik)$ fro this case is
described in details \cite{borkir1} and partially in Math.App.(\ref{JOSTASscalar}). Finally we obtain the result for $ l>0$ in the form
\be
f_l(ik)\sim \frac{V_0}{2 }\int\limits_{r_1}^{r_2}rdr \sum\limits_{n,j} X_{n,j}\frac{t^j}{l^n}
\ee
in terms of polynomials of the variable $t=\frac{1} {\sqrt{ 1+(\frac{kr}{l})^2 } }$.  
Here  $ X_{n,j}$ are:
\bea
&X_{1,1}=1,\ \ & X_{3,3}=\frac{1}{8}(1-2r^2V_0),\nn\\
&X_{3,5}=-\frac{6}{8},\ \ & X_{3,7}=\frac{5}{8}.
\label{Xnj_scalar}
\eea

For $l=0$ we have additionally  at $k\ra \infty$ 
\be f_0(ik)\sim -\frac{V_0}{2k}(r_1-r_2),\ee
the derivation is presented in Math.App. (\ref{K0_I0_asymp}-\ref{lnfas0}).


\subsubsection{Spinor field in cylindric magnetic background}

\ \ \ To obtain the uniform asymptotic expansion of the \jf, we proceed in much the
same way as in the previous section for the scalar field.

It starts again with the integral \lseq for the general solution $\phi(r)$
\be
 \phi(r)=\phi^{0}(r)+\int\limits_0^r g(r,r')V(r')\phi(r')r' dr',
\ee
where the $\phi(r)$ is a two-component spinor \f, satisfying the \eqq
\be
{\cal D}\phi(r)=V(r)\phi(r)
\label{\eqq}
\ee
with
\be
{\cal D}=\left\{
\begin{array}{cc}
p_0-m & \pd_r-\fr{l}{r}\\ -\pd_r-\fr{l+1}{r} & p_0+m
\end{array}\right\}, \ \ \ V(r)=-\fr{\Omega a(r)}{r}\left\{\begin{array}{cc}0&1\\ 1&0\end{array}\right\}.
\ee

The Green \f  $g(r,r')$ is chosen in the form
\be
 g(r,r')=-\fr{\pi}{2i}[\phi^0_J(r) \phi^{0\ T}_H(r')- \phi^0_J(r') \phi^{0\ T}_H(r)].
\label{spinor_green_\f}
\ee

Here the $\phi^0_J(r)$ and $\phi^0_J(r)$ are solutions which behave at $r\ra
0$ as ones of the free \eqq, it means the \eqq (\ref{\eqq}) with $V=0$:

\bea
\phi^0_J(r)=\left(\begin{array}{cc}\sqrt{p_0+m} & J_{l+1}(kr)\\
\sqrt{p_0-m} &J_{l}(kr) \end{array}\right) \ for\  l>0,\\ \phi^0_J(r)=\left(\begin{array}{cc}\sqrt{p_0+m} & J_{-l-1}(kr)\\
-\sqrt{p_0-m} &J_{-l}(kr) \end{array}\right) \ for\  l<0,
\eea
and the corresponding $\phi^{0\ T}_H(r')$ is obtained from $\phi^0_J(r)$ by the
substitution $J_{\rho}(kr)$ through $H_{\rho}^{(1)}(kr)$ and transposition.
It leads to the \eqq for the real \jf $f_l(k)$

\be f_l(k)=1-\fr{\pi}{2i} \int\limits_0^\infty r\ dr \phi^{0\ T}_H(r) V(r) \phi(r), \ee
where $\phi(r)$ obeys the full \eqq (\ref{\eqq}).

The orbital quantum number $l$ is substituted by $\nu$ by means of
(\ref{redef_l_to_nu}),
Sec \ref{hamiltonians}, and the radial momentum $k$ is turned to $ik$ respectively.
In contrary to the recently considered scalar field problem, in the case
of spinor field we have to retain the terms up to $\ln f_\nu^{(4)} $ in the expansion 
(\ref{lnf_of_ik}, $l\leftrightarrow\nu $), in order to take into account all
powers of $\nu$ up to $\nu^{-3}$ needed for the minimal subtraction. 
 This procedure can be look up in details in \cite{borkir1} especially for
the problem of the spinor field in magnetic background. (The similar
calculation is given in the present paper in details for the spinor field in
background of NO-vortex, see the next subsection and Math.App.\ref{jostUA_NO}).

As a result, the expansion in powers of $\nu$ for logarithm of asymptotic
Jost function of imaginary momentum is obtained in the form (see\cite{borkir1})\\
\begin{equation}
ln \boldmath f_{as}\unboldmath(ik)=\sum\limits_{n=1}^3\sum\limits_{j=1}^9
\int\limits_{0}^{\infty} \fr{dr}{r}X_{nj}\fr{t^j}{\nu^n},\\
\label{lnfas_x}
\end{equation}
where $X_{nj}$ are represented in terms of $ r, \Omega, a(r) $ and
their derivatives:

\bea
&&X_{1,1}= -X_{1,3} =X_{2,6} =\half (a\Omega)^2\nn\\
&&X_{2,2}= \fr{1}{4}\Omega^2(a^2-raa')\nn\\
&&X_{2,4}= \fr{1}{4}\Omega^2 (-3a^2+raa')\nn\\
&&X_{3,3}= \fr{1}{4}\Omega^2(a^2-raa'+\half r^2 aa''-\half \Omega^2 a^4 )\nn\\
&&X_{3,5}= \fr{1}{8}\Omega^2(-\fr{39}{2}a^2+7raa'-r^2aa''+6\Omega^2 a^4 )\nn\\
&&X_{3,7}= \fr{1}{8}\Omega^2(35 a^2 -5raa'-5\Omega^2 a^4 )\nn\\
&&X_{3,9}= \fr{-35}{16}\Omega^2 a^2,
\label{X_nj_spinor}
\eea
 $t=(1+[\fr{kr}{v}]^2)^{-\half}$. 

This expansion had been obtained in \cite{borkir2} by iterations of
Lippmann-Schwinger equation up to the order $\nu^{-3}$. In general it
is possible to obtain higher orders using this formalism. However as
the calculations \cite{borkir2} showed, the complication of the
involved expressions increases very fast. It is remarkable that this
expression does not contain a term with power $\nu^{-2}$. In the
finite part of the energy the corresponding term is cancelled in the
sum of terms corresponding to positive and negative orbital momenta
$\nu$ as well. The absence of the power $\nu^{-2}$ is a succession of
the zero heat kernel coefficient $a_{3/2}$.  Also it is a non-trivial
fact that both, the
fourth power of the magnetic flux $\Omega$ and the second one is present.

It can be checked numerically that $ln f^{as}$ (\ref{lnfas_x}) is indeed
the uniform asymptotic expansion of logarithm of (\ref{f_v_of_ik+}, \ref{f_v_of_ik-}) for
$k,\nu$- large, $k/ \nu=z$-fixed. The numerical examples will be shown in
Sec.(\ref{results}).
 
\subsubsection{Spinor field in arbitrary abelian Higgs background with\\ cylindric symmetry}
\ \ \  As it was pointed out in Sec.\ref{hamiltonians}, the abelian Higgs
 background distinguishes from the case of pure magnetic background
 considered above, since the spinor field has a space-dependent mass,
coming from interaction with the scalar Higgs condensate. It gives reason
to discuss this problem more detailed, following in principle \cite{Bordag:2003at}.

  We start with the \lseq in the form (\ref{lipp_schw_gen}),
 and in order to proceed with the expansion for \jf in the form (\ref{lnf_of_ik},
 we have to define an expression for the Green function in (\ref{lipp_schw_gen}) in a suitable form.

 Consider the equations of motion for the spinor field $\Psi$. In
Sec.\ref{hamiltonians} they have been reduced for the special case of
cylindrical symmetry in cylindrical coordinates $(r,\varphi,z)$ to the form

\bea
\left\{\begin{array}{cc}
    p_0+\hat{L}-m \sigma_3 & p_3 \sigma_3 \\ \\
    p_3 \sigma_3 & p_0+\hat{L}+m \sigma_3
 \end{array}\right\}
\left(\begin{array}{cc}
     \Phi \\ \\ \chi 
 \end{array}\right)  =0
\label{dirac_eq_4comp}
\eea
with
\be
\hat{L}=i \sum\limits_{k=1}^2 \sigma_k (\fr{\pd}{\pd x^k}+ieA_k)
\label{dirac_op_L}
\ee
for a wave function $\psi(r)$ of the radial coordinate only. Now let the
potential $V(r)$ be the solution of the Nielsen-Olesen vortex equation, (that
we namely need) and is expressed in terms of functions $v(r)$ and $f(r)$
respectively.

For the radial equations for two-component spinor $\Phi=(\psi^u(r)
,\psi^l(r))$ we obtained:

\be
\left\{\begin{array}{cc} p_0-\fr{f_e\eta}{\sqrt{2}}f(r) & \fr{\pd}{\pd r}-\fr{l-n
 v(r)}{r}\\ \\-\fr{\pd}{\pd r}-\fr{l+1-n v(r)}{r} & p_0+\fr{f_e\eta}{\sqrt{2}}f(r)
 \end{array}\right\}\Phi(r)=0
\label{dirac_NO}
\ee
with
\be
\Phi(r)=\left\{\begin{array}{c} \psi^{u}(r)\\ \\ \psi^{l}(r)  
\end{array}\right\}.
\label{Spinor_2comp}
\ee

Indices $l$ and $u$ denote ``upper'' and ``lower'' components of spinor respectively.
These equations can be solved numerically for any specified functions $v(r)$
and $f(r)$. The integer $n$ is the vorticity of the vortex configuration
(Sec.\ref{hamiltonians}). All further calculations are given for $n=1$.

For boundary values of potentials $f(r)$ introduce the notations:

\be
\fr{f_e\eta}{\sqrt{2}}f(r)=m_0,\  r\ra 0\ \ \ \
\fr{f_e\eta}{\sqrt{2}}f(r)=m_e,\  r\ra\infty,
\ee
 and for the $v(r)$ potential is still valid

 \be
v(r)=0,\ r\ra 0\ \ \ \ \ v(r)=1,\ r\ra\infty .
 \ee

Further we introduce the notation for scalar potential
\be
\fr{f_e\eta}{\sqrt{2}}f(r)=\mu(r).
\label{def_of_mu}
\ee

Now consider the construction
\be
g(r,r')=-\fr{\pi}{2i}\{ \Phi^0_J(r) \Phi^{0\ T}_H(r')- \Phi^0_J(r') \Phi^{0\ T}_H(r)\},
\label{green_func_ew}
\ee
where

\be
\Phi_H^0= \left[\begin{array}{cc} \sqrt{p_0+m_0} &
 H^{(1,2)}_{l+1}(p r)
 \\ \\\sqrt{p_0-m_0}  & H^{(1,2)}_{l}(p r)
 \end{array}\right],\ \ 
\Phi_J^0= \left[\begin{array}{cc} \sqrt{p_0+m_0} &
 J_{l+1}(p r)
 \\ \\\sqrt{p_0-m_0}  & J_{l}(p r)
 \end{array}\right]
\label{phi_H_phi_J_spinors}
\ee
 with $p=\sqrt{p_0^2-m_0^2}$,
and assure, that it satisfies the definition of Green function for the \eqq
 (\ref{dirac_NO}) at the limit of $r\ra 0$, which we assume to be a
 homogeneous \eqq by means of (\ref{hom_eq}), Sec.{\ref{lippsw}}.  
 Here we must take into consideration that the integral over $r'$ in the
 general solution for the inhomogeneous equation (\ref{dirac_NO}) in view of (\ref{inhom_eq}), Sec.{\ref{lippsw}} is to be taken
 
 \be
 \Phi^{gen}(r)=\Phi^{0}(r)+\int\limits_0^r dr'\ g(r,r') \hat{V}(r')\Phi^{gen}(r'),
 \label{gen_solution}
 \ee
in accordance with (\ref{lipp_schw_gen}), Sec.{\ref{lippsw}}, where the
integration is to carry on in limits from $r'=0$ to $r'=r$. This is equivalent to the choice of the Green function in the form
 \be
 g(r,r')=-\fr{\pi}{2i}\{ \Phi^0_J(r) \Phi^{0\ T}_H(r')- \Phi^0_J(r') \Phi^{0\ T}_H(r)\}\theta(r-r').
 \label{green_func_teta}
 \ee

Basically the method presupposes, that the \lseq does not ``know'' about the behaviour of the
solution $\phi$ of the (\ref{dirac_NO}) outside of the close neighbourhood
to the point $r=0$, where the functions (\ref{phi_H_phi_J_spinors}) satisfy the
\eqq, respectively thereby the constructed Green function (\ref{green_func_ew})
as well. These informations are all to be contained in the integral kernel $\hat{V}(x)$
of (\ref{gen_solution}) that is even the scattering potential of \lseq.

Then one can proceed with iterations of the \lseq in the same way as it has
been done for the spinor field in magnetic background (previous subsection)
and represent the logarithmic \jf immediately as the series
\be
\ln f_\nu( ik )= \ln f^{(1)}_\nu( ik ) +\ln f^{(2)}_\nu( ik ) +\ln f^{(3)}_\nu( ik
) +\ln f^{(4)}_\nu( ik ).
\label{lnf_series}
\ee

Here, as well as in the previous case, we need at least four initial terms
of the uniform asymptotic series for $\ln f_\nu( ik )$ in order to keep all powers
of $\nu$ up to $\nu^{-4}$.

In order to use the \lseq  properly we solve the \eqq of motion (\ref{dirac_NO})
 with a little bit generalization. Suppose, the scalar potential $\mu(r)$, representing
 the local space depending mass of the spinor field $\Phi(r)$ to acquire the
 value $m_0$ (not necessary equal zero) at the coordinate origin $r=0$ and
 $m_e$ outside of the vortex core at $r\ra \infty$ respectively.

 Here we must handle the solutions of (\ref{dirac_NO}) carefully. First of all rewrite
 this \eqq for spinor function $\Phi$ as

\be
\{\hat{P}-\hat{V}\}\Phi = 0,
\label{lseq_of_P_V}
\ee
with the separated "source part" $\hat{V}$ which contain the dependence of
the potentials $\mu(r)$ and $v(r)$.
Introducing the redefinition of $\mu(r)$
\be
\mu(r)=\mu(r)-m_0+m_0= \tilde{\mu}(r)+m_0
\ee
we rewrite the (\ref{dirac_NO}) in form (\ref{lseq_of_P_V}) with 
\be
\hat{P}=\left[\begin{array}{cc} p_0-m_0 & \pd_r-\fr{l}{r}\\ \\-\pd_r-\fr{l+1}{r}
    &  p_0+m_0  \end{array}\right]
\label{operator_P}
\ee

\be
\hat{V}=\left[\begin{array}{cc} -\tilde{\mu}(r) & -\fr{a}{r}\\ \\-\fr{a}{r} &
    \tilde{\mu}(r) \end{array}\right]
\label{potential_V}
\ee

Then we can assure that for vanishing matrix potential $\hat{V}$ the
following spinor functions 

\bea
\Phi_H= \left[\begin{array}{cc} \sqrt{p_0+m_0} &
 H^{(1,2)}_{\nu+\half}(q_0 r)
 \\ \\\sqrt{p_0-m_0}  & H^{(1,2)}_{\nu-\half}(q_0 r)
 \end{array}\right]
\\
\Phi_J= \left[\begin{array}{cc} \sqrt{p_0+m_0} &
 J_{\nu+\half}(q_0 r)
 \\ \\\sqrt{p_0-m_0}  & J_{\nu-\half}(q_0 r)
 \end{array}\right],
\label{phi_H_phi_J_spinors}
\eea
with $q_0=\sqrt{p_0^2-m_0^2} $,
really satisfy the homogeneous "free" equation at $r\ra 0$.
Indeed the 2-component spinor functions $\Phi_H,\Phi_J$ satisfy exactly the \eqq
 (\ref{lseq_of_P_V})  without the right-hand side, i.e. at $\hat{V}=0$

\be
\hat{P}\Phi_H = \hat{P}\Phi_J=0.
\label{P_0_free_equation}
\ee

 Thus the form (\ref{lseq_of_P_V}) with $\hat{P}$ and $\hat{V}$ defined by (\ref{operator_P},
 \ref{potential_V}) is obviously appropriable to carry on iterations of \lseq.
 Further steps are represented in Math.App \ref{jostUA_NO}. The final expression for $\ln f_\nu(ik)$ can be represented again in the form
\be
\ln f_\nu(ik)=\int\limits_0^\infty \fr{dr}{r}\sum\limits_{n=1}^3
\sum\limits_{j=1}^{3n}
X_{nj}\fr{t^j}{\nu^n}
\label{lnfas1} 
\ee
with $t=\fr{1}{\sqrt{1+(kr/\nu)^2}}$ and the coefficients $X_{nj}$:
\bea\label{Xnj}
X_{11}&=&\frac12 v(r)^2+\frac12 r^2(\mu(r)^2-m_e^2),\nn \\
X_{13}&=&-\frac12 v(r)^2,\nn \\
X_{33}&=&\frac14  v(r)^2-\frac18 r^2 v'(r)^2-\frac18 v(r)^4-\frac14
r^2 v(r)^2(\mu(r)^2-m_e^2)\nn \\
&&-\frac18 r^4 (\mu'(r)^2+(\mu(r)^2-m_e^2)^2),\nn \\
X_{35}&=&-\frac{39}{16} v(r)^2+\frac18 r^2 v'(r)^2+\frac{3}{4} v(r)^4
+\frac34 r^2 v(r)^2(\mu(r)^2-m_e^2)\nn \\
&&-\frac{3}{16}r^2 (\mu(r)^2-m_e^2) ,\nn \\
X_{37}&=&\frac{35}{8}v(r)^2-\frac{5}{8} v(r)^4+\frac{5}{16}r^2
(\mu^2-m_e^2),\nn \\
X_{39}&=&-\frac{35}{16}v(r)^2.
\label{repres_xij}
\eea

The dependence of $X_{nj}$ on $v(r)$ is the same as in the previous problem
of magnetic flux tube (obtained in \cite{borkir1}, where it was denoted by
$a(r)$),
the dependence on $\mu(r)$ is a new feature. We use
integration by parts in Eq.( \ref{lnfas1}) in order to get the
shortest representation for the coefficients $X_{nj}$.

\subsection{The representation of the renormalized ground state energy}
\label{efineas}

\ \ \ The results of Sec.\ref{spectralsum} allows to represent the
regularized \gse in terms of \jf for corresponding scattering problem.
 These representations (\ref{e0_of_lnf}, \ref{e_o_scalar_of_f_l}) for each of considered models are constructed as a sum over orbital quantum numbers of integrals over
 radial momenta from logarithmic derivative of the \jf. The infinite
 contributions of the empty Minkowski space are dropped through this
 performance. However the regularized energy still remains infinite through
 the \uv divergences.

 Further in the Sec.\ref{zfreg} the asymptotic expansion for the regularized
\gse in terms of global \hkc was obtained (\ref{large_mass_expansion}). It
corresponds to the asymptotic of large mass $m\ra \infty $ for massive
quantum fields. It was also pointed out that this expansion allows to
identify all \uv divergences of energy.
 
Proceeding from the (\ref{large_mass_expansion}) we define the terms with non
negative powers of $m$ (all the contributions up to $a_2$) as divergent part
$E^{div}$ and subtract it out. 
Thus the renormalized energy is defined

\be
E^{ren}=E_0^{reg}-E^{div },
\label{E_ren_def}
\ee
that now is finite, since all the present \uv divergences are contained in the
$E^{div }$.
\bea
{  E}^{div}&=&-\frac{m^4}{64\pi^2}\left(\frac{1}{s}+\ln\frac{4\mu^2}{m^2}-\half\right)a_0
-\frac{m^3}{24\pi^{3/2}}a_{1/2}+\nn\\&& \frac{m^2}{32\pi^2}\left(\frac{1}{s}+\ln\frac{4\mu^2}{m^2}-1\right)a_1+\frac {m}{16\pi^{3/2}}a_{3/2}\nn\\
&-&\frac{1}{32\pi^2}\left(\frac{1}{s}+\ln\frac{4\mu^2}{m^2}-2\right)a_2.
\label{E_div_def}
\eea

 It should be noticed that such a subtraction removes not only divergences
 being contained in the \hke but also several finite
contributions not related immediately to the \uv divergences at ($s\ra 0$). 
  This action is dictated by the physical requirement, that the energy of
 quantum fluctuations must tend to zero for large mass limit.
 This approach differs from the minimal subtraction
 (\ref{minimal_subtraction}) and is valid for massive fields only.

 \subsubsection{The divergent part of \gse}
 
 \ \ \ For certain models considered here the coefficients $a_n$ entering in
 (\ref{E_div_def}) and therefore the divergent part of energy are calculated.\\
 \\

{\bf  1.Scalar field $\phi$ with the scalar step-function background $V(x)$ }\\

\ \ \  The problem is formulated in Sec.\ref{hamiltonians} and the corresponding
 non zero \hkc  $a_1,a_2$ are derived in Sec.\ref{heatkernel},
 Eq.(\ref{a1_scalar}, \ref{a1_scalar}). Thus all the coefficients $a_n$ in
 (\ref{E_div_def}) are known:
\bea
&&a_0=1\nn\\
&&a_{1/2}=a_{3/2}=0\nn\\
&&a_1= 2\pi\int -V(r)\ r\ dr\nn\\
&&a_2= 2\pi\int\left( -\fr{1}{6}\Delta V(r)+\half V^2(r)\right)\ r\ dr
\eea

 For the considered potential $V(r)$ Sec.\ref{hamiltonians},
 (\ref{scalar_potential}) the relevant coefficients contributing in $E^{div}$
 are

\bea
a_1=\pi V_0(R_1^2-R_2^2),\nn \\
a_2=-\half \pi V_0^2(R_1^2-R_2^2),
\eea
what provides

\be
E^{div}=\fr{(R_1^2-R_2^2)}{32\pi}\left\{ \half V_0^2\left( \fr{1}{s}+\ln   \fr{4\mu^2}{m^2}-2\right)+   V_0 m^2 \left( \fr{1}{s}+\ln   \fr{4\mu^2}{m^2}-1\right) \right\}.
\label{edivscalar}
\ee\\
\\
{\bf 2.Spinor field $\psi$ with vector (magnetic) background $A_\mu(x)$}\\

\ \ \ The \hkc for the Dirac spinor field in background of vector filed
are calculated in Sec.\ref{hamiltonians}, Eq.(\ref{a1_a2_spinor}). In fact
only the coefficient $a_2$ contributes to the $E^{div}$:
\bea
&&a_0=1 \nn\\
&&a_{1/2}=a_{3/2}=a_1=0\\
&&a_2=\Tr\ \int d^3x\left[-\fr{1}{12 }F_{\mu\nu}F^{\mu\nu}+\fr{1}{8}\sigma^{\mu\nu}F_{\mu\nu})^2 \right] =\fr{2}{3}\int d^3x F_{\mu\nu}F^{\mu\nu}\nn\\
\eea

For the magnetic rectangular shaped tube (\ref{potential_A},\ref{a_r1}) a
constant cylindric magnetic flux $\Omega$ Sec.\ref{hamiltonians}, (\ref{delta}) it means:

\be
a_2=\fr{2}{3}\int d^2x F^2_{\mu\nu}
=\fr{8 \pi}{3}\Omega^2\int\limits_{0}^{\infty}dr r h(r)^2 = \fr{16 \pi\Omega^2}{3}\fr{1}{R_2^2-R_1^2}.  
\label{spinor_a2_explicit}
\ee
 
Moreover in the section (\ref{a52}) the next non-zero heat kernel coefficient
$a_{5/2}$ has been calculated (that however does not contribute to the
divergent part of energy ) 
\be
a_{5/2}= \fr{5}{8\sqrt{\pi}}e^2 \Omega^2 \fr{(R_1+R_2)}{(R_1^2-R_2^2)^2}\nn\\
\ee

This coefficient (and further half-integer numbered coefficients) is present
in the \hke because of non-analyticity of the background  $A(r)$ (the remark
at end of Sec.\ref{heatkernel}).

The corresponding $E^{div }$ for this problem is therefore
\be
  E^{div}=\fr{\Omega^2}{6\pi(R_2^2-R_1^2)}(\fr{1}{s}-2+\ln \fr{4\mu^2}{m^2}).
\ee\\
\\

{\bf 3. Spinor field with background of the Nielsen-Olesen vortex}\\

 The only non zero coefficients contributing to the divergent part of the
 ground state energy are (\ref{Bordag:2003at}):
 
\bea
&&a_1=-\Tr\ \int d^2 x V(x),\nn\\
&&a_2= \Tr\ \int d^2 x \left( -\fr{1}{12}F_{\mu\nu}F^{\mu\nu} +\half V(x)^2
  -\fr{1}{6}\Delta V(x)\right),
\eea
where the trace is to take over spinor indices.

All the $a_n$ with half-integer $n$ are zero since the background
potentials $f(r), v(r)$ are supposed to be smooth (the solutions of the NO-equations).

In terms of $\mu(r)\e \fr{f_e\eta}{\sqrt{2}}f(r),\ v(r)$ the heat kernel coefficients read:

\bea
&&a_1=-8\pi \int\limits_0^\infty r\ dr(\mu(r)^2-m_e^2),\nn\\
&&a_2= 8\pi \int\limits_0^\infty r\ dr\left\{ \fr{1}{3}\fr{v'(r)^2}{r^2}+\half[\mu'(r)^2+(\mu(r)^2-m_e^2)^2]  \right\}
\label{a1_a2_NO}
\eea

Thus the $E^{div }$ reads:
\bea
&&E^{div }=- \fr{1}{4\pi}
\left\{
  \left(\fr{1}{s}+\ln \fr{4\mu^2}{m^2}-1\right)\int\limits_0^\infty r\ dr
  \fr{f_e^4 \eta^4}{4} (f(r)^2-1)
  +\right.\\&&\left. \left(\fr{1}{s}+\ln \fr{4\mu^2}{m^2}-2\right)\int\limits_0^\infty r\ dr
  \left[ \fr{1}{3} \fr{v'(r)^2}{r^2}+\fr{f_e^2 \eta^2}{4} f'(r)^2+\fr{f_e^4 \eta^4}{8} (f(r)^2-1)^2\right]
\right\}\nn
\label{NO_ediv}
\eea


\subsubsection{The finite and the asymptotic parts of energy}

\ \ \ $E^{ren}(s)$ defined by (\ref{E_ren_def}) is though finite at $s\rightarrow0,$ but it is not possible to carry out this limit analytically. Indeed, we have merely the finite sum
 of two terms, each of them is infinite and represented in distinguished form:
 the first one is an infinite sum of infinite integrals, and the second one
 is generally speaking an algebraic polynomial of powers $s^{-1}$. The possible way to solve this problem would be to find
 some intermediate object possessing both the features of two summands in
 $E^{ren}$, i.e. it should be on the one hand a sum of infinite
 integrals, and on the other hand it must be a polynomial of $s$ containing in
 particular the powers $s^{-1}$.
 
 Furthermore we know that the divergence of the expression for $E_0^{reg}$
is of the \uv type. In terms of variables $l$ and $k$ (orbital and radial
momenta) it means that the divergence appears at large values of them. It suggests that the
needed object is some kind of asymptotics for $k,l\ra \infty$

One can construct such an object proceeding from the structure of
 $E_0^{reg}$. The following discussion is restricted on the case of scalar
 field. It will be applied later for other models. 

 Let we have an expression of kind (\ref{e0_of_lnf}). Consider the orbital
 momenta $l$ large enough, that it can be treated as a continuous variable.
Suppose the \f $\ln f_l(ik)$ entering here has an asymptotic expansion in the
combined limit
 \be k\ra \infty,\  l\ra \infty,\ 
 k/l= z=const,\label{combined_limit}
 \ee
the so called ``uniform asymptotic expansion'', known e.g. for Bessel \fs \cite{abrsteg}.
 
Then the \jf is expected to have the asymptotic behaviour
\be f_l(ik)\sim 1+\fr{c_1}{l}+\fr{c_2}{l^2}+\dots,\ee
that follows from the general properties of the \jf \cite{taylor}.
For $\ln f_l(ik)$ we have under assumption of (\ref{combined_limit})
\be
\ln f_l(ik)\sim \fr{C_1(z)}{l}+\fr{C_2(z)}{l^2}+\fr{C_3(z)}{l^3}+\dots
\ee

Substituted into the expression

 \be
E^{reg}\sim C_s \sum\limits_{l=0}^\infty \int\limits_m^\infty dk (k^2-m^2)^{1-s} \fr{\pd}{\pd k}\ln f_l(ik),
\ee
the (\ref{combined_limit}) provides \be
E^{reg}\sim C_s \sum\limits_{l=0}^\infty l^{2(1-s)} \int\limits_m^\infty dz
(z^2-\fr{m^2}{l^2})^{1-s}
\left(\fr{C'_1(z)}{l}+\fr{C'_2(z)}{l^2}+\fr{C'_3(z)}{l^3}+\dots \right)
\label{dots}
 \ee
 Now, as far as the parameter $s$ is not fixed, we suppose for a moment that $s$ is large enough that the integral in (\ref{dots}) converges. Then we can
 interchange the integration with the sum in brackets. It allows to
 establish that the first three terms in the \rg limit $s\ra 0$.
\be
  C_s \sum\limits_{l=0}^\infty l^{2(1-s)} \int\limits_m^\infty dz
(z^2-\fr{m^2}{l^2})^{1-s}\left(\fr{C'_j(z)}{l}\right),\ \ j=1,2,3
\label{divergent}
 \ee
can be divergent, and for all further ones up from $j=4$ the sum over $l$
converges.

We denote the divergent terms of the uniform asymptotics (\ref{divergent}) as
${\cal E}^{asym}_\infty$
\bea 
{\cal E}^{asym}_\infty=
C_s \sum\limits_{l=0}^\infty l^{2(1-s)} \int\limits_m^\infty dz
(z^2-\fr{m^2}{l^2})^{1-s}\left( \sum_{j=1,2,3}\fr{C'_j(z)}{l}\right)=\nn\\
 C_s \sum\limits_{l=0}^\infty \int\limits_m^\infty dk (k^2-m^2)^{1-s}
 \fr{\pd}{\pd k}\ln f_l^{as}(ik),
 \label{easym_infty}
\eea
 where we come back to the notations of $k$, and $f_l^{as}(ik)$ denotes now
 the first several terms of the uniform asymptotics of the \jf in the limit
 (\ref{combined_limit}) which cause the divergence in (\ref{divergent}),
 namely the terms up to $l^{-3}$ power.
  Thus we arrive at the statement that the value of the expression
  \be
 E^{f}= E^{reg}-{\cal E}^{asym}_\infty
 \label{efin_def}
  \ee
  is finite at the limit $s\ra0$. The analytical representation of
  $f_l^{as}(ik)$ has been obtained recently in Sec.\ref{lippsw}.
 
   Since the $\ln f_l^{as}(ik)$ is defined we make a factitious manipulation
   with $E^{ren}$ (\ref{E_ren_def})
namely, we add the identical zero expression ${\cal E}^{asym}_\infty-{\cal E}^{asym}_\infty$ to the $E^{ren}$,
 \be
E^{ren}=(E_0^{reg} - {\cal E}^{asym}_\infty) + ({\cal E}^{asym}_\infty-E^{div })
 \ee
 and denote the term in the second bracket as the "asymptotic'' part of
 energy $E^{as}$
 \be 
E^{as}={\cal E}^{asym}_\infty-E^{div }
\label{eas_def}
\ee

Thus we have finally the renormalized energy $E^{ren}$ consistent of two
parts - the finite and the asymptotic one:
\be E^{ren}=E^{f}-  E^{as}\ee
Note, that the both parts are finite. Consider the calculation of each of
these parts in details.

\subsubsection{The finite part $E^f$ of the \gse}

\ \ \ It is supposed that the finite energy $E^f$, defined (for scalar field) by (\ref{efin_def}) as
\bea\label{efin_explicit}
&&E^f= C_s \sum\limits_{l=-\infty}^\infty\int\limits_m^{\infty} dk (k^2-m^2) \fr{\pd}{\pd k}[\ln f_l(ik)-\ln
f_l^{as}(ik)] =\nn\\&& C_s \sum\limits_{l=0}^\infty\int\limits_m^{\infty} dk(k^2-m^2)^{1-s} \fr{\pd}{\pd k} \ln f_l^{sub}(ik)\eea
remains finite in the limit $s\ra 0$ since all \uv divergences are removed.
  From the explicit construction of $\ln f_l^{sub}(ik)$ the asymptotic
  behaviour of this function at $l\ra\infty$ is expected to be at least
  \be\ln f_l^{sub}(ik)\ \sim \ O(\fr{1}{l^4}) \ee
In particular, for the spinor field problem the uniform asymptotics of the \jf
 does not contain in fact any even powers of $\nu$  because of the symmetry properties of this scattering problem (see Sec.\ref{jostfunc}). Thus
 the asymptotic behaviour of $\ln f_l^{sub}(ik)$ is really of order $1/l^5$,
 that is also confirmed by numerical evaluations.
  
 The correct subtraction provides indeed a well convergent integral over $k$
 for each $l$ with following convergence of the sum over $\nu  ($or$\  l)$

A couple of numerical examples (Fig.\ref{bundle},\ref{contrint}) in Sec.\ref{results} illustrates how it works.
\\

{\bf 1. $E_{f}$ for the scalar field in rectangular scalar background}\\

 We start with the definition of $E^f$ (\ref{efin_def}) given explicit
 in (\ref{efin_explicit})
 
Using the symmetry of \jf with respect to sign of $l$ ,Sec.\ref{jostfunc}(\ref{symmetry_l}), we can rewrite the $E_0^{reg}$

\bea
&&E^{f}= C_s \fr{1}{\sqrt{\pi}} \int\limits_m^\infty (k^2-m^2)^{1-s} \fr{\pd}{\pd k}
\ln f_0^{sub}(ik)+
2 C_s \fr{1}{\sqrt{\pi}} \sum\limits_{l=1}^\infty \int\limits_m^\infty (k^2-m^2)^{1-s} \fr{\pd}{\pd k} \ln f_l^{sub}(ik)\nn\\&&=E_0^f+E_l^f,
\label{scalarEF}
\eea

\be
E_0^{f}=
 C_s \fr{1}{\sqrt{\pi}}\int\limits_m^\infty
(k^2-m^2)^{1-s} \fr{\pd}{\pd k} [\ln f_0(ik)-\ln f_0^{as}(ik) ],
\label{scalarEF0}
\ee

\be
E_l^{f}=
2 C_s \fr{1}{\sqrt{\pi}} \sum\limits_{l=1}^\infty \int\limits_m^\infty
(k^2-m^2)^{1-s} \fr{\pd}{\pd k} [\ln f_l(ik)-\ln f_l^{as}(ik) ].
\label{scalarEFl}
\ee

These expressions are suitable to calculate immediately.
To this end we need the uniform asymptotic expansions $\ln f_l^{as}(ik)$ of $\ln f_l(ik),\ l>0$
and for $\ln f_0(ik)$ respectively.
 The asymptotic expansion for $\ln f_l^{as}(ik)$ has been obtained from the
 iterations of \lseq, Sec.\ref{lippsw}. For $\ln f_0(ik)$ it can be done
 immediately through the substitution of the uniform asymptotics of Bessel
 \fs $K,I$ in the exact \jf (\ref{scalar_jost_exact}), as given in Math.App (\ref{K0_I0_asymp}).\\

{\bf 2. $E^{f}$ for spinor field in magnetic flux $\Omega$}\\

The sum over orbital momenta $l$ in (\ref{efin_explicit}) is performed for
 the spinor problem by means of the substitution $l\ra\nu$, Sec.\ref{hamiltonians} (\ref{redef_l_to_nu}). We recall the representation
Sec.\ref{spectralsum}, (\ref{e0_of_lnf}):
\bea
E_0(s)=C_s
\sum\limits_{\nu=1/2,3/2,\dots} \int\limits_m^{\infty} dk
(k^2-m^2)^{1-s} \partial_k [\ln f^+_\nu(ik)+\ln f^-_\nu(ik)],
\label{e0_of_lnf_C_s}
\eea
where \be \label{const_Cs} C_s=-\mu^{2s}\frac{1}{\sqrt{\pi}}
\frac{\Gamma(s-1)}{\Gamma(s-\half)}\frac{\cos \pi s}{\pi},\ee
and define with the help of (\ref{efin_explicit}) the $E^{f}$ as:
\bea
&&E^{f}=C_s\sum\limits_{\nu=1/2,3/2,\dots} \int\limits_m^{\infty} dk
(k^2-m^2)^{1-s} \partial_k [\ln f^+_\nu(ik)+\ln f^-_\nu(ik)-
\ln f^{as+}_\nu(ik)-\ln f^{as-}_\nu(ik)]=\nn\\
&&C_s\sum\limits_{\nu=1/2,3/2,\dots} \int\limits_m^{\infty} dk
(k^2-m^2)^{1-s} \partial_k \ln f^{sub}_\nu(ik).
\label{Ef_spinor}
\eea

 It is important to notice, that the \jf $\ln f_\nu(ik)$ in contrary to the
 case $\ln f_l(ik)$ of the scalar problem, possesses no symmetry properties like
 (\ref{symmetry_l}), so the exact Jost \fs
 (\ref{f_v_of_ik+}, \ref{f_v_of_ik-}), Sec.\ref{jostfunc} for spinor field are different.
 However it has been shown in \cite{borkir1} by explicit calculations, that
 the uniform asymptotics for (\ref{f_v_of_ik+}, \ref{f_v_of_ik-}) coincides with each other, so that
 \be
 \ln f^{as+}_\nu(ik)=\ln f^{as-}_\nu(ik)\e \ln f^{as}_\nu(ik),\
 k, \nu\ra\infty,\ k/\nu=const
 \label{lnfas+-}
 \ee
 and thus we can define \be
 \ln f^{sub}_\nu(ik)=\ln f^+_\nu(ik)+\ln f^-_\nu(ik)-2\ln f^{as}_\nu(ik).\ee

 The numerical evaluations Sec.\ref{results} confirms (\ref{lnfas+-}) and
 provides finite results for $E^f$.

\subsubsection{The asymptotic part $E^{as}$ of the \gse}

\ \ \  Consider the asymptotic part $E^{as}$ of the renormalized energy, defined by (\ref{eas_def}).
 
  The divergent part $E^{div}$ has been defined from the heat kernel expansion and will
 be not further simplified. The ${\cal E}^{asym}_\infty$ allows however advanced transformations.
  We start with the definition of ${\cal E}^{asym}_\infty$ (\ref{eas_def}) and
 perform the analytical continuation of ${\cal E}^{asym}_\infty$ as a complex \f of $s$ in the limit $s\ra 0$.
So we have
 \be {\cal E}^{asym}_\infty=
 C_s \sum\limits_{l=-\infty}^\infty \int\limits_m^\infty dk (k^2-m^2)^{1-s}
 \fr{\pd}{\pd k}\ln f_l^{as}(ik).
 \label{easym_infty2}
\ee

 First of all we suppose that initial value of $s$ is large enough to provide convergence
 of the integral over $k$ and of the sum over $l$ as well, so that the sum
 and the integral can be interchanged.
  In the section \ref{lippsw} it has been shown that the uniform
 asymptotics of the logarithm of \jf for all the problems considered here can
 be represented in the form:
 \be
\ln f_l^{as}(ik)=\int\limits_{0}^{\infty}\fr{dr}{r}\sum\limits_{n=1}^3\sum\limits_{j=1}^{3n}
\int\limits_{0}^{\infty}X_{nj}\fr{t^j}{l^n}
\ee

 The functions $X_{nj}=X_{nj}(r)$ entering here are calculated in the section
 ( \ref{lippsw}).\\
\\

{\bf 1. $E_{as}$ for the scalar field in rectangular scalar background}\\

Let the asymptotic part of the energy be written in the form

\be
{\cal E}^{as}_\infty={\cal E}^{as}_0+{\cal E}^{as}_\pm,
\ee
where the contribution of the $l=0$ in the sum over $l$ is separated, 

\bea
&&{\cal E}^{as}_0=-\fr{1}{4\sqrt{\pi^3}} \fr{\Gamma(s-1)\sin \pi
  s}{\Gamma(s-\half)}\int\limits_m^\infty dk (k^2-m^2)^{1-s} \fr{\pd}{\pd k}
\ln f_0^{as}(ik)
\eea
and
\be
{\cal E}^{as}_\pm=-\fr{1}{2\sqrt{\pi^3}} \fr{\Gamma(s-1)\sin \pi
  s}{\Gamma(s-\half)} \sum\limits_{l=1}^\infty  \int\limits_m^\infty dk (k^2-m^2)^{1-s} \fr{\pd}{\pd k}
\ln f_l^{as}(ik).
\label{e_pm}
\ee
 Here the symmetry property (\ref{symmetry_l}, established in Sec.\ref{jostfunc}) of the
 \jf $\ln f_l^{as}(ik)$ is used.
 The further calculations are given in details in Math.App \ref{easSCALAR}.  

Thus the finite part of $E^{as}$ is represented in terms of integrals not complete calculable analytically, but well convergent numerically.\\
\\ 

 {\bf 2. $E^{as}$ for spinor field in square shaped magnetic flux $\Omega$}\\

  The procedure is in principle the same as in the case of the scalar field.
 Starting with the explicit formula for the ${\cal E}^{as}_\infty$
 
\be
E^{as}=2C_s\sum\limits_{v=1/2, 3/2, . . . }^{}\int\limits_{m}^{\infty}dk(k^2-m^2)^{1-s}
\fr{\pd}{\pd k}\int\limits_{0}^{\infty}\fr{dr}{r}\sum\limits_{n=1}^3\sum\limits_{j=1}^9
\int\limits_{0}^{\infty}X_{nj}\fr{t^j}{\nu^n} - E^{div}\\
\label{eas-ediv}
\ee
with the constant $C_s =\fr{1}{2\pi}[1+s(-1+2ln2\mu)]$, what is indeed the
constant $C_s$ introduced by (\ref{const_Cs}), expanded at $s\ra 0$.
 The explicit substitution of $X_{nj}$ (\ref{X_nj_spinor}) in
 (\ref{eas-ediv}) provides under application of the Abel-Plana formula, but
 in this case it must be summarise over half integer values of $\nu$
 (Math.App. \ref{abelplan}), and the sum is represented by integrals as
 \be
 E^{as}=E^{as}_1+E^{as}_2,
 \ee

where the resulting expression is separated again into finite $E^{as}_1$ and
singular (divergent) parts $E^{as}_2$. The derivation of $E^{as}_1,
E^{as}_2$ coincides formally with the one for the case of spinor field in
Higgs background, given in Math.App.(\ref{jostUA_NO})
(up to the coefficients $X_{nj}$, which are given by (\ref{X_nj_spinor}))

The finite part of $E^{as}$ ( $E^{as}_2$) is performed in the similar way as described
recently for the scalar problem.

 After the partial integrations over $\nu$ several times, that is necessary
 to get rid of divergent powers, the finite part  $E^{as}_2$ results again in
the expression, represented in terms of integrals not complete
calculable analytically, but well convergent numerically: for spinor field in
square shaped magnetic flux $\Omega$ it reads

\bea
&&E^{as}=\nn\\&& \fr{-4}{\pi}\int\limits_{0}^{\infty}\fr{dr}{r^3}[\Omega^2 a(r)^2
f_1(rm)-\Omega^2 r^2 a'(r)^2 f_2(rm)+\Omega^4 a(r)^4 f_3(rm)]\nn\\&&
=\Omega^2 e_1(R_1,R_2)+\Omega^4 e_2(R_1,R_2)
\label{easspinor}
\eea
the $g_i(x)$ are given in Math.App. (\ref{g_i_of_x}). The coefficients $e_1,
e_2$, entering here are to evaluate numerically, the corresponding results
are represented in Sec.\ref{results}, (Fig.\ref{fig1})\\
\\

{\bf 3. $E^{as}$ for abelian Higgs model with NO-vortex}\\
 
 The calculation procedure for $E^{as}$ is sufficiently the same as in the
 case of spinor field in magnetic background considered above. Since there is
 no explicit analytical form of potentials $f(r)$, $v(r)$, the integration
 over $r$ is included in the double integral being executed numerically using
 the same package "NIntegrate" of {\it Mathematica}.

 The analytical formulae for numerical evaluations is derived and represented
 in Math.App. \ref{NO_Eas} sufficiently in the similar way as it has been done for the previous case.


 \subsubsection{Cancellation of divergent contributions in the
  asymptotic\\ energy}

\ \ \  As it was described above, the asymptotical part of the renormalized
 \gse is constructed as a difference of two parts $E^{div}$ and ${\cal E}^{as}_\infty$, each of them is
 infinite in the limit $s\ra 0$. On the other hand the complete
 expression of the renormalized energy (\ref{E_ren_def}) was presupposed
 in the top of Sec.\ref{efineas} to be finite.
  Since the part $E^f$ defined above (\ref{efin_def}) remains finite due
 to the cancellation of divergences between $E^{asym}$ and $E^{reg}$  ,
  (what has been also confirmed numerically), the same must take place between
 divergent contributions in  $E^{div}$ and $E^{asym}$ in $E^{as}$.

Fortunately one can see how does it come about analytically, that is
a non-trivial fact.
It could seem to be surprising that the divergent part of the asymptotic expansion
of energy ${\cal E}^{as}_\infty$ coincides exactly with the divergent energy
$E^{div}$ defined from the heat kernel expansion, which are originally
different objects. 

The minimal necessary condition, which is to expect is the coincidence of
terms with negative powers of $s$ in ${\cal E}^{as}_\infty$ and $E^{div}$ in
the limit $s\ra 0$ (speaking exactly, there is only the first order pole
$1/s$). It assures that our renormalized result is finite.

This feature can be also used for explicit calculation of the \hkc $a_n$ if
the uniform asymptotics of the corresponding \jf is known. Such an example is
demonstrated in Sec. \ref{a52}. The $E^{div}$ (\ref{E_div_def}) contains not
only pole contributions, which are definitive to be cancelled, but also
finite terms, which appears also in ${\cal E}^{as}_\infty$ and provide
partial or full subtraction with the corresponding ones in $E^{div}$.

We consider the mechanism of this subtraction on the example of the {\bf spinor field in the magnetic background.}
For this case we have the single term contributing to the divergent energy:
\be
 E^{div}=\fr{a_2}{32 \pi^2}(\fr{1}{s}-2+\ln\fr{4 \mu^2}{m^2}), 
 \label{e_div_def}
\ee
where $a_2$ is the only non-zero heat kernel coefficient $a_n,\ 0 <
n\le2$, contributing to the $E^{div}$ (\ref{E_div_def}). We recall the
$a_2$, obtained in Sec. \ref{heatkernel} Eq.(\ref{spinor_a2_explicit}):
\be
a_2=\fr{8\pi}{3} \Omega^2 \int\limits_0^\infty r\ dr h(r)^2,
\label{a2_now}
\ee
$\Omega$ and $h(r)$ are the same as introduced in Sec. \ref{hamiltonians},
Eq.(\ref{delta}, \ref{h_r1}).
 Now we construct the ${\cal E}^{as}_1$ in ${\cal E}^{as}_\infty$
 (\ref{eas_intfy}) using (\ref{terms_eas1}) as
\be
{\cal E}^{as}_1=\int\limits_0^\infty \fr{dr}{r} \sum\limits_{n=1}^3
\sum\limits_{j=1}^{3n} X_{nj} \I_{ij}.
\label{calc_eas1}
\ee
 Here for the problem of pure magnetic string the corresponding coefficients
 $X_{nj}$ from Sec. \ref{lippsw} Eqs. (\ref{X_nj_spinor}) are to use.
 By performing of the power expansion at the limit $s\ra 0$ and collecting
 of the powers $s^{-1}$ and $s^0$ we obtain:
 \be
 {\cal E}^{as}_1= \fr{\Omega^2}{12\pi} \left[\fr{1}{s}+\ln \fr{4\mu^2}{m^2}-2 \right] \int\limits_0^\infty \fr{dr}{r^2} (a(r)a''(r)r-a(r)a'(r)).
\label{expression}
 \ee

Then, taking into account, that
\be
\left(\fr{aa'}{r}\right)'= \fr{r a a''-a a'}{r^2}+\fr{(a')^2}{r},\nn
\ee
we rewrite the integral in the (\ref{expression}) as
\be
\left.\int\limits_0^\infty \fr{dr}{r^2} (a(r)a''(r)r-a(r)a'(r))=a(r)h(r)\right|_0^\infty-
\int\limits_0^\infty r\ dr h(r)^2.
\ee
 From the explicit shape of the potential (\ref{a_r1}, \ref{h_r1}),
 Sec.\ref{hamiltonians} it follows, that the surface term is zero and the
 $ {\cal E}^{as}_1 $ is equal exactly to $E^{div}$ (\ref{e_div_def}), thus
 the both terms in $E^{as}$ cancels each other.

 In particular it confirms that the subtraction was executed properly and the terms containing pole at $s\ra 0$ cancel each other.

 We could also consider the contribution in (\ref{calc_eas1}) of the
 coefficients $X_{nj}$ with even $n$ ($X_{22}, X_{24}, X_{26}$), which are
 also present in (\ref{X_nj_spinor}).
  The immediate calculations provide the term
  \be
  {\cal E}^{as}_{1\ even}=-m \fr {\Omega^2}{16}\left. \int\limits_0^\infty r\ dr
 \fr{a^2(r)-2a'(r)r}{r^2}= -m \fr {\Omega^2}{16} \fr{a^2(r)}{r}\right|_0^\infty.
  \ee
The power of mass $m^1$ suggests the correspondence to the \hk coefficient $a_{3/2}$.  
It shows in particular that if the potential $a(r)$ decreases at $r\ra 0$ as
power $r^{1/2}$ or slower, the non zero contribution is provided. For our
potential (\ref{a_r1}) this is not the case. Generally speaking, we can
 derive two conditions for the potential $a(r)$ at $r=0$ to keep the ${\cal
 E}^{as}_{1\ even}$ to be zero:
\be a(r)|_{r=0}=\left.\fr{\pd}{\pd r} a(r)\right|_{r=0}= 0 \ee
  The first requirement is clear since $a(r)$ corresponds to the tangential
  component $A_{\vp}$ of the vector field $\vec{A}$, thus the condition
 $a(0)\not=0$ would mean for this vector potential to have a singularity at
 $r=0$. If we have
$\left.\fr{\pd}{\pd r} a(r)\right|_{r=0}\not= 0$, then $a(r)$ being viewed as
a surface in cylindric coordinates $\{ r, \vp, a(r)\}$ has a conical
 non-smoothness at $r=0$. In both cases the coefficient $a_{3/2}$ in the \hk
 expansion is different from zero. Thus we have an implicit suggestion, that
 even powers of the orbital momenta does not contribute to the \gse because
 of  $a_{3/2}=0$.  
  

 It could be suitable to remark, that the exact cancellation takes place
 only for special choice of the uniform asymptotic expansion of the \jf,
 namely here the choice of the half-integer orbital momenta $\nu$ is crucial.
  If we would have considered the \jf depending on the original integer
 momenta $l$, the ${\cal E}^{as}_1 $ would not coincide with
 the $E^{div}$ in finite part.

Further we restrict us on short remarks to the two other models.

For the {\bf spinor field in the electroweak vortex background} we proceed in the similar way as it has been carried out recently, but
 in the expression of ${\cal E}^{as}_1 $ (\ref{calc_eas1}) we have to use
 the coefficients $X_{nj}$ from (\ref{repres_xij}). It brings the ${\cal
 E}^{as}_1$ according to (\ref{eas1_eas2}, \ref{contrib_eas1}) to be exactly
 the same expression as $E^{div}$ defined from (\ref{e_div_def}) with the non
 zero \hkc $a_1, a_2$ (\ref{a1_a2_NO}), thus the cancellation comes about
 completely.

In the case of the {\bf scalar field in the scalar background}, the divergent
part of $ {\cal E}^{as}_\infty $ is the ${\cal E}^{as}_1 $ (\ref{Eas1-Eas3}),
which subtract out exactly from the $E^{div}$ (\ref{edivscalar}). The three
other constituent parts ${\cal E}^{as}_0, {\cal E}^{as}_2, {\cal E}^{as}_3$
are pure finite and contribute to the finite part of  $E^{as}$.
 
 \subsection{Renormalization and interpretation}\label{renormalize}

We defined in Sec.(\ref{efineas}) the renormalized ground state energy as
\begin{equation}
 E^{ren}=E_0-E^{div },
 \label{e0-ediv}
\end{equation}
where $E^{div }$ has been obtained from the heat kernel expansion
Sec.(\ref{heatkernel}, \ref{zfreg}).

 The main point of the \rn in quantum field theory is the redefinition of the
 theory constants through the appearing counter terms. Certainly, here we
 take in mind renormalizable models. Generally speaking, if one obtains a
 finite effective action, one needs to give an interpretation to appearing
 (possible new) effective constants.
 
 The physically consequent interpretation of subtraction of 
 the terms contained in $E^{div }$ is sometimes not quite well defined and
 not unique.

 For the renormalized ground state energy the asymptotic dependence
on powers of $m$ at $m\rightarrow \infty$ follows from
Eq.(\ref{large_mass_expansion}) to have the form
 \begin{equation}
 E_0^{ren}\stackrel{\sim }{_{m\rightarrow\infty}} \sum\limits_{n > 2} \fr{e_n}{m^{2n}}  
 \end{equation}
with some coefficients $e_n$.

In accordance with the interpretation of $ E^{ren}$ it must vanish in
the limit of $m\rightarrow \infty$ since it is the energy of vacuum
fluctuations so that $E^{ren}$ fulfils the normalization condition \cite{BKEL}
 \begin{equation}
 \lim E^{ren}=0\ \ at\ \ m \rightarrow \infty .
 \label{norm_cond}
 \end{equation}

Through the subtraction of terms containing all
non-negative powers of $m^2$ (which are the terms of heat kernel
expansion up to $a_2$) the condition (\ref{norm_cond}) is satisfied
automatically.
 
Some comments on the subtraction scheme are in order.
The
scheme had been used in a number of Casimir energy calculations.  In
\cite{Bordag:2002dg} it had been shown to be equivalent to the so
called ``no tadpole'' normalization condition which is common in field
theory.  It should be noticed that this scheme does not apply to
massless fluctuating fields, for a discussion of this point see
\cite{Bordag:1998vs}.
 
 We consider below some further examples of \rn caused by this scheme.
\subsubsection{The counter terms}
 
\ \ \ Proceeding from the HKE we defined the terms with non negative powers of m
(all the contributions up to $a_2$) as $E^{div}$ and subtract it out.
Thus the renormalized energy is defined by (\ref{e0-ediv}) and is now finite.
The subtraction of $E^{div}$ implies the addition of corresponding
counter terms to the initial lagrangian.
 Since the whole energy of the field system consists of the \gse of quantum
 field and classical background energy, the form of counter terms should be
 suitable to redefine certain parameter of the model, that enters in the classical energy.\\
 \\

{\bf 1. The counter terms in the scalar field model with scalar step-function
  background}

The counter term contains the contributions of the \hkc $a_1$ and $a_2$ only since
all other are zero for this background (Sec.\ref{efineas},\ref{heatkernel}).
\be
{\cal
  E}^{div}= \frac{m^2}{32\pi^2}\left(\frac{1}{s}+\ln\frac{4\mu^2}{m^2}-1\right)a_1-\frac{1}{32\pi^2}\left(\frac{1}{s}+\ln\frac{4\mu^2}{m^2}-2\right)a_2.
\ee

 The coefficients  $a_1$ and $a_2$ were calculated in
 Sec.\ref{heatkernel} ,Eq. (\ref{a1_scalar},\ref{a2_scalar}) in terms of the scalar potential V(x). To make the \rn descriptive we remember that $V(x)$
was introduced in Sec.\ref{hamiltonians} as

\be  V(x)=\alpha \Phi(x)^2 \ee
to be quadratic in the background $\Phi(x)$, the interaction constant is
thereby dimensionless.

Now on comparing the classical energy
\be
 {\cal E}^{cl}=\half\{ \partial_0\Phi\partial^0\Phi+(\nabla\Phi)^2+ M^2\Phi^2
 + \la\Phi^4 \}
\ee
with the resulting counter term

\be
{\cal
  E}^{div}=
-\frac{m^2\alpha \Phi^2}{32\pi^2}(\frac{1}{s}+\ln\frac{4\mu^2}{m^2}-1)-
\frac{\alpha^2\Phi^4 }{64\pi^2}(\frac{1}{s}+\ln\frac{4\mu^2}{m^2}-2)
\ee
one can ascertain that the parameter $\la$ and M that are the
self-interaction coupling and the mass of $\Phi$ respectively, become
redefined in accordance with:

\bea
\la\ \ \ra \ \ \tilde{\la} = \la +\frac{\alpha^2 }{32\pi^2}(\frac{1}{s}+\ln\frac{4\mu^2}{m^2}-2)\\
M^2\ \ \ra \ \ \tilde{M}^2 =M^2+\frac{m^2\alpha }{16\pi^2}(\frac{1}{s}+\ln\frac{4\mu^2}{m^2}-1)
\eea

so that both parameters has been subjugated the infinite \rn.
\\

{\bf 2. The counter term for Dirac field interacting to the magnetic background
$A_\mu(x)$}

 The single counter term reads:
 
\be\label{counter}
  E^{div}=
-\fr{1}{32\pi^2}(\fr{1}{s}-2+\ln \fr{4\mu^2}{m^2})a_2, \ee
where $a_2$ for this problem (\ref{dirac_a1_a2})
\be a_2=\fr{8\pi}{3}\Omega^2 \int r\ dr\ h(r)^2 \e e^2\fr{2\phi^2}{3\pi} \int
  r\ dr\ h(r)^2\ee

On the other hand we have for the classical energy of the magnetic background
\cite{bogolubov}:
\be E^{class}= \fr{1}{4} F^{\mu\nu}F_{\mu\nu}
=\half\int dx\ {\bf B(x)}^2 = e^2\fr{\phi^2}{4\pi}\int r\ dr\ h(r)^2
\label{magnetclass}
\ee

Thus the addition of (\ref{counter}) to the (\ref{magnetclass}) implies
the redefinition of the coefficient in front of the Fermi term as follows
\be \fr{1}{4}\ra \fr{1}{4}+e^2\fr{2}{3}\left( \fr{1}{s}-2+\ln \fr{4\mu^2}{m^2}
\right), 
\ee
that corresponds to redefinition of the cosmological constant by the factor
$e^2$ and to the \rn of the electric charge \cite{bensant} 

$e\ra e', $ so that \be\fr{1}{e'^2}=  \fr{1}{e^2}+\fr{8}{3} \left( \fr{1}{s}-2+\ln \fr{4\mu^2}{m^2}
\right)\ee

 The electric charge becomes in such a way finitely renormalized.\\
\\

{\bf  3. The Dirac field with Yukawa-type coupling to the Nielsen - Olesen
string}\\
\\
\ Consider the counter term provided by the divergent energy $E^{div}$ is
 represented
 in terms of NO-vortex
 solutions $f(r), v(r)$. Sec.\ref{efineas}, (\ref{NO_ediv}).

 On the other hand we have the classical energy of the NO background (recall the
 results of Sec.\ref{hamiltonians}):
 
 \be
 {\cal E}^{class}=\pi\int r\ dr\left\{\fr{1}{q^2r^2}v'(r)^2+\eta^2
 [f'(r)^2+\fr{f(r)^2}{r^2}(1-v(r))^2 ]+ \fr{\la\eta^4}{2}(f(r)^2-1)^2
 \right\}
 \label{class_energy}
 \ee
 
  On comparing these two expressions we have to establish that the counter term
  (\ref{NO_ediv}) does not allows to carry out the complete \rn which could
  be consequently interpreted as in two previous cases. Indeed, the term
  proportional to $v'(r)^2$ provides the standard \rn of the electric charge $q$
 in the classical action. The divergences proportional to $f'(r)^4$ can be
  absorbed through the infinite \rn of the scalar coupling constant $\la$.
   Then the remaining freedom is the change of the condensate $\eta$, that
  cannot be redefined properly because the structure of terms proportional to
  $\eta$ in $E^{div}$ distinguishes obviously from the one in the $
  E^{class}$.
    This is not surprising since the model given by (\ref{lagr_with_interaction}) in
  Sec.\ref{hamiltonians} contains a non-polynomial interaction and therefore
  is algebraically non renormalizable.
    There is however some possibilities to give an interpretation of this \rn
    procedure by considering the classical background fields to satisfy the
    \eqqs of motions. An example of such on-shell \rn is given in Math.App. \ref{NOrenorm}.
    

 \subsection{Higher orders of the uniform asymptotic expansion of the \jf\\ and the \hk coefficient $a_{5/2}$}
\label{a52}
The magnetic background considered in Sec.(\ref{hamiltonians})
 has singular surfaces where the magnetic field jumps. Here we represent the
 calculations have been done in \cite{Drozdov:2002um}. The heat kernel expansion for the case of
singularities concentrated at surfaces has been considered in
[\cite{borvass}, \cite{moss}, \cite{gkv}, \cite{gkv2}].  Although the
general analysis of \cite{gkv} is valid for our background, an
explicit expression for $a_{5/2}$ has not been calculated before.
 
We can use our obtained Jost function (\ref{f_v_of_ik+}) to calculate
 the coefficient $a_{5/2}$ in the heat kernel expansion.
 
Suppose we have obtained the value of $E_0(s)$ (\ref{e0_of_lnf})in the
 point $s=-1/2$. It follows from (\ref{zeta_d_expansion})
 that
 \begin{equation}
 E_0(s)\sim - \fr{\mu^{2s}}{2(4\pi)^{3/2} } \fr{m^{4-2s}}{\Gamma(s-\half)}
 \sum\limits_{n=0, \half, 1, . . . }\fr{a_n}{m^{2n}} \Gamma(s-2+n)
 \end{equation}
and at the limit of $s \rightarrow -\half$ only the term containing
 $a_{5/2}$ remains to be nonzero in the sum.
\begin{equation}
 E_0(\half)= - \fr{\mu^{2s}}{2(4\pi)^{3/2} } a_{5/2}
\end{equation}

From the other hand we have (\ref{e0_of_lnf})
\begin{equation} 
 E_0(s)= C_s \sum\limits_{\nu=1/2, 3/2, . . . } \int\limits_m^\infty
 dk(k^2-m^2)^{1-s} \fr{\pd}{\pd k} f_\nu(ik)=C_s h(s).
\end{equation} 
Here we substitute the exact Jost function by its uniform asymptotic
represented in the form
\begin{equation} 
 f^{ua}_\nu(ik)=\sum\limits_{n=1,3,4,5...} \fr{h_n(t_1, t_2)}{\nu^n},
 \label{unif_as_f}
\end{equation}
where the coefficients $h_n$ are functions of
$t_1=(1+(\fr{kR_1}{\nu})^2)^{-1/2}$,\\
$t_2=(1+(\fr{kR_2}{\nu})^2)^{-1/2}$, the power $\nu^{-2}$ is absent,
as noticed above (Sec.\ref{lippsw}),
\begin{equation} 
C_s= -\fr{\mu^{2s}\Gamma(s-1)}{4 \sqrt{\pi}\Gamma(s-1/2)}\fr{-4 \sin(\pi s)}{\pi}
\end{equation}
At the limit $s \rightarrow -\half$ it yields 
\begin{equation} 
 E_0(-\half)=\fr{4}{3\pi} Res_{s \rightarrow -\half} h(s) 
\end{equation}
and therefore we obtain for $a_{5/2}$
\begin{equation}
 a_{5/2}=-\fr{64\sqrt{\pi}}{3} Res_{s \rightarrow -\half} h(s).
\end{equation}

To obtain the explicit form of $Res_{s \rightarrow -\half} h(s)$ we
 use the uniform asymptotic expansion of the Jost function
 (\ref{f_v_of_ik+}).
 The terms $h_n(t_1,t_2)$ in uniform asymptotic expansion can
 be obtained either by iterations of Lippmann-Schwinger equation (see
 \cite{borkir2}, \cite{borkir1}) or by using the explicit form of the
 Jost function as well. All the further terms up from $h_4$ are
 produced from the explicit form of the Jost function
 (\ref{f_v_of_ik+}) because of complication of the first way for higher
 orders $n$ (see the remark to \ref{X_nj_spinor} in Sec.\ref{lippsw}).  Namely, we
 obtain several higher orders $1/\nu$ of uniform asymptotic expansion
 for special functions $I_{\nu},K_{\nu},_{1}F_1$ which the exact Jost
 function (\ref{f_v_of_ik+}, \ref{f_v_of_ik-}) consists of (it can be done starting with
 the explicit form for two first orders and executing the recursive
 algorithm several times\cite{abrsteg}), then after substitution of
 each of functions $I ,K, _{1}F_1$ by its corresponding uniform
 asymptotic expansion and separation of powers of $\nu$ we arrive at
 the form (\ref{unif_as_h}). The coefficients $h_n(t_1, t_2),
 n=1,...,4$ are given in the Appendix (\ref{unif_as_h}).
 
 If the function $h_n(t_1, t_2)$ is a polynomial over $t_1, t_2$, so
 we can consider some term $t^j , (t=t_i, i=1, 2 ) $ of it.  Notice,
 that for $h_1(t_1,t_2),h_3(t_1,t_2)$ it is not the case, but we can
 treat the terms of kind $\fr{1}{1+t}$ and $\fr{1}{(1+t)^k}$,
 $k$-integer, as an infinite sum of powers $t$.
 
 Apart from the construction of $t= t_1, t_2$ (\ref{unif_as_f}) they
 are strictly positive and less than 1, therefore each of the series in $
 \fr{1}{1+t}=\sum\limits_{i=0}^{\infty}t^{2i} - \sum\limits_{i=0}^{\infty} t^{2i+1}$ converges regular and
 uniform and respecting that we have only finite integer powers $k$ so
 that $\left( \sum\limits_{i'=0}^{\infty} t^{i'} \right)^{k'}$ converges as well
 (the $k',i'$ are integer). The sum over $i'$ can be interchanged with the one over $\nu$
 (Princeheim's Theorem) thus the following procedure for some integer power
 $t^j$ is valid.
 
 Performing the sum
\begin{equation}
 h(s)=\sum\limits_{\nu=1/2, 3/2,\dots} \int\limits_m^\infty
 dk(k^2-m^2)^{1-s} \fr{\pd}{\pd k} \sum\limits_{n=1, 3, 4, 5} h_{n,j}(R_1,R_2) \fr{t^j}{\nu^n}
\end{equation}
by meaning of (\ref{abelplan}) we obtain the sum of two parts
\begin{eqnarray}
h(s)&=&\int\limits_0^\infty d\nu\int\limits_m^\infty dk(k^2-m^2)^{1-s}\fr{\pd}{\pd
 k}\sum\limits_n h_{n,j}(R_1,R_2)\fr{t^j}{\nu^n}+ \\\nn
&&\int\limits_0^\infty \fr{d\nu}{e^{2\pi\nu}+1}\fr{1}{i}\left[ \int\limits_m^\infty
 dk(k^2-m^2)^{1-s}\fr{\pd}{\pd k}\sum\limits_n h_{n,j}(R_1,R_2)\fr{t^j}{\nu^n}
\right]_{\nu=i\nu}^{\nu=-i\nu},
\label{two_parts}
\end{eqnarray}
where $t=t_{1, 2}$. The first summand of (\ref{two_parts}) gives for each power $j$ of $t=\{ t_1, t_2 \}$ 
(using the (\ref{ident1}))
\begin{equation}
 -\fr{m^3}{2}
  \fr{\Gamma(2-s)\Gamma(\fr{1+j-n}{2})\Gamma(s+\fr{n-3}{2})}{(Rm)^{n-1}\Gamma(j/2)}
\label{first_part}  
\end{equation}
(where R denotes $R_1$ and $R_2$ respectively)
at the limit $s \rightarrow -\half$ only the terms corresponding to $n=2$
and $n=4$ (which will be calculated below) have a pole.

For the second part of the Abel-Plana formula (\ref{two_parts}) we have
\begin{eqnarray}
&&\fr{1}{i}\int\limits_0^\infty\fr{d\nu}{1+e^{2\pi\nu}}
\fr{\Gamma(2-s)\Gamma(s+j/2-1)}{jR^{4-2s}\Gamma(j/2)} \\
&\times&\left[(i\nu)^{2+j-n}[(mR)^2+(i\nu)^2]^{1-j/2-s}-(-i\nu)^{2+j-n}[(mR)^2+(-i\nu)^2]^{1-j/2-s}\right]\nn
\label{2_part}
\end{eqnarray}

It can be seen using the Laurent expansions at $s \rightarrow -\half$
of Gamma functions entering the (\ref{2_part}), that only the terms
containing -1, 1, and 3 powers of $t_i$ can contribute to the residuum
at $s \rightarrow -\half$. But the expression in square brackets in
(\ref{2_part}) can be performed as
\begin{equation}
 -\nu^2(\nu^2-(mr)^2)[(i\nu)^{j-n}e^{i\pi (1-j/2-s)}-(-i\nu)^{j-n}e^{-i\pi (1-j/2-s)}]
\end{equation}
thus for $j-n$ even the expression in square brackets produces (dropping the
non sufficient coefficient $\pm 2$ or $\pm 2i$):\\
\hspace*{2cm}for $j=-1, 1 :\ \ \ \ \sin \pi(1-j/2-s)=\cos \pi s $\\
\hspace*{2cm}for $j=3 :\ \ \ \ \sin \pi(1-j/2-s)= -\cos \pi s $\\
\\
and for $j-n$ odd respectively\\
\hspace*{2cm}for $j=\pm 1 :\ \ \ \ \cos \pi(1-j/2-s)=\pm\sin \pi s $\\
\hspace*{2cm}for $j=3 :\ \ \ \ \cos \pi(1-j/2-s)= -\sin \pi s $\\
\\
But for $s \rightarrow -\half$ these functions behave as\\
\hspace*{2cm}$\sin \pi s \sim -1+\fr{\pi^2}{2}(\half+s)^2, \\ \cos \pi
s \sim \pi(\half+s)\\ $ and it means that only the contribution of
terms with odd powers $j-n$ of $i\nu$ could survive and these
corresponds for the possible values of $j$ to $\nu^{-2}, \nu^{-4},
\nu^{-6}, . . . $. In fact the coefficient $h_2(s)$ at $\nu^{-2}$ is
zero, and all other possible terms up from $\nu^{-4}$ does not contain
any powers of $t_i$ lower than 4; one can see it for example in the
explicit form of the uniform asymptotic expansion of special
functions entering the $\ln f_v(ik)$ (\ref{unif_as_h}, Sec.\ref{high_ord}).
Therefore the
second summand of (\ref{two_parts}) does not produce any contribution to
the $Res_{s \rightarrow -\half} h(s)$. Thus we have that only the
contribution from the first summand of (\ref{two_parts}) remains, and
since the term of $n=2$ is zero the searched residuum resulting from
the term of $n=4$ is:
\begin{equation}
 Res_{s \rightarrow -\half} h(s)=-\half.\fr{3 \sqrt{\pi}}{4}
 \fr{\Omega^2}{4(R_1^2-R_2 ^2)^2} (R_1+R_2)\sum\limits_{j=4, 6}\fr{\Gamma(\fr{1+j-n}{2})}{j/2},
 \end{equation} 
 where the (\ref{first_part}) and the explicit form of $ h_4(t_1, t_2)=\fr{\Omega^2}{4(R_1^2-R_2 ^2)^2}[R_1^4(t_1^4-t_1^6)+R_2^4(t_2^4-t_2^6)]$ have been used. 
Finally we have
 \begin{equation}
 Res_{s \rightarrow -\half} h(s)=\fr{15\pi}{128}\fr{\Omega^2(R_1+R_2)}{(R_1^2-R_2^2)^2}
 \end{equation}
 and therefore
 \begin{equation}
 a_{5/2}= \fr{5\pi^{3/2}}{2} \fr{\Omega^2(R_1+R_2)}{(R_1^2-R_2^2)^2}
 \label{coeff_a52}
 \end{equation}
This is the heat kernel coefficient $a_{5/2}$ for the configuration of
the magnetic background field as given by
Eqs.(\ref{potential_A}-\ref{h_r1}).
  
We can calculate the heat kernel coefficient $a_{5/2}$ in a more
general situation when the magnetic field jumps on an {\it arbitrary}
surface $\Sigma$. The coefficients $a_n$ for $n=1/2,...,2$ can be read
off rather general expressions of the paper \cite{gkv}.  Let $B^\pm$
be values of the magnetic field on two sides of $\Sigma$.  According
the analysis of \cite{gkv}, the coefficient $a_{5/2}$ must be an
integral over $\Sigma$ of a local invariant of canonical mass
dimension 4, which is symmetric under the exchange of $B^+$ and $B^-$
and which vanishes if $B^+=B^-$ (i.e. when the singularity
disappears). There is only one such invariant which gives rise to the
following expression:
 \begin{equation}
   a_{5/2}= \xi \int\limits_{\Sigma} (\vec{B}^+ -  \vec{B}^-)^2 d\mu(\Sigma),
   \label{general_a52}
 \end{equation}
where the integration goes over the surface $\Sigma$ and
$\vec{B}^{\pm}$ are the values of the magnetic field on both sides of
$\Sigma$ in the given point. The yet undefined constant $\xi $ can be
found using Eq.(\ref{coeff_a52}) which constitutes a special case of
(\ref{general_a52}). Here the surface $\Sigma$ consists of two circles
in the $ (\vec{X},\vec{Y})$-plane so that
 \begin{equation}
 \int\limits_{\Sigma} (\vec{B}^+ -  \vec{B}^-)^2 d\mu(\Sigma)=2\pi(R_1+R_2)\vec{B}^2,   
 \end{equation}
where the jump $(\vec{B}^+ - \vec{B}^-)$ is just the value of
$\vec{B}$ at $r \in [R_1,R_2]$. With this, Eq.(\ref{general_a52})
takes the form
\begin{equation}
   a_{5/2}=2\pi e^2 \xi\vec{B}^2(R_1+R_2).
 \label{agen}
\end{equation} 
On the other hand, from Eq.(\ref{h_r1}, \ref{delta}) we have
\begin{equation}
 \Phi= B\int\limits_{r \in[R_1,R_2]}B d^2 x=\pi B(R_2^2-R_1^2)
\label{phiofb}
\end{equation}
and with Eq.(\ref{delta}) from Eq.(\ref{coeff_a52}) it follows that
 \begin{equation}
 a_{5/2}= \fr{5}{8\sqrt{\pi}} e^2 \Phi^2 \fr{R_1+R_2}{(R_2^2-R_1^2)^2}.
 \label{asp}
 \end{equation}
comparing (\ref{agen}) with (\ref{asp}) we get
\begin{equation}
  \xi = \fr{5}{16\pi^{3/2}}.
\label{ksi}
\end{equation}  


\section{Graphics and numerical results}
\label{results}

\subsection{Scalar string}

\subsubsection{Finite and asymptotic vacuum energy}

{\bf 1. Finite energy}\\

\begin{figure}[h] 
\epsfxsize=8cm\epsffile{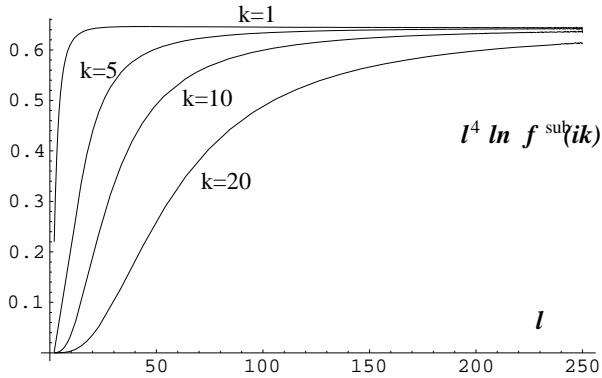}
\caption{\small\bf $\ln f_l^{sub}(ik)$ multiplied by $l^4$ at $R_1=1,R_2=2, V_0=1.1$ for $k=1,5,10,20$}
 \label{lnfsub_plot}
\end{figure}

\begin{figure}[h] 
\epsfxsize=8cm\epsffile{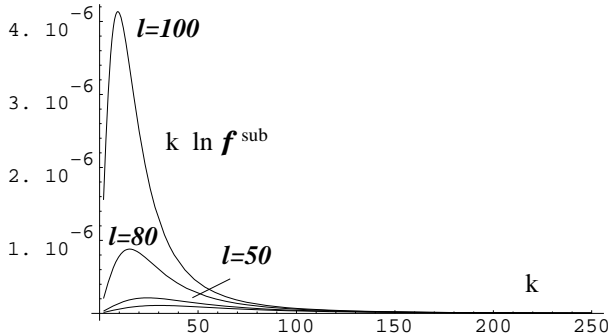}
\caption{\small\bf $k \ln f_l^{sub}(ik)$ at $R_1=1,R_2=2, V_0=1.1$ for $l=30,50,80,100$}
 \label{integrand_plot}
\end{figure}

 The finite \gse for scalar field is calculated numerically by using
 of (\ref{scalarEF}). The plot of the $\ln f_l^{sub}(ik)$ entering here
 is shown on Fig.(\ref{lnfsub_plot}) for several values of momentum $k$. It is
 multiplied by $l^4$ to make the asymptotic behaviour at $l\ra\infty$
 prescribed in Sec.\ref{efineas} visible. The Fig.\ref{integrand_plot} shows
 some plots of the integrand over $k$ in (\ref{e_pm}) for several orbital momenta $l$. It can be seen that the shapes of curves provide well
 convergent integrals for each of $l$.\\

{\bf 2. Asymptotic energy}\\

\begin{figure}[h] 
\epsfxsize=8cm\epsffile{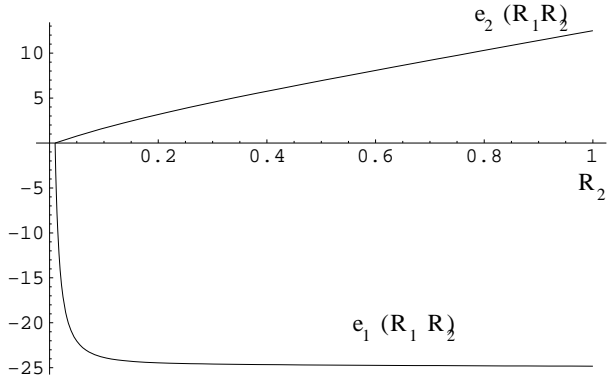}
\caption{\small\bf $e_1, e_2$  at $V_0 = 33.1$; $R_1=0.01,\ 0.010001< R_2<
  0.9999$}
\label{outaway}
\end{figure}

\begin{figure}[h] 
\epsfxsize=8cm\epsffile{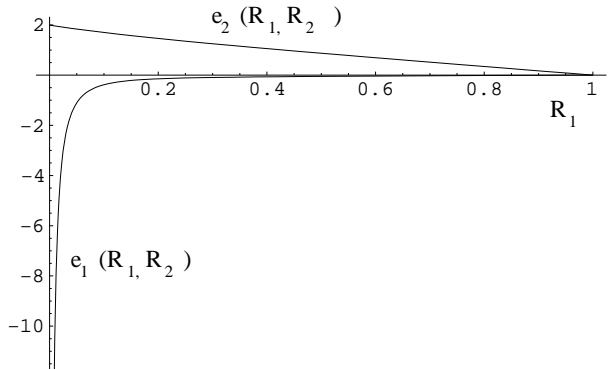}
\caption{\small\bf $e_1, e_2$   at $\ V_0=33.1$, $R_2=1.,\  0.010001< R_1< 0.9999$}
 \label{inout}
\end{figure}

The asymptotic energy $E^{as}$ contains contributions of orders $V_0$ and $V_0^2$ of
the potential respectively. We represent the $E^{as}$ to be consistent of
these two terms \be E^{as}= e_1(R_1,R_2)V_0 +e_2(R_1,R_2)V_0^2,  \ee
and the coefficients $e_1(R_1,R_2), e_2(R_1,R_2)$ are shown in
Fig.\ref{inout} - Fig.\ref{outaway} for fixed value of potential $V_0=33.1$.
 On the first plot the outer radius is fixed and the inner one moves from
 inside to them - the asymptotic energy shows physically "well'' behaviour,
 both contributions vanish at infinitely thin potential wall. 
The second plot demonstates the growth of both contributions if the inner
radius is fixed and the outer increases.

\subsubsection{Parametrical dependence of vacuum energy}

\ \ \ We are interesting for the dependence of the \gse on the external parameter,
such as the inner/outer radia $R_1,R_2$ of the background and the height
$V_0$ of potential as well.

%
%
%
%
The remarkable properties occur, if the height of scalar potential $V_0$
becomes small. Generally, the \gse remains positive almost for all values of
$V_0$ as shown in Fig.\ref{posenergy}. However, it exists a very  thin region
of potential values between $V_0 \sim 0.01$ and $V_0 \sim 0.065$ where the
$E^{ren}$ is negative (Fig.\ref{negenergy}) and possesses generally a global minimum (Fig.\ref{globmin})

\begin{figure}[h] 
\epsfxsize=10cm\epsffile{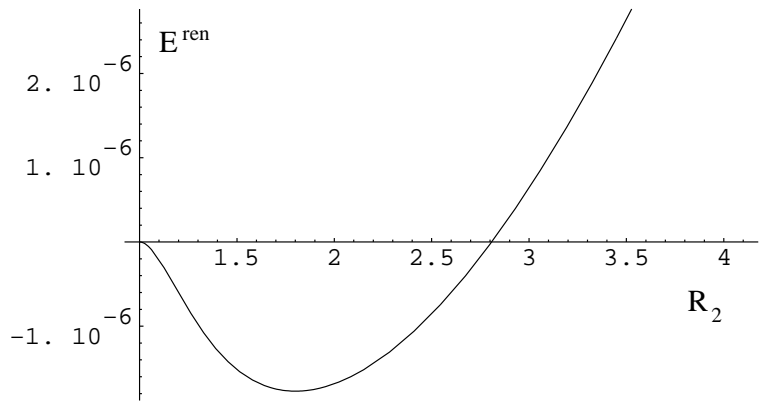}
\caption{\bf The deepest minimum of the \gse $E^{ren}$ at $V_0 = 0.062$; $R_1=1$.}
 \label{deepmin}
\end{figure}

\begin{figure}[h] 
\epsfxsize=10cm\epsffile{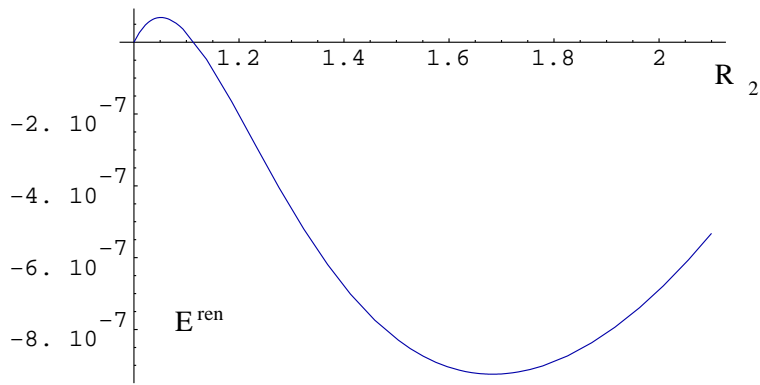}
\caption{\bf The critical behaviour of $E^{ren}$ with two extrema at $V_0 = 0.067$; $R_1=1$.}
 \label{critical}
\end{figure}

\begin{figure}[h] 
\epsfxsize=10cm\epsffile{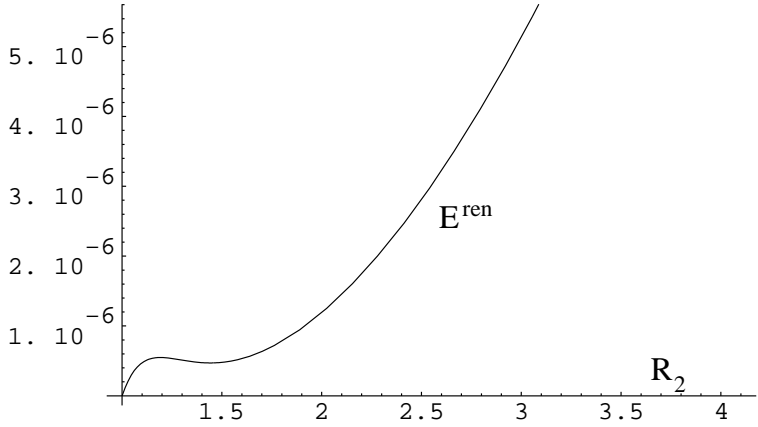}
\caption{\bf Both extrema of $E^{ren}$ vanish at $V_0 = 0.071$; $R_1=1$.}
 \label{minvanish}
\end{figure}

\begin{figure}[h] 
\epsfxsize=10cm\epsffile{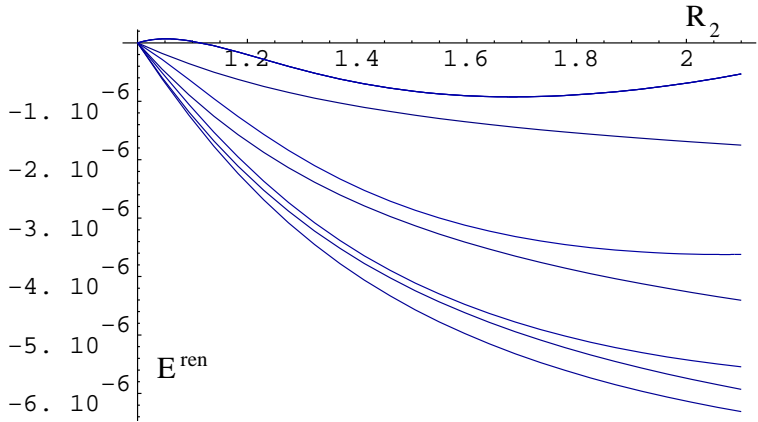}
\caption{\bf The $E^{ren}$ is negative for potential $V_0$ between $\sim 0.01$
  and $\sim 0.065$}
 \label{negenergy}
\end{figure}

\begin{figure}[h] 
\epsfxsize=10cm\epsffile{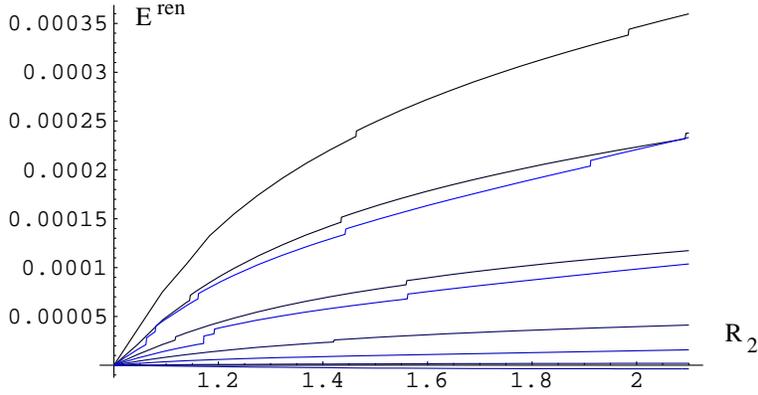}
\caption{\bf Samples of behaviour of $E^{ren}$ for $0.1 < V_0< 1.1$}
 \label{posenergy}
\end{figure}

\begin{figure}[h] 
\epsfxsize=10cm\epsffile{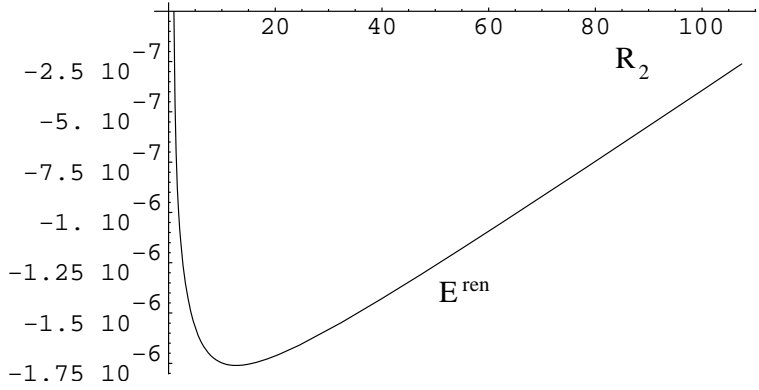}
\caption{\bf For small $V_0=0.003$ $E^{ren}$ demonstrates a global minimum.}
 \label{globmin}
\end{figure}

 While the \gse changes the sign from negative to positive one, it shows the
 non-trivial behaviour which is represented by Fig.\ref{critical}. The
 critical curve has two extrema: at $R_2\sim 1.067$ (minimum) and $R_2\sim
 1.05$ (maximum).
 At  $V\sim 0.062$ the critical energy approaches the deepest minimum
 (Fig.\ref{deepmin}), and at $V\sim 0.071$ the minimum vanishes
 (Fig.\ref{minvanish}).
 
Since the dependence of classical energy of the background is pure quadratic
 on $R_2$, this behaviour is kept also for total energy, the positions of
 extrema will be shifted.

\begin{figure}[h] 
\epsfxsize=10cm\epsffile{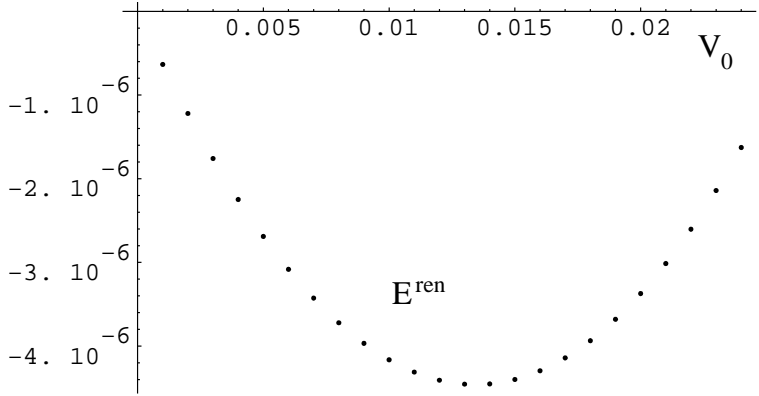}
\caption{\bf For fixed $R_1=1, R_2=11.5$, $E^{ren}$-dependence on $V_0$ has a slight minimum.}
 \label{slight}
\end{figure}

\begin{figure}[h] 
\epsfxsize=10cm\epsffile{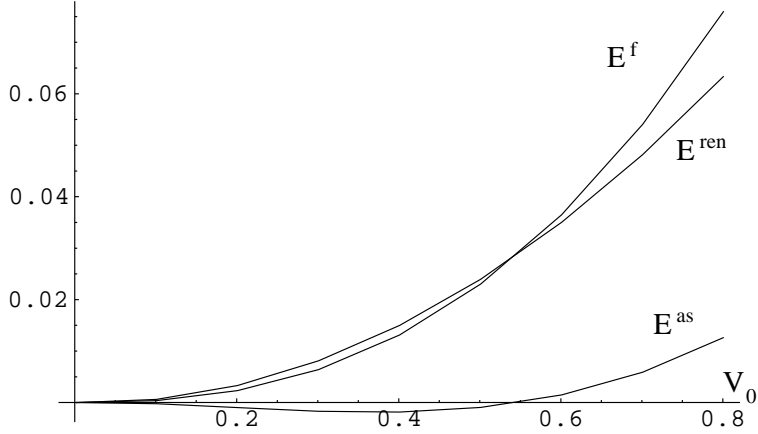}
\caption{\bf $E^{f}$,  $E^{as}$, $E^{ren}$ of $V_0$ for $R_1=1, R_2=11.5$}
 \label{grow}
\end{figure}

\begin{figure}[h] 
\epsfxsize=10cm\epsffile{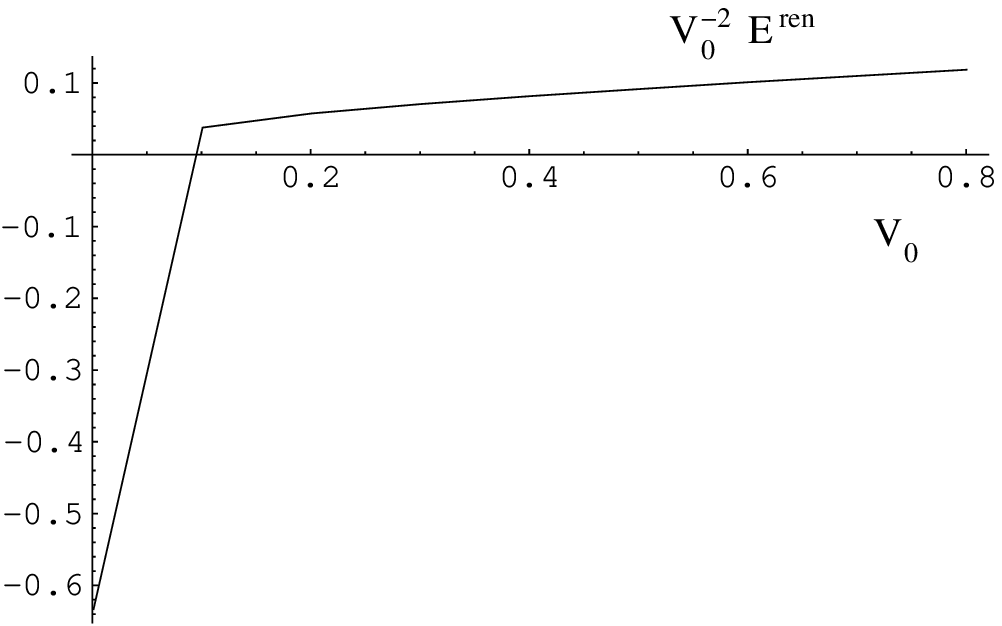}
\caption{\bf $E^{ren}$ divided by $V_0^2$.}
 \label{growV2}
\end{figure}

\begin{figure}[h] 
\epsfxsize=9cm\epsffile{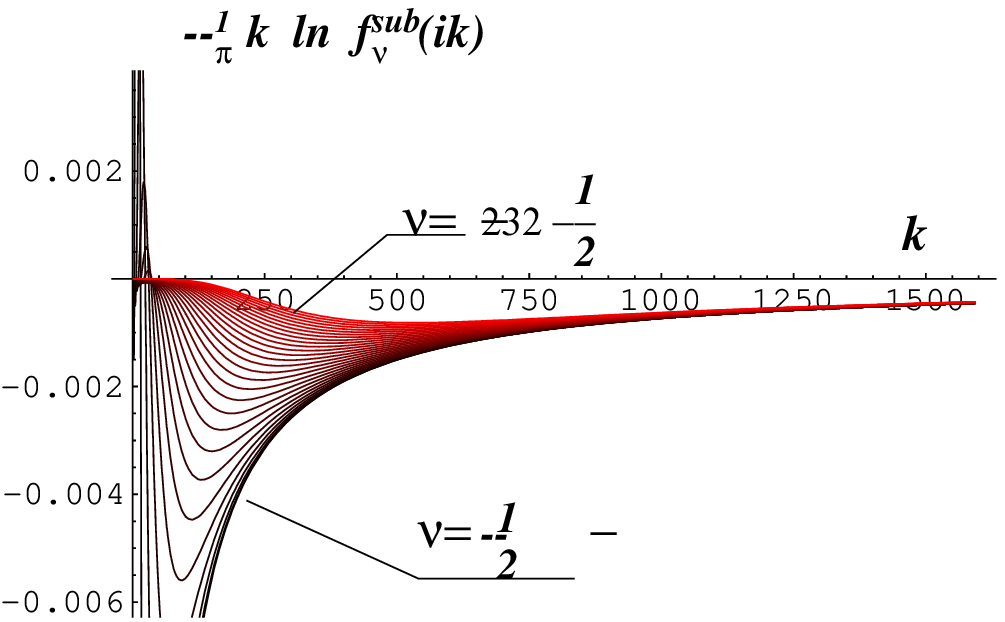}
\caption{
  for $ R_1=0.0001, R_2=1.,\  \Omega=3.$ the bundle of $-\fr{1}{\pi}k\ln
f_{\nu}^{sub}(ik); \half < \nu < 250\half$, multiplied with $k^2$}
 \label{bundle}
\end{figure}

\begin{figure}[h]
 \epsfxsize=9cm\epsffile{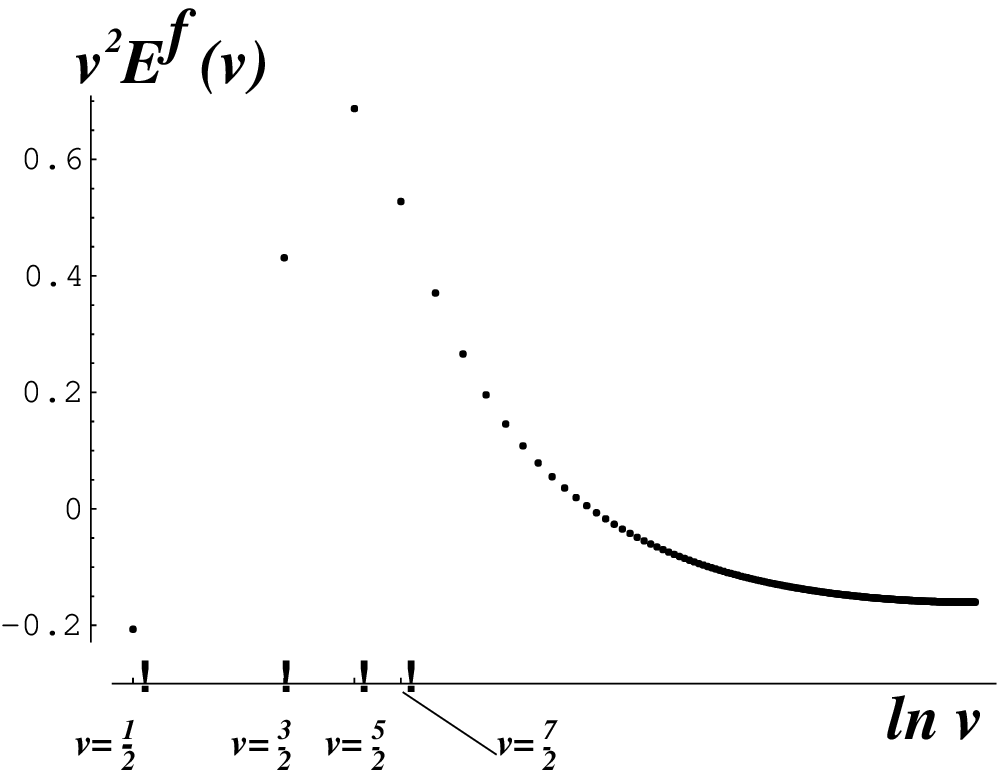}
 \caption{Contribution of each integral $-\fr{1}{\pi}\int\limits_m^\infty k\ln f_{\nu}^{sub}(ik) dk$ (multiplied
with $\nu^2$}
 \label{contrint}
\end{figure}

Now consider the dependence of $E^{ren}$ on the potential height $V_0$ if
$R_1, R_2$ are fixed.
 The Fig.\ref{slight} shows the slight minimum at $R_1=1,
 R_2=11.5$, that corresponds approximatively the global minimum of $E^{ren}$
 at $V_0=0.003$. If the potential  $V_0$ grows, the dependence becomes
 obviously quadratic in the leading order, (see Table I and Fig.\ref{grow},\ref{growV2}).
Since the classical energy is linear on potential, this behaviour is kept
also for total energy as well.\\
\\
{\bf Table I}
\begin{tabular}{c||c|c|c}\hline
$V_0$ & $E^{f}$ & $E^{as}$ & $E^{ren}$ \\
\hline\hline
  0.001  &   3.7765 $10^{-6}$     &      -4.41055 $10^{-6}$    &        -6.3405 $10^{-7}$\\ 
  0.101  &   -0.00022342  &     0.000609497  &       0.000386072\\
  0.201  &   -0.00098331  &     0.00331244   &      0.00232913\\
  0.301  &   -0.0017050   &    0.00810441   &      0.00639944\\
  0.401  &   -0.0018532   &    0.0149854    &     0.0131322\\
  0.501  &   -0.0009434   &    0.0239555  &      0.0230121\\
  0.601  &   0.0014908   &    0.0350145   &    0.0365053\\
  0.701  &   0.0058864   &    0.0481626   &    0.0540491\\
  0.801  &   0.0126312   &    0.0633998   &    0.076031\\
\\
\hline \hline
\end{tabular}\\
\newpage 

\newpage
\newpage

\newpage

\newpage

\newpage

\subsection{Magnetic string}
\subsubsection{Finite and asymptotic vacuum energy}

{\bf 1. Finite energy}\\

\begin{figure}[h] 
\epsfxsize=9cm\epsffile{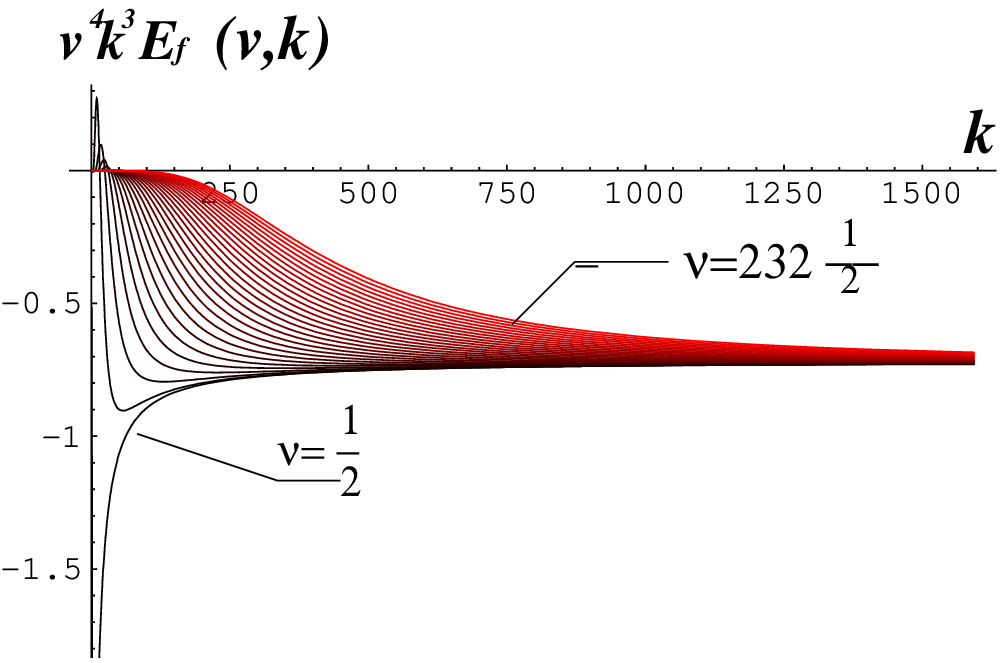}
\caption{The contribution $E_f(\nu,k)$ multiplied by $\nu^4 k^3$ of the
individual radial momenta for several orbital momenta to the finite
part of the ground state energy for $R_1=0.0001$, $R_2=1$, $\Omega=3$.}
 \label{fig2}
\end{figure}

The finite part of the ground state energy, $E_f$, is used in the form
as given by Eq. (\ref{Ef_spinor}). Let $E_f(\nu,k)$ denote the function to
be integrated and summed over in that expression. In Fig.\ref{fig2} it
is shown as function of $k$ for several values of the orbital momentum
$\nu$. In order the make the behaviour better visible it is multiplied
there by $\nu^4 k^3$. These functions are smooth for all values of
$k$, starting from some finite values at $k=0$.\\


\begin{figure}[h]   
\epsfxsize=9cm\epsffile{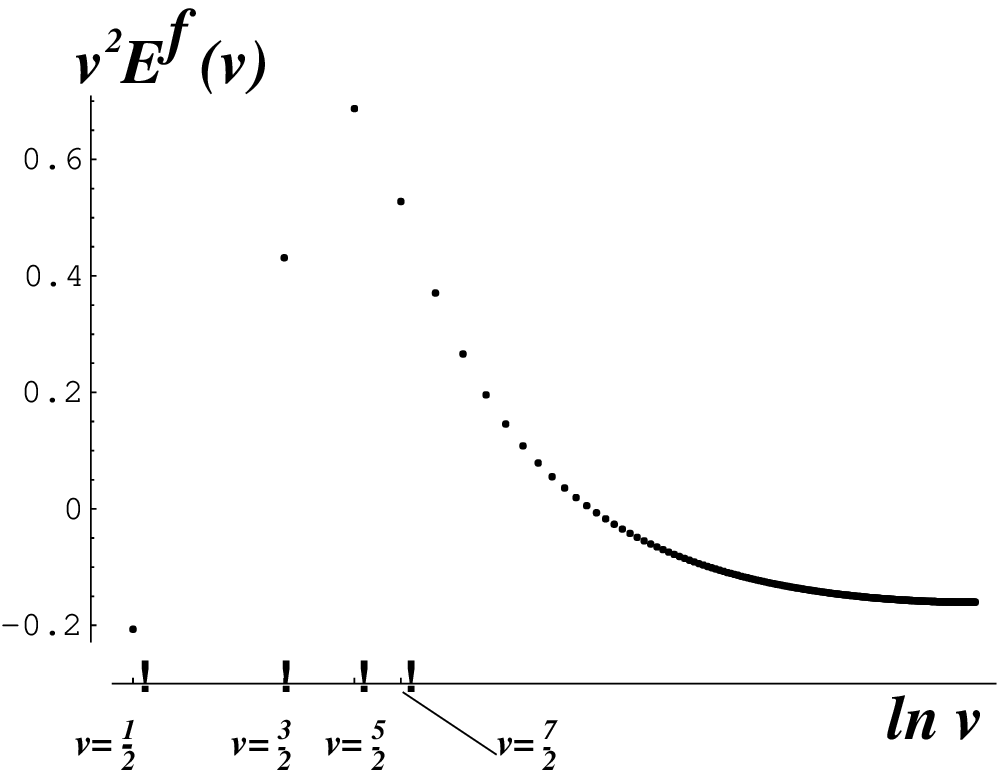}
\caption{The contribution $E_f(\nu)$ of the individual orbital momenta
to the finite part of the ground state energy multiplied by $\nu^2$
for $R_1=0.0001$,$R_2=1$ and $\Omega=3$.} \label{fig3}
\end{figure}

For large $k$, the
functions $\nu^4 k^2 E_f(\nu,k)$ shown here tend to a constant thus
the integral over $k$ is convergent.  All integrals have been
truncated at $k=1500$. The error caused by this is quite small and
does not change the results shown in the Table II. These integrals we
denote by $E_f(\nu)$. They are shown as function of $\nu$ in
Fig.\ref{fig3} in a logarithmic scale. Again, we multiplied by a power
of argument, here by $\nu^2$, in order to make the behaviour for large
$\nu$ visible. It is seen that $\nu^2 E_f(\nu)$ tends to a constant so
that the sum over $\nu$ is convergent. The sum is taken up to
$\nu=232.5$ and again the remainder is small.
The calculations have been performed for several values of the
parameters. The results are displayed in Table.II. The computations are
performed with an adapted arithmetical precision. In intermediate
steps compensations between sometimes very large quantities
appeared. The precision was adapted accordingly. For example, for
$R_1=0.99,\nu=250.5,k=1600$ as much as 1404 decimal positions have been
necessary to get at least 16 digits precision of the integrand $E_f(\nu,k)$.
This was a factor causing large computation time.

{\bf 2. Asymptotic energy}

\begin{figure}[h]
 \begin{picture}(60,100)(0,15) 
\epsfxsize=7cm\epsffile{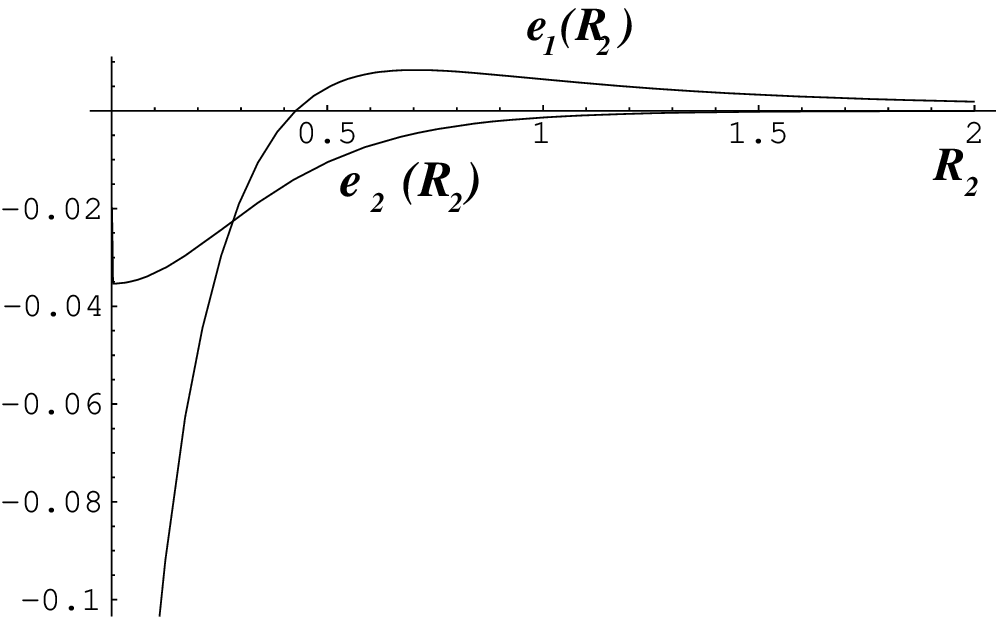}
\epsfxsize=7cm\epsffile{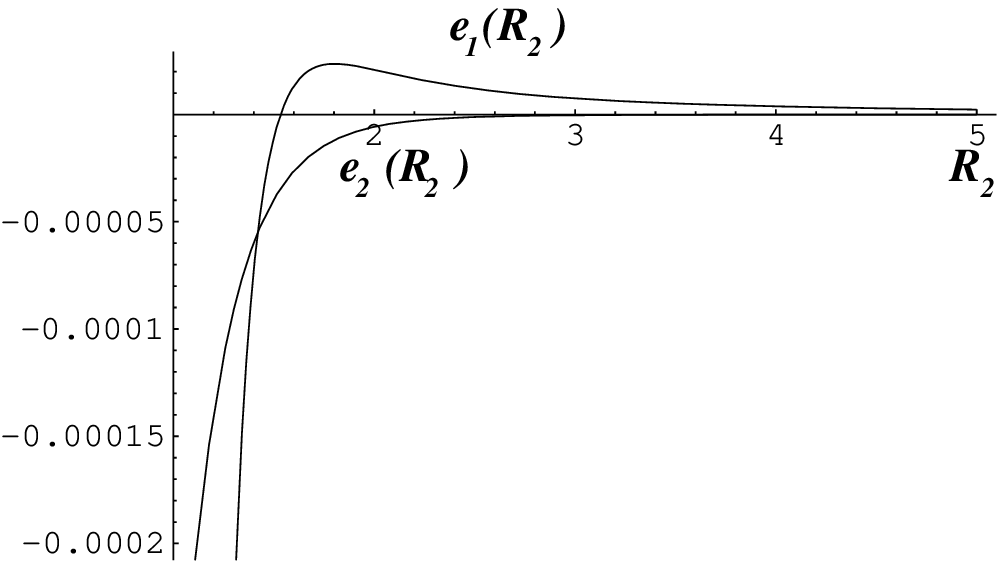} 
 \end{picture} 
\caption{The asymptotic part of energy as function of the outer
radius for $R_1=0.00001$ and $R_1=1$ resp.}\label{fig1}
\end{figure}

The asymptotic part of the ground state energy as given by
Eqs.(\ref{easspinor}) together with (\ref{g_i_of_x}) can be represented as sum
of a part proportional to the second power of the flux and one proportional
to the fourth power,
\begin{equation}
E^{as}= -\fr{4}{\pi}[e_1(R_1,R_2)\Omega^2+e_2(R_1,R_2)\Omega^4]
\label{eas_of_e1_e2}.
\end{equation}
The corresponding coefficients $ e_1(R_1,R_2)$ and $e_2(R_1,R_2)$ can
be calculated numerically without problems. They are shown as
functions of $R_2$ for fixed $R_1$ in Fig.\ref{fig1}  (multiplied by
$R_2^2$). 

\subsubsection{Parametrical dependence of vacuum energy}

{\bf Table II} The numerical evaluations for several values of $R_1$ and
$\Omega$\\ \\ $R_1$=0.0001\\
\begin{tabular}{c||c|c|c|c|c}\hline
$\Omega$ & $E^{f}$ & $E^{as}$ & $E^{ren}$ & $E^{class}$ & $E^{tot}$\\
\hline\hline
0.5& -0.0130152& 0.00152886& -0.0114863& 3.14159& 3.13011\\
1.& -0.052403& 0.00509591& -0.0473071& 12.5664& 12.5191\\
3.& -0.450093& -0.052012& -0.502105& 113.097& 112.595\\
6.& -0.91953& -1.52936& -2.44889& 452.389& 449.94\\
10.& 4.85954& -12.9483& -8.08871& 1256.64& 1248.55\\
15.& 46.3333& -67.3661& -21.0328& 2827.43& 2806.4\\
21.& 215.273& -261.526& -46.2534& 5541.77& 5495.52\\
30.& 987.62& -1095.29& -107.666& 11309.7& 11202.1\\
\hline \hline
\end{tabular}\\
\\
$R_1$=0.9\\
\begin{tabular}{c||c|c|c|c|c}\hline
$\Omega$ & $E^{f}$ & $E^{as}$ & $E^{ren}$ & $E^{class}$ & $E^{tot}$\\
\hline\hline
0.5 &-0.278679 &-0.00105651 &-0.279736 & 16.5347 & 16.255  \\ 
1. &-1.11431 &-0.00451479 &-1.11882 &66.1388 & 65.02\\
3. &-10.0633 &-0.0683525 &-10.1317 &595.249 &  585.117\\
6. &-40.4372 & -0.647622&-41.0848 &2381. &  2339.92\\
10. &-112.17 &-4.2629 &-116.433 &6613.88 &  6497.45\\
20. &-423.279 &-63.2506 &-486.53 &26455.5 & 25969.\\
40. &-1079.05 &-992.187 &-2071.24 &105822. & 103751.\\
\hline \hline
\end{tabular}\\
\\
$R_1$=0.95\\
\begin{tabular}{c||c|c|c|c|c}\hline
$\Omega$ & $E^{f}$ & $E^{as}$ & $E^{ren}$ & $E^{class}$ & $E^{tot}$\\
\hline\hline
0.5& -0.72166& -0.00212769& -0.723788& 32.2215& 31.4977\\
1.& -2.86633& -0.00877711& -2.87511& 128.886& 126.011\\
3.& -25.8293& -0.104564& -25.9339& 1159.97& 1134.04\\
10.& -287.834& -4.39364& -292.228& 12888.6& 12596.4\\
\hline \hline
\end{tabular}\\
\\
$R_1$=0.99\\
\begin{tabular}{c||c|c|c|c|c}\hline
$\Omega$ & $E^{f}$ & $E^{as}$ & $E^{ren}$ & $E^{class}$ & $E^{tot}$\\
\hline\hline
0.5& -5.59211& -0.0106385& -5.60275& 157.869& 152.266\\
1.& -23.2975& -0.0428038& -23.3403& 631.476& 608.136\\
3.& -202.057& -0.409235& -202.466& 5683.28& 5480.82\\
\hline \hline
\end{tabular}\\
\\
$R_1$=0.997\\
\begin{tabular}{c||c|c|c|c|c}\hline
$\Omega$ & $E^{f}$ & $E^{as}$ & $E^{ren}$ & $E^{class}$ & $E^{tot}$\\
\hline\hline
1.& -91.996& -0.142036& -92.138& 2097.54& 2005.4\\
\hline \hline
\end{tabular}\\
\\
$R_1$=0.999\\
\begin{tabular}{c||c|c|c|c|c}\hline
$\Omega$ & $E^{f}$ & $E^{as}$ & $E^{ren}$ & $E^{class}$ & $E^{tot}$\\
\hline\hline
1.& -311.182& -0.425555& -311.608& 6286.33& 5974.72\\
\hline \hline
\end{tabular}

\begin{figure}[h] 
\epsfxsize=10cm\epsffile{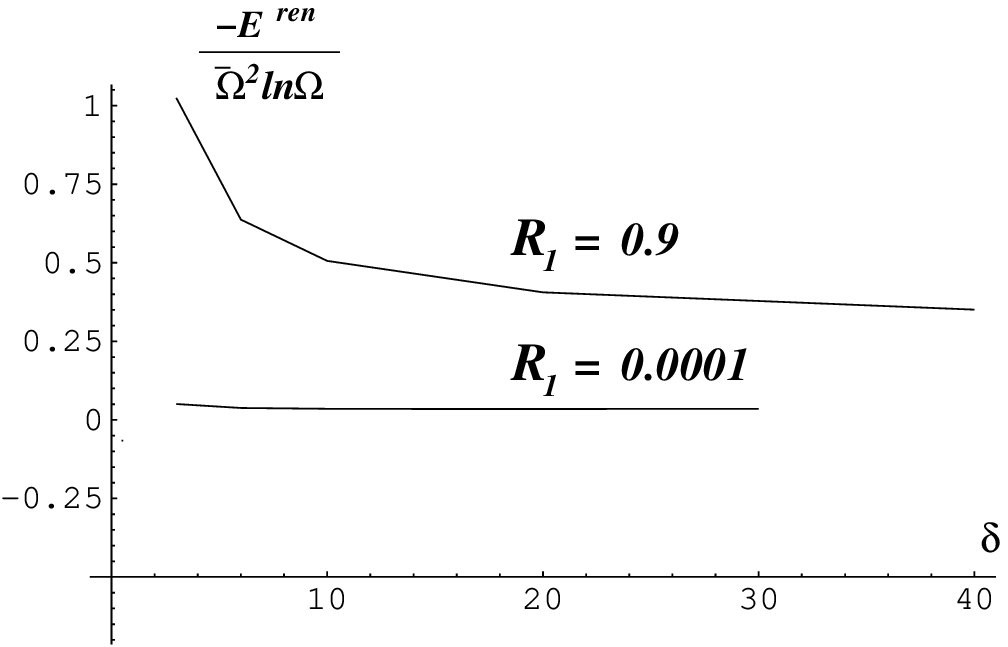}
\caption{The ground state energy divided  by $\Omega^2\ln\Omega$  as function of $\Omega$.}\label{fig4}
\end{figure}

The results
are displayed mainly in Table II. For small inner radius of the flux
the results are close to them of \cite{borkir2} where the same problem
for a flux tube with homogeneous magnetic field inside, which
corresponds to $R_1=0$ here, was considered. Especially, it is seen
that for large flux $\Omega$ there is a compensation of the
$\Omega^4$-contribution between the finite and the asymptotic parts
of the ground state energy leaving a behaviour proportional to
$\Omega^2\ln\Omega$ as shown in Fig.\ref{fig4}. Here the asymptotic
part gave an essential contribution. The ground state energy remains
negative, but numerically small. Only for very large flux it could
overturn the corresponding classical energy, but these values of the
flux are clearly unphysical.

For values of the inner radius close to the outer one, $R_1\to1$
(where we have put $R_2=1$) the picture changes. Here the vacuum
energy grows faster than the classical one. Generally, both diverge
proportional to $(1-R_1)^{-1}$, the classical energy is equal to
$E_{\rm class}=\Omega^2 2\pi (1-R_1)^{-1}$. The vacuum energy,
multiplied by $(1-R_1)$, is shown in Fig.\ref{fig5} in a logarithmic
scale. It is negative and growing a bit faster than the classical one
which would be a constant in this plot. Here the asymptotic part of
the ground state energy becomes increasingly unimportant (see Table II).

\subsubsection{Some remarks on the numerical difficulties}

The calculations have been performed for several values of the
parameters. The results for magnetic string are displayed in Table.1. The computations are
performed with an adapted arithmetical precision. In intermediate
steps compensations between sometimes very large quantities
appeared. The precision was adapted accordingly. For example, for
$R_1=0.99,\nu=250.5,k=1600$ as much as 1404 decimal positions have been
necessary to get at least 16 digits precision of the integrand $E_f(\nu,k)$.
This was a factor causing large computation time.
\newpage

\subsection{NO vortex}

\begin{figure}[t]  \label{fig5}
\epsfxsize=12cm
\epsffile{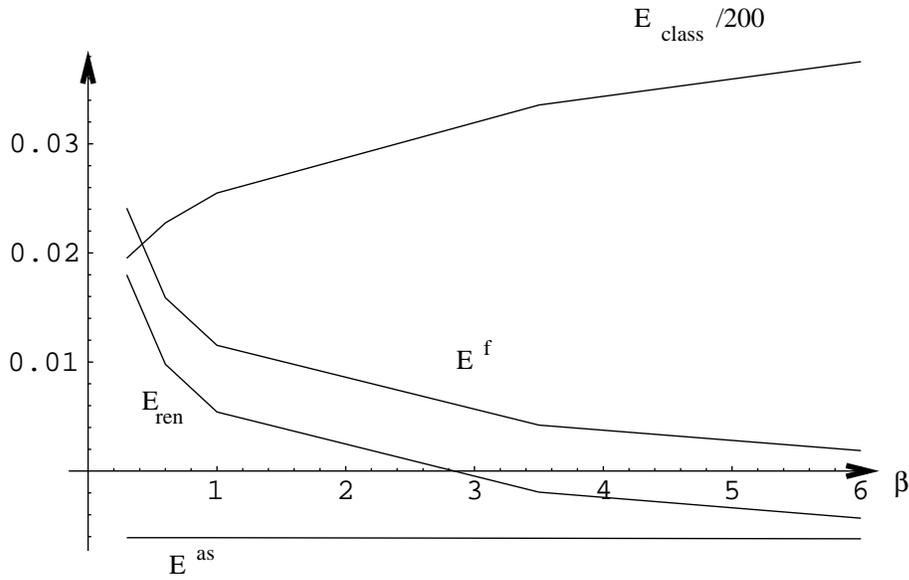}
\caption{The vacuum energy as a function of $\beta$ for the $q=0.5$,
  $f_e=1$, $\eta=1$. In order to represent all quantities within one
  plot the classical energy is divided by 200.}
\end{figure}
\begin{figure}[t]  \label{fig6}
\epsfxsize=12cm
\epsffile{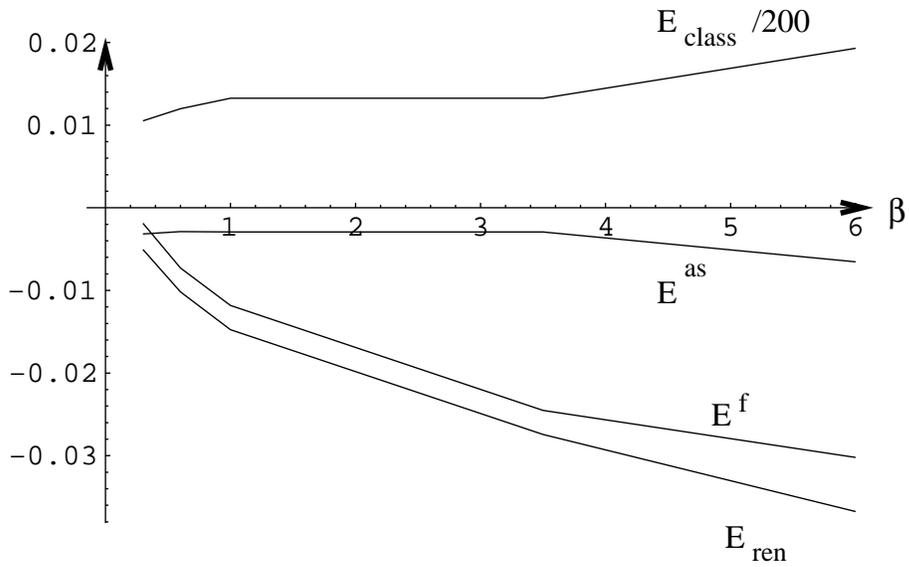}
\caption{The same as in figure \ref{fig5}  but for $q=2$.}
\end{figure}

We investigated numerically the vacuum energy for values of
$\beta$ ranging from $\beta=0.3$ to $\beta=6$ for some choices of
the  parameters $f_e$ and $\eta$. The results are represented in
Table III. In the Figs.  \ref{fig5} and  \ref{fig6} two examples for the dependence of
the individual parts of the vacuum energy on the parameter $\beta$
are displayed.

{\bf The constituent parts of the vacuum energy\\ Table III}
\begin{tabular}{llllllll}\hline
$q$ &$\beta$& $f_e$&$\eta$& $E_{\rm  class}$& $\E^{\rm as} $ &
  $\E^{\rm f} $ & $\E_{\rm ren}=\E^{\rm as} +\E^{\rm f} $ \\\hline
 0.5 & 0.3 & 1& 1 & 3.902 & -0.006102 & 0.02410 & 0.01800\\
0.5 & 0.6 & 1& 1 & 4.549 & -0.006102 & 0.01590 & 0.009795\\
0.5 & 1. & 1& 1 & 5.099 & -0.006104 & 0.01153 & 0.005421\\
0.5 & 3.5 & 1& 1 & 6.710 & -0.006146 & 0.004224 & -0.001922\\
0.5 & 6. & 1& 1 & 7.507 & -0.006208 & 0.001884 & -0.004325\\  \hline
2. & 0.3 & 1& 1 & 2.105 & -0.003170 & -0.001883 & -0.005053\\
2. & 0.6 & 1& 1 & 2.399 & -0.002891 & -0.007285 & -0.01018\\
2. & 1. & 1& 1 & 2.652 & -0.002924 & -0.01182 & -0.01475\\
2. & 3.5 & 1& 1 & 2.652 & -0.002924 & -0.02451 & -0.02743\\
2. & 6. & 1& 1 & 3.861 & -0.006539 & -0.03022 & -0.03676\\  \hline
0.5 & 0.3 & 0.1& 0.1 & 0.03902 & -0.00006102 &-9.589 $10^{-6}$ &-0.00007061\\
0.5 & 0.6 & 0.1& 0.1 & 0.04549 & -0.00006102 & -0.00001214 & -0.00007316\\
0.5 & 1. & 0.1& 0.1 & 0.05099 & -0.00006104 & -0.00001421 & -0.00007525\\
0.5 & 3.5 & 0.1& 0.1 & 0.06710 & -0.00006146 & -0.00001975 & -0.00008121\\
0.5 & 6. & 0.1& 0.1 & 0.07507 & -0.00006208 & -0.00002216 & -0.00008425\\  \hline
\end{tabular}

The vacuum energy $E^{ren}$ decreases for large values of the vortex
parameter $\beta$. The contribution of $E^{ren}$ to the total energy of
vortex can be positive as well as negative, dependent on the magnitude of the
electric charge $q$. Nevertheless it remains sufficiently smaller
compared to the classical energy $E_{class}$.


\section{ Conclusions and Discussions }

\ \ \ In the preceding sections the \gse for a spinor quantum field in the
background of a rectangular shaped magnetic flux  and of the Nielsen-Olesen
vortex, and for a scalar field in the rectangular scalar background has been calculated.

The used \rg procedure is a kind of zeta functional \rg. The sum over
spectrum of hamiltonian is performed in terms of the \jf for the
corresponding scattering problem. 

The \jf for each model has been written down explicitly, also its asymptotic
 part as well. The numerical calculation for several models (spinor field) required work with high arithmetic precision.
The obtained results presume some discussion\\

\subsection{Scalar field in scalar background.}

This model is the simplest construction which allows to carry out
analytic calculations as well as numeric evaluations relative easy.

The \rg procedure with the \rn of the \gse and its interpretation
is also quite transparent. Nevertheless this model looks up a number
of interesting results concerning the behaviour of the vacuum energy.
 The vacuum energy can provide positive as well as negative contributions
 to the total energy. The sign of it depends on the magnitude of the background
 potential $V$. Moreover, there is a thin range of values $V$, whereat the
 \gse shows the critical behaviour for its dependence on the outer radius
 $R_2$ possessing one maximum and one minimum. It gives reason to suppose an
 existence of stable states of the scalar string, generated by the vacuum
 corrections. \\

\subsection{Spinor field in magnetic background.}

 The spinor vacuum energy for the constructed magnetic string is investigated
 for several values of potential $\Omega$ and inner radius $R_1$. It shows no
 extrema, so the magnetic string constructing in such a way must be unstable.
 The known asymptotic behaviour of the one one-loop vacuum energy for large
 magnitude of background potential in the form $\Omega^2 \ln \Omega$ is also
 confirmed fro this model.  
 
 The question
whether the vacuum energy for sufficiently small thickness of magnetic
flux may become larger than the classical one cannot be answered by the numerical
results obtained. The problem is that the computations become very
time consuming because of the increasing precision which is
required. Also, one has to take higher $k$ and $\nu$ into account.
The weakening of the growth for $R_1=0.997$ and $R_1=0.999$ (while $R_2$ is
fixed to be equal 1) seen in
Fig.\ref{fig5} may be caused just by this reason. Here one has to note
that the integrand is for large $k$ and $\nu$ always negative (see
Fig.\ref{fig2}) so that dropping some part (as we did within the
numerical procedure) diminishes the result. So, as a result, we cannot
exclude from the given calculation that the vacuum energy grows for a
strong background field faster than the classical energy.

\begin{figure}[h] 
\epsfxsize=12cm\epsffile{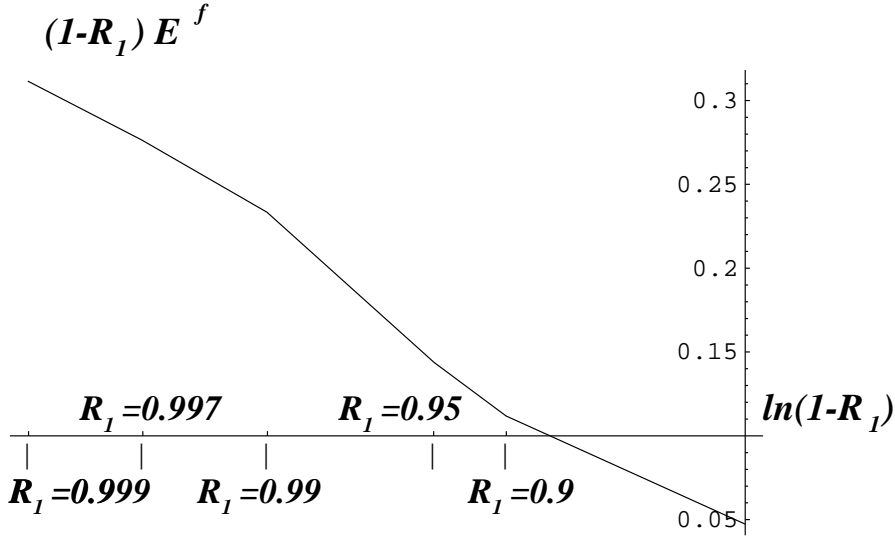}\ \ 
\caption{The vacuum energy multiplied by $(1-R_1)$ in a logarithmic
scale for $\Omega=1$.}\label{MAGfig5}
\end{figure}

Further work is necessary to better understand these questions. An
improvement of the numerical procedure is certainly desirable. It
could go along two lines. First, in the calculation of the Jost
function the compensation of large exponentials should be avoided by
taking them into account analytically. Second, in the compensation
between the logarithm of the Jost function and its asymptotic
expansion in the integrand of $E_f$ in Eq.(\ref{Ef_spinor}) higher
orders of the asymptotic expansion could be used. However, for this
reason one would have to continue the procedure invented in
\cite{borkir2} for this expansion using the Lippmann-Schwinger equation
or to find some equivalent procedure.
\\

\subsection{Spinor field in NO vortex background.}

During the analytical \rg of the \gse in the model with the special
interaction term is occurred. Namely the standard zeta functional \rg determines the counter terms from the
first \hkc (up to $a_2$). It turned out that to
renormalize the vacuum energy one has to introduce a counter term which
is not present in the initial action and represents a non polynomial
interaction. This is caused by the non polynomial interaction present
in the model itself Sec.(\ref{hamiltonians}).

In the considered case the background is given purely numerically.
This is the crucial technical
difference to the previous two models where it allowed an analytically
calculation.
However, it has been demonstrated that the similar computational scheme used here is able to
work successfully with such backgrounds. This is a step forward to physically
really interesting problems. In the considered model the stability of
the background is given by topological arguments and for a realistic
choice of the parameters the quantum contribution is small. This
smallness has two sources.  The one is the smallness of the coupling
constants which appears in front of the quantum contribution relative
to the classical one. The second one is a purely numerical smallness
of about two orders of magnitude. It is present even if the parameters
and couplings are all of order one, see for instance the Fig.\ref{fig5}.

As seen from the table and from the Figs.\ref{fig5} and \ref{fig6}, for the
dependence of the vacuum energy on the various parameters no
general rule can be seen so far. Even the sign of the vacuum
energy changes. The same applies to the relative weight of the
individual contributions.  Sometimes the asymptotic part is
dominating, in other cases, however, the 'finite' part is larger.
In general, they are of the same order. From this one can conclude
only that in the given background there is no small parameter
which could allow for some approximation scheme. For instance,
if the asymptotic part is dominating one could hope to get a good
approximation by including higher orders of the uniform asymptotic
expansion of the Jost function into the asymptotic part of the
vacuum energy and neglect the 'finite' part of it which is the
numerically much harder part of the problem. The examples in
\cite{bordagcolor} have been a kind of suggesting this way in
contrast to the example in the present paper.

For the considered model of a spinor in the background of the NO
vortex, the vacuum energy due to its smallness has only a very negligible 
influence on the dynamics of the background. Hence in the considered
case the vacuum energy is of limited physical importance. However, its
calculation gave new insights into the structure of such calculations
and demonstrated the power of the methods used.  A next step could be
to apply them to the $Z$ and electroweak strings where the stability
is not guaranteed by topological arguments and where the stability
issue is not finally settled with respect to the fermion contributions
\cite{acc_vach, Groves:1999ks}.\\

\subsection{Summary}

Basically, the contribution of the quantum ground state has an influence on
the macroscopic behaviour of the field system. The degree of this influence
depends on the nature of the interacting background and fluctuations as well
as on the kind of interaction and coupling parameters.
 The vacuum corrections can provide both negative and positive contributions
 to the total energy.
 
It has been shown for scalar and magnetic strings with a simple rectangular shape and for the NO
vortex.

In all three cases the general dependence of the vacuum contributions on the
background properties is not to be formulated certainly. 

For the scalar string, even the alternation of the stability properties is possible.
 The large-potential asymptotics known for homogeneous fields is
 confirmed for scalar and spinor strings.

 In the technical aspect, the known methods using the zeta functional
 \rg and \hke have been applied successfully to more complicated
 shapes of backgrounds as before, and in particular to the shape being given
 pure numerically (NO-vortex).

 The obtained results suggest the applicability of the developed methods and
 technics to calculations of the vacuum energy in more complicated
 background configurations, such as monopoles, electroweak strings, skyrmions,
 sphalerons etc. 

 These future investigations would lead to better understanding of the
 phenomena of vacuum energy and its relation to the geometry of
 classical backgrounds.

\section*{Acknowledgements}

\ \ \ I thank first of all my supervisor M. Bordag for advice and fruitful
collaboration during my research, completing of the present thesis and
publication of the results.

I thank D.V.~Vassilevich for valuable discussions,
helpful remarks and numerous corrections.

I thank P.M.~Lavrov and A.V.~Yurov for corrections, and A.~Ivanov
for helpful discussions.

I thank the Graduate college {\it Quantenfeldtheorie} at the University
of Leipzig and its speaker B. Geyer for support and friendly environment.

I am especially grateful to the {\it Naturwissenschaftlich Theoretisches
  Zentrum (NTZ)} at the University of Leipzig and its director W. Janke for the
necessary support during the final part of my work.


 %
\section{Appendix}
\subsection{The uniform asymptotic expansions for logarithm of \jf}
\subsubsection{The uniform asymptotics of modified Bessel \fs\\ K and I}
\label{besselasym}
The uniform asymptotic expansion of modified Bessel \fs $K_{\nu\pm\half}(kr),
I_{\nu\pm\half}(kr)$ in the combined limit $\nu,k\ra\infty, \ k/\nu\e z=const
$, can be obtained as series of the variable $t=1/\sqrt{1+z^2 r^2}$ \cite{abrsteg}:
\bea
K_{\nu+\half}(\nu z\ r) \sim  \sqrt{\fr{\pi}{2\nu}} \exp\{ \sum\limits_{i=-1}^\infty \fr{S_i^-[\half,t]}{\nu^i}\}   \e  \sqrt{\fr{\pi}{2\nu}}  e^{-\nu\eta}\sqrt{t} \left(
\fr{1+t}{1-t}\right)^{1/4}  \exp\{ \sum\limits_{i=1}^\infty \fr{S_i^-[\al,t]}{\nu^i}
\}\nn\\
K_{\nu-\half}(\nu z\ r)  \sim  \sqrt{\fr{\pi}{2\nu}} \exp\{ \sum\limits_{i=-1}^\infty \fr{S_i^-[-\half,t]}{\nu^i}\}  \e  \sqrt{\fr{\pi}{2\nu}}  e^{-\nu\eta}\sqrt{t} \left(
\fr{1-t}{1+t}\right)^{1/4}  \exp\{ \sum\limits_{i=1}^\infty \fr{S_i^-[\al,t]}{\nu^i}
\}\nn\\
I_{\nu+\half}(\nu z\ r)  \sim \sqrt{\fr{1}{2\pi\nu}}  \exp\{ \sum\limits_{i=-1}^\infty \fr{S_i^-[\half,t]}{\nu^i}\}  \e  \sqrt{\fr{1}{2\pi\nu}}  e^{\nu\eta}\sqrt{t} \left(
\fr{1-t}{1+t}\right)^{1/4}  \exp\{ \sum\limits_{i=1}^\infty \fr{S_i^+[\al,t]}{\nu^i}
\}\nn\\
I_{\nu-\half}(\nu z\ r) \sim \sqrt{\fr{1}{2\pi\nu}}  \exp\{ \sum\limits_{i=-1}^\infty \fr{S_i^-[-\half,t]}{\nu^i}\}   \e  \sqrt{\fr{1}{2\pi\nu}}  e^{\nu\eta}\sqrt{t} \left(
\fr{1+t}{1-t}\right)^{1/4}  \exp\{ \sum\limits_{i=1}^\infty \fr{S_i^+[\al,t]}{\nu^i}
\}\nn
\label{KI_asymp}
\eea
where $S_i^\pm[\al,t]$ are:\\
for $K_{\nu+\al}$, $\al=\pm\half$ :
\bea
&& S_{-1}^-[t,\al]= -\fr{1}{t}-\half \ln \fr{1-t}{1+t}\nn\\
&& S_{0}^-[t,\al]= \fr{\ln t}{2}+\fr{\al}{2} \ln \fr{1+t}{1-t}\nn\\
&& S_{1}^-[t,\al]=\fr{t}{24}( -3+12 \al^2 -12 \al t+ 5t^2)\nn\\
&& S_{2}^-[t,\al]=\fr{1}{48} t^2 \left[ -8\al^3 t + 12\al^2(-1+2t^2)+\al(26t -30t^3) +3(1-6t^2+5t^4) \right]\nn\\
&& S_{3}^-[t,\al]= \fr{1}{5760}\left\{ t^3\left[ -375 + 4779 t^2 - 9945t^4+5525 t^6 +
240 \al^4 (3t^2-1) - \right.\right.\nn\\
&&\left.\left. 960 \al^3 t(4t^2-3)- 720 \al t(7-22 t^2+15t^4) -120 \al^2(13-84 t^2+ 75 t^4)\right]  \right\}\nn\\
&& S_{4}^-[t,\al]=\nn\\&& \fr{1}{1920} \left\{ t^4\left[-48\al^5 t(5t^2-3)+240
\al^4(1-8t^2+8t^4)-40 \al^3 t (69 -240 t^2 + 175 t^4 )+ \right.\right.\nn\\
&&\left.\left. 120\al^2 (-7+ 94t^2 -207t^4 +120 t^6)+ \al( 3561 t- 21225t^3-342255 t^5 -16575 t^7)\right.\right.\nn\\
&&\left.\left. +15(13-284 t^2+1062 t^4- 1356t^6+ 565 t^8)\right] \right\}
\label{S_asymp_K}
\eea

for $I_{\nu+\al}$, $\al=\pm\half$ :

\bea
&& S_{-1}^+[t,\al]= \fr{1}{t}+\half \ln \fr{1-t}{1+t}\nn\\
&& S_{0}^+[t,\al]= \fr{\ln t}{2}-\fr{\al}{2} \ln \fr{1+t}{1-t}\nn\\
&& S_{1}^+[t,\al]= -\fr{t}{24}( -3+12 \al^2 +12 \al t+ 5t^2)\nn\\
&& S_{2}^+[t,\al]=\fr{1}{48} t^2 \left[ 8\al^3 t + 12\al^2(-1+2t^2)+\al(-26t +30t^3) +3(1-6t^2+5t^4) \right]\nn\\
&& S_{3}^+[t,\al]=-\fr{1}{5760}\left\{ t^3\left[ -375 + 4779 t^2 - 9945
t^4+5525 t^6 +240 \al^4 (3t^2-1) +\right.\right. \nn\\ &&\left.\left. 
960 \al^3 t(4t^2-3)+ 720 \al t(7-22 t^2+15t^4) +120 \al^2(13-84 t^2+ 75 t^4)\right]  \right\}\nn\\
&& S_{4}^+[t,\al]=\nn\\&& \fr{1}{1920} \left\{ t^4\left[48\al^5 t(5t^2-3)+240 \al^4
(1-8t^2+8t^4)+ 40 \al^3 t (69 -240 t^2 + 175 t^4 )+\right.\right.\nn\\&&
\left.\left. 120\al^2 (-7+ 94t^2 -207t^4 +120 t^6)- \al( 3561 t- 21225
t^3-342255 t^5 -16575 t^7)\right.\right.\nn\\&&\left.\left.+15(13-284 t^2+1062 t^4- 1356t^6+ 565 t^8)
\right]  \right\},
\label{S_asymp_I}
\eea
what leads in particular to
\be
\eta\e\eta(t)=\fr{1}{t}+\ln\fr{1-t}{1+t}
\ee
in (\ref{KI_asymp}, \ref{S_asymp_K})
\\
\subsubsection{The saddle point expansion}
\label{saddlepoint}

The evaluation of integrals of type
\be
\int\limits_0^r e^{\nu\eta(r')}\phi(r')dr',
\ee
entering in the expressions of $\ln f_\nu^{(n)}$ Sec.\ref{lippsw} (\ref{lnf_of_ik}) has been done by using the
saddle point approach (See. for example \cite{ryder}). The derivation is
described in details in \cite{borkir1}. We use the results given in
App.\cite{borkir2} as a starting point:
\be \int\limits_0^r e^{\nu\eta(r')}\phi(r')dr'=
e^{\nu\eta(r)}\sum\limits_{k=1}^\infty h_{k-1}\nu^{-k}\label{saddle_exp}\ee
with the first several coefficients of the expansion
\bea
&& h_0=\fr{\phi(r)}{\eta'(r)}, \ \ \ \ \ \ \ \ h_1=\fr{\phi(r)\eta''(r)}{\eta'(r)^3} -
\fr{\phi'(r)}{\eta'(r)^2}\nn\\
&& h_2=\fr{\phi''(r)}{\eta'(r)^3}- \fr{3\phi'(r)\eta''(r)}{\eta'(r)^4}+ \fr{3\phi(r)\eta''(r)^2}{\eta'(r)^5}- \fr{\phi(r)\eta'''(r)}{\eta'(r)^4}\label{saddle_coef}
\eea
\\

\subsubsection{Scalar field with scalar background} \label{JOSTASscalar}

 We proceed from the expansion of the logarithmic \jf Sec.\ref{lippsw}
 (\ref{lnf_of_ik}) obtained by iterations of the \lseq (\ref{lipp_scw_of_k_i}). 

 The terms $\ln f_l^{(1)}(ik)$ are obtained as described in Sec.\ref{lippsw}
by substitution of (\ref{second_iter}) in (\ref{jost_func_ik_of_LS}):

{\small
\bea
&& \ln f_l^{(1)}(ik)=\int\limits_0^\infty r dr V(r)K_l(kr)I_l(kr)\nn\\
&& \ln f_l^{(2)}(ik)=\int\limits_0^\infty r dr V(r)K_l(kr)\int_0^r r'dr' V(r')\left[
 I_l(kr) K_l(kr')-I_l(kr') K_l(kr)\right] I_l(kr')-\nn\\ &&\half \int\limits_0^\infty r dr V(r)K_l(kr)I_l(kr)\int\limits_0^\infty r' dr' V(r')K_l(kr')I_l(kr')\nn\\
&& \ln f_l^{(3)}(ik)=\int\limits_0^\infty r dr V(r)K_l(kr)\int_0^r r'dr'
V(r')\left[ I_l(kr) K_l(kr')-I_l(kr') K_l(kr)\right]\times\nn\\&&\times  \left(
\int\limits_0^{r'} r''dr''V(r'')I_l(kr')K_l(kr'')I_l(kr'')-
\int\limits_0^{r'} r''dr''V(r'') K_l(kr')I_l(kr'')^2
 \right)\nn\\
&&-\half\left( \int\limits_0^\infty r dr V(r)K_l(kr)I_l(kr)\int\limits_0^\infty
  r' dr' V(r')K_l(kr')\int\limits_0^{r'}r''dr'' V(r'')[I_l(kr') K_l(kr'')-\rl
\hspace*{5cm}  I_l(kr'')  K_l(kr') ]I_l(kr'')+\right.\nn\\&&\left.
\int\limits_0^\infty
  r dr V(r)K_l(kr)\int\limits_0^{r} V(r')[I_l(kr) K_l(kr')-I_l(kr')
  K_l(kr) ]I_l(kr') \times\rl \hspace*{4cm} \int\limits_0^\infty r'' dr'' V(r'')K_l(kr'')I_l(kr'')\right)+\\
  &&\fr{1}{3} \int\limits_0^\infty r dr V(r)K_l(kr)I_l(kr)
  \int\limits_0^\infty r' dr' V(r')K_l(kr')I_l(kr')
  \int\limits_0^\infty r'' dr'' V(r'')K_l(kr'')I_l(kr'')\nn
  \label{lnf1-3_roh}
 \eea
}

 The expression $\ln f_l^{(1)}(ik)$ has already a usable form and $\ln f_l^{(2-3)}(ik)$ can be simplified if we use its symmetry properties in respect of the change
 $r\leftrightarrow r'$.

 The procedure is illustrated by the $\ln f_l^{(2)}(ik)$.
We start with the expression (\ref{lnf1-3_roh})

\bea
&& \ln f_l^{(2)}(ik)=\nn\\
&& \int\limits_0^\infty r dr V(r)K_l(kr) I_l(kr)   \int_0^r r'dr' V(r')
 K_l(kr') I_l(kr')-\nn\\
&& \int\limits_0^\infty r dr V(r)K_l(kr) K_l(kr)   \int_0^r r'dr' V(r')
 I_l(kr') I_l(kr')\nn\\
 &&-\half \int\limits_0^\infty r dr V(r)K_l(kr)I_l(kr)\int\limits_0^\infty r' dr' V(r')K_l(kr')I_l(kr').\nn
\eea

Then we represent the last term as
\bea
&&\half \int\limits_0^\infty r dr V(r)K_l(kr)I_l(kr)\int\limits_0^\infty r' dr'
V(r')K_l(kr')I_l(kr')=\nn\\
&&\half \int\limits_0^\infty r dr
V(r)K_l(kr)I_l(kr)\left[\int\limits_0^r+\int\limits_r^\infty   \right] r' dr'
V(r')K_l(kr')I_l(kr'),\nn
\eea
and take into account the rule of change the integration order:
\be \int\limits_0^\infty f(r) dr \int\limits_0^r g(r')dr'\e 
\int\limits_0^\infty g(r) dr \int\limits_r^\infty f(r')dr', \ee
which allows to see, that the last term is equal exactly to the first one but
with the sign minus in front of them. Thus these terms cancel each other, and
we have:
\bea
 \ln f_l^{(2)}(ik)=-\int\limits_0^\infty r dr V(r)K_l(kr) K_l(kr)   \int_0^r r'dr' V(r')
 I_l(kr') I_l(kr').
\eea

We do not discuss here the simplification of $\ln f_l^{(3)}(ik)$, because it is not
needed for minimal subtraction, as pointed out in Sec. \ref{lippsw}.

Since the contribution of $l=0$ in the sum over $l$ in (\ref{scalarEF},
\ref{scalarEF0}, \ref{scalarEFl}) Sec.\ref{efineas}, is separated,
the $\ln f_0(ik)$ is to perform additionaly. Instead of the uniform
asymptotic expansion (because $l\e 0\e const$) the usual asymptotics at $k\ra
\infty$ is used.

 We use entirely the corresponding asymptotics of modified Bessel \fs
\bea
I_0(z)\sim  \fr{e^z}{\sqrt{2\pi z}}\left\{1+ \fr{1}{8z}+\fr{9}{2(8z)^2}+\dots\right\}\nn\\
K_0(z) \sim  e^{-z}\sqrt{ \fr{\pi}{2 z} }\left\{1-\fr{1}{8z}+\fr{9}{2(8z)^2}+\dots \right\},
 \label{K0_I0_asymp}
\eea
which are to substitute into (\ref{jost_func_of_LS}-\ref{lnf_of_ik},Sec.\ref{lippsw}) with the argument $z$ having
the values $kR_1, kR_2, qR_1, qR_2 $.

Recall, that $q=\sqrt{k^2+V_0}$, that provides in the considered limit
$k\ra\infty$
\be q\sim k+\fr{V_0}{2k}-\fr{V_0^2}{4k^3}+O(k^{-4}). \ee

Substituted into (\ref{scalar_jost_exact}) it provides the leading order for
$\ln f_0^{as}(ik)$:
\be \ln f_0^{as}(ik) \sim -\fr{V_0}{2k}(R_1-R_2).\label{lnfas0}\ee

\subsubsection{Spinor field with magnetic background}

( To \ref{a52},  {\bf Higher orders of the \jf })

\label{high_ord}

The asymptotic of the logarithmic Jost function can be obtained in the form:
$\ln f_{\nu}(ik)= \sum\limits_{n}h_n(t_1,t_2) $,(see \ref{unif_as_f})
with:
\\
\begin{eqnarray}
h_1&=&\frac{4}{3}\lambda^2[
{\Mvariable{R1}}^4 \Mvariable{t1}( 2 + \Mvariable{t1}) {( 1 + \Mvariable{t2}) }^2
-3{\Mvariable{R1}}^2{\Mvariable{R2}}^2 {(1 +\Mvariable{t1}) }^2 \Mvariable{t2} ( 1 + \Mvariable{t2})+
{\Mvariable{R2}}^4 { ( 1 + \Mvariable{t1}) }^2\Mvariable{t2}( 1 +
  2\Mvariable{t2})
  ]\nn\\
\nn\\
h_2&=&0\nn\\
\nn\\
h_3&=&-\frac{1}{6}{\lambda}^2 [ 2{\Mvariable{R1}}^4{\Mvariable{t1}}^3 {( 1 + \Mvariable{t2}) }^2 +{\Mvariable{R1}}^2 {\Mvariable{R2}}^2
{( 1 + \Mvariable{t1}) }^2 {\Mvariable{t2}}^3{( 1 + \Mvariable{t2}) }^2(3 {\Mvariable{t2}}^2 - 4)-\nn\\&&
{\Mvariable{R2}}^4 {( 1 + \Mvariable{t1}) }^2{\Mvariable{t2}}^3
(3{\Mvariable{t2}}^4 + 6 {\Mvariable{t2}}^3 -{\Mvariable{t2}}^2 -
8\Mvariable{t2} - 2 ) ]+\nn\\
&& \frac{2}{15} {\lambda}^4 [ 4 {\Mvariable{R1}}^8
{\Mvariable{t1}}^3( 4 + \Mvariable{t1}){( 1 + \Mvariable{t2}) }^4-
5 {\Mvariable{R1}}^6 {\Mvariable{R2}}^2{( 1 + \Mvariable{t1}) }^4
{\Mvariable{t2}}^3{( 1 + \Mvariable{t2}) }^4 +\nn\\
&&15 {\Mvariable{R1}}^4 {\Mvariable{R2}}^4 {( 1 + \Mvariable{t1}) }^4 {\Mvariable{t2}}^3 {( 1 +
  \Mvariable{t2} ) }^2 ( {\Mvariable{t2}}^2 + 2 \Mvariable{t2} - 1)-\nn\\&&
5 {\Mvariable{R1}}^2 {\Mvariable{R2}}^6 { ( 1 + \Mvariable{t1}) }^4
{\Mvariable{t2}}^3( 3 {\Mvariable{t2}}^4 +12{\Mvariable{t2}}^3 + 6
{\Mvariable{t2}}^2 - 4 \Mvariable{t2} - 1 )+\nn\\&&
{\Mvariable{R2}}^8 {( 1 + \Mvariable{t1} ) }^4{\Mvariable{t2}}^3 (5
{\Mvariable{t2}}^4 +20 {\Mvariable{t2}}^3 - 4\Mvariable{t2} -1 ) ]\nn\\
\nn\\
h_4&=&\frac{\delta^2 [ {\Mvariable{R1}}^4 {\Mvariable{t1}}^4 ( 1 - {\Mvariable{t1}}^2)  + 
         {\Mvariable{R2}}^4 {\Mvariable{t2}}^4 ( 1 - {\Mvariable{t2}}^2 ) ]}
          {4 {( \Mvariable{R1}^2 - \Mvariable{R2}^2) }^2 }\nn\\
\nn\\
\lambda&=&  \frac{\delta}{( \Mvariable{R1}^2 - \Mvariable{R2}^2 )
     ( 1 + \Mvariable{t1}) ( 1 + \Mvariable{t2})}
 \label{unif_as_h}
\end{eqnarray}   
\\

\subsubsection{Spinor field with NO vortex background}
\label{jostUA_NO}
{\bf Uniform asymptotics for \jf in abelian Higgs Background}
We start with the expansion  (\ref{lnf_of_ik}) obtained in \cite{borkir1} in
the form

\bea 
\ln f^{(1)}_{\nu}(k)&=&\left(\fr{\pi}{2i}\right)\int\limits_{0}^{\infty}d r\
r\  \Phi^{0\ T}_H(r)\hat{V}(r)\Phi^0_{J}(r),\label{lne}\\
\ln f^{(2)}_{\nu}(k)&=&-\left(\fr{\pi}{2i}\right)^{2}
\int\limits_{0}^{\infty}d r\ r\ \int\limits_{0}^{r}d r'\ r'\ 
  \Phi^{0\ T}_H(r)\hat{V}(r)\Phi^0_H(r)\  \nn\\
&&\hspace{1cm}\times  \Phi^{0\ T}_{J     }(r')\hat{V}(r')\Phi^0_{J    }(r'),
\label{lnz}\\ 
\ln f^{(3)}_{\nu}(k)&=&2\left(\fr{\pi}{2i}\right)^{3}
\int\limits_{0}^{\infty}d r\ r\ \int\limits_{0}^{r}d r'\ r'\ 
\int\limits_{0}^{r'}d r''\ r''\ 
  \Phi^{0\ T}_H(r ) \hat{V}(r  )\Phi^0_H(r)\   \nn \\
&& \hspace{1cm}\times \Phi^{0\ T}_H(r' )\hat{V}(r' )\Phi^0_{J      }(r' )
  \Phi^{0\ T}_{      J}(r'')\hat{V}(r'')\Phi^0_{J      }(r''),
\label{lnd}\\ 
\ln f^{(4)}_{\nu}(k)&=&-\left(\fr{\pi}{2i}\right)^{4}
\int\limits_{0}^{\infty}d r\ r\ \int\limits_{0}^{r}d r'\ r'\
\int\limits_{0}^{r'}d r''\ r''\ \int\limits_{0}^{r''}d r'''\ r'''\  \nn\\ 
&&\hspace{1cm}\times
\left[ ( 4 
  \Phi^{0\ T}_H(r ) \hat{V}(r  )\Phi^0_H(r)\ 
  \Phi^{0\ T}_H(r' )\hat{V}(r' )\Phi^0_{J      }(r' )     \right.   \nn    \\
&&\hspace{1.4cm}
  \Phi^{0\ T}_H(r'' )\hat{V}(r'' )\Phi^0_{J      }(r'' )
  \Phi^{0\ T}_{      J}(r''')\hat{V}(r''')\Phi^0_{J      }(r''') \nn\\ 
&&\hspace{1.4cm}
+2
  \Phi^{0\ T}_H(r ) \hat{V}(r  )\Phi^0_H(r)\ 
  \Phi^{0\ T}_H(r' )\hat{V}(r' )\Phi^0_H(r' )       \nn       \\
&&\left.\hspace{1.4cm}
  \Phi^{0\ T}_{J      }(r'' )\hat{V}(r'' )\Phi^0_{J      }(r'' )
  \Phi^{0\ T}_{      J}(r''')\hat{V}(r''')\Phi^0_{J      }(r''')  \right],
\label{lnv}
\eea

As a matter of fact we have to treat three kinds of products, entering in
the integrals \ref{lnf_series}, namely
\be
 \Phi^{0\ T}_H(x) \hat{V}(x) \Phi^0_H(x)\nn,\ \ 
 \Phi^{0\ T}_J(x) \hat{V}(x) \Phi^0_J(x)\nn,\ \ 
 \Phi^{0\ T}_H(x) \hat{V}(x) \Phi^0_J(x),\nn
\label{products_HVH}
 \ee
and we denote them as $\G_{\{ H V H \}}(x),\ \G_{\{ J V J\}}(x)$ and $\G_{\{H
  V J\}}(x)$ for the first, second and third product respectively.

Performing the products  (\ref{products_HVH}) explicitly we obtain the
expressions as follows:
\bea
&&\G_{\{H V H \}} = \mu(r) ( C^- H_{-} H_{-}- C^+ H_{+} H_{+})-\fr{2 v(r)}{r}\sqrt{C^+C^-}(H_{+}
H_{-} )\\
&&\G_{\{J V J \}} = \mu(r) ( C^-  J_{-} J_{-}-C^+J_{+} J_{+})-\fr{2 v(r)}{r}\sqrt{C^+C^-}(J_{+}
J_{-} )\\
&&\G_{\{H V J\}} = \mu(r) ( C^-  H_{-} J_{-}-C^+ H_{+} J_{+})-\fr{v(r)}{r}\sqrt{C^+C^-}(H_{+}
J_{-}+ H_{-} J_{+})\nn\\
\label{products}
\eea

The notations has been used here are

\bea
&&C^\pm \e p_0\pm m_0\nn\\
&&J_{\pm}\e J_{\nu\pm\half}(q_0 r)\nn\\
&&H_{\pm}\e H^{(1)}_{\nu\pm\half}(q_0 r)\nn\\
&&\sqrt{C^+C^-}\e \sqrt{p_0^2- m_0^2 }\nn\\
\label{denoted_plus_minus_k} 
\eea

 Recall that we have also to construct the \jf of imaginary momentum $ik$. This
 procedure leads to substitution of Bessel \fs by modified ones. As a
 result we have for the products defined by (\ref{products}):
  \bea
&&\G_{\{H V H \}} \ra \G_{\{K V K \}}\e ie^{-\pi \nu}\left( \fr{2}{i\pi}\right)^2 \times\nn\\&&\left\{ \tilde{\mu}[ ip(
  K_{-}K_{-}+ K_{+}K_{+})+m_0(K_{+}K_{+}-K_{-}K_{-})]-\fr{2
    v(r)}{r}\sqrt{p^2+m_0^2} K_{-}K_{+}  \right\}\nn\\
&&\G_{\{J V J \}} \ra \G_{\{I V I \}} \e \\&&\ \ \ \ \ -i\ e^{-\pi \nu} \left\{\tilde{\mu} [ ip(
  I_{-}I_{-}+ I_{+}I_{+})+m_0(I_{+}I_{+}-I_{-}I_{-})]+\fr{2
    v(r)}{r}\sqrt{p^2+m_0^2} I_{-}I_{+}  \right\}\nn\\
&&\G_{\{H V J\}} \ra \G_{\{K V I\}} \e  \fr{2}{i\pi}\times\nn\\ &&\left\{\tilde{\mu} [ ip(K_{-}I_{-}- K_{+}I_{+})-m_0
  (K_{-}I_{-}+ K_{+}I_{+})]-\fr{v(r)}{r}\sqrt{p^2+m_0^2}(K_{+}I_{-}-K_{-}I_{+})
  \right\}\nn
 \label{products_HHtoKK_etc} 
\eea

 The notation of indices $\pm$ means as previous: 
 \bea
I_{\pm}\e I_{\nu\pm\half}(q r)\\
K_{\pm}\e K_{\nu\pm\half}(q r)
\label{denoted_plus_minus_ik} 
\eea

The substitution $ k\ra ik $ implies also the following redefinition of $q_0, p_0$

\bea
&&k\ra ik\\
&&p_0 \ra i\ p =i\ \sqrt{k^2 -m_e^2}\\
&&q_0 \ra i\ q =i\ \sqrt{p^2+m_0^2}= i\ \sqrt{k^2-m_e^2+m_0^2}
\eea
respectively.

It can be seen from(\ref{products_HHtoKK_etc})
that the Bessel \fs which are to be integrated in (\ref{lne}-\ref{lnv}) are ordered in products of pairs of kind
\be
r\ K_\pm K_\pm,\  r\ I_\pm I_\pm,\  r\ K_\pm I_\pm,\  r\ K_\mp K_\pm,\  r\ I_\mp
I_\pm,\  r\ K_\mp I_\pm
\label{pairs}
\ee

 Now we substitute the uniform asymptotic expansions represented as (\ref{products_HHtoKK_etc}) in the  (\ref{lne}-\ref{lnv}) and rewrite each of
 integrands (\ref{pairs}, indices are omitted) as
 
 \bea
 I\ I\ &\ra& \ r\ e^{2\nu\eta} \tG_{II}\nn\\
 K\ I\ &\ra& \ r \tG_{KI}\nn\\
 K\ K\ &\ra& \ r\ e^{-2\nu\eta} \tG_{KK}\nn\\
 \eea
The application of the expansion (\ref{saddlepoint}) to (\ref{lnv}) leads to 

\bea
&&\ln f_\nu^{(1)}\sim  -\left( \fr{\pi}{2i} \right)  \int\limits_0^\infty rdr \tG_{KI}\nn\\
&&\ln f_\nu^{(2)} \sim  -\left( \fr{\pi}{2i} \right)^2 \int\limits_0^\infty rdr
\tG_{KK} \left\{ \fr{1}{\nu} \fr{r \tG_{II} }{2\eta'}+ \fr{1}{\nu^2} \left(
\fr{2r\tG_{II}\eta''}{(2\eta')^3}- \fr{\tG_{II}+r\tG_{II}' }{(2\eta')^2
}\right)+  \right.\nn\\&&\left.  \fr{1}{\nu^3}\left( \fr{r\tG_{II}'' +2 \tG_{II}'  }{(2\eta')^3} -
\fr{6(\tG_{II}+r\tG_{II}' )\eta''}{(2\eta')^4}+ \fr{3r\tG_{II}(2\eta'')^2 }{(2\eta')^5} -\fr{2r\tG_{II}\eta'''   }{(2\eta')^4}  \right)
\right\}\nn\\
&&\ln f_\nu^{(3)}\sim -2 \left( \fr{\pi}{2i} \right)^3 \int\limits_0^\infty rdr\tG_{KK}
\left\{
\fr{1}{\nu^2} \fr{r^2 \tG_{KI}\tG_{II} }{4(\eta')^2}+\fr{1}{\nu^3} \left(
\fr{3r^2\tG_{KI}\tG_{II}\eta'' }{2\eta'} \right.\right.\nn\\&&\left.\left. - \fr{3r\tG_{KI}\tG_{II}}{2}- r^2\tG_{KI}\tG_{II}'- r^2\tG_{KI}'\tG_{II}\right)
\right\}\nn\\
&&\ln f_\nu^{(4)}\sim -\left( \fr{\pi}{2i} \right)^4 \int\limits_0^\infty
rdr\tG_{KK} \fr{1}{\nu^3} \fr{r^3}{(\eta')^3}\left(  \half\tG_{KI}^2\tG_{II}
-\fr{1}{8} \tG_{KK} \tG_{II}^2 \right)
\label{G_saddle_exp}
\eea

Here primes mean the derivation  with respect to $r$, and following notations
are introduced:
{
\small
\bea
&&\tG_{II}(r)=\nn\\&& \fr{t}{2 \pi \nu}
 \left\{\mu(r)
    \left[
 i p \left( \sqrt{\fr{1+t}{1-t}} \exp ( 2\sum\limits_{j=1}^3 \fr{S^+_j[t,-1/2]}{\nu^j} ) + \sqrt{\fr{1-t}{1+t}} \exp (2 \sum\limits_{j=1}^3 \fr{S^+_j[t,1/2]}{\nu^j} )  \right) \rrll
-m\left( \sqrt{\fr{1+t}{1-t}} \exp (2\sum\limits_{j=1}^3
 \fr{S^+_j[t,-1/2]}{\nu^j} ) -  \sqrt{\fr{1-t}{1+t}} \exp (2\sum\limits_{j=1}^3 \fr{S^+_j[t,1/2]}{\nu^j} )
 \right) \right]\rl
+ 2 v(r) z_0 \nu/(r^2) \exp (\sum\limits_{j=1}^3 \fr{S^+_j[t,1/2]+S^+_j[t,-1/2]}{\nu^j}  )
\right\};
\\
&&\tG_{KI}(r)=\nn\\&&\fr{t}{i \pi \nu}\left\{
   \mu(r)\left[
i p\left(  \exp \sum\limits_{j=1}^3 \fr{S^-_j[t,-1/2]+S^+_j[t,-1/2]}{\nu^j}  -
   \exp \sum\limits_{j=1}^3 \fr{S^-_j[t,1/2]+S^+_j[t,1/2]}{\nu^j}  \right) \rrll  
-m\left( \exp \sum\limits_{j=1}^3 \fr{S^-_j[t,-1/2]+S^+_j[t,-1/2]}{\nu^j} + \exp \sum\limits_{j=1}^3 \fr{S^-_j[t,1/2]+S^+_j[t,1/2]}{\nu^j} \right)\right]\rl
- v(r) z_0 \nu/(r^2) \left( \sqrt{\fr{1+t}{1-t}}  \exp \sum\limits_{j=1}^3
   \fr{S^-_j[t,1/2]+S^+_j[t,-1/2]}{\nu^j}  -\rrll\ \ \ \ \ \ \ \ \ \ \ \  \ \
   \ \ \ \ \ \ \ \ \ \ \ \ \ \ \ \ \ \
   \sqrt{\fr{1-t}{1+t}} \exp \sum\limits_{j=1}^3 \fr{S^-_j[t,-1/2]+S^+_j[t,1/2]}{\nu^j} \right)
   \right\};
\\   
&&\tG_{KK}(r)=\nn\\&&-\left(\fr{2}{i \pi}\right)^2 \fr{t \pi}{2 \nu}\left\{ 
   \mu(r)\left[
i p\left( \sqrt{\fr{1-t}{1+t}} \exp (2 \sum\limits_{j=1}^3 \fr{S^-_j[t,-1/2]}{\nu^j}  ) + \sqrt{\fr{1+t}{1-t}} \exp (2\sum\limits_{j=1}^3 \fr{S^-_j[t,1/2]}{\nu^j}  )  \right)\rrll 
+m\left( \sqrt{\fr{1+t}{1-t}} \exp (2 \sum\limits_{j=1}^3 \fr{S^-_j[t,1/2]}{\nu^j} ) - \sqrt{\fr{1-t}{1+t}} \exp (2 \sum\limits_{j=1}^3 \fr{S^-_j[t,-1/2]}{\nu^j} ) \right) \right]\rl
- 2 v(r) z_0 \nu/(r^2) \exp ( \sum\limits_{j=1}^3 \fr{S^-_j[t,1/2]+S^-_j[t,-1/2]}{\nu^j} ) \right\};
\label{tilde_G}
\eea
}
                                %

Since the \fs entering here depends not on $r$ explicit, but on $t(r)$ only, we use the derivation with respect to $t$ instead of
$r$, keeping in mind the following obvious relations for derivatives of $t$
with respect to $r$ (index $r$ denotes the derivation):

\bea
&&t_r=-\fr{t}{r}(1-t^2)\nn\\
&&t_{rr}=\fr{t}{r^2}(1-t^2)(2-3t^2) \nn\\
&&t_{rrr}=\fr{t}{r^3}(1-t^2)^2( 15t^2-6)
\eea


 Then we can carry out the simple but very laborious procedure, like  the case of spinor field
 in the magnetic background. Namely we use the general form of the uniform
 asymptotics by means of \be q,\nu\ra\infty,\ q/\nu =z_0 = const
 \label{uniform}\ee of the
 logarithmic \jf (\ref{lnf_of_ik}).
To this end we proceed from the expressions of $\ln f^{(j)}_\nu(ik),\
 j=1,2,3,4$ obtained in \cite{borkir2} and given above in this section of
 Math.App.(\ref{lne}-\ref{lnv}). They
 are already expressed in terms of products (\ref{products_HHtoKK_etc}). One
 has now to insert the uniform asymptotics (\ref{uniform}) for each of the
 modified Bessel \fs entering here (\ref{denoted_plus_minus_ik}),
\bea
 q/\nu =z_0 = const\\
I_{\nu\pm\half}(q r) \ra I_{\nu\pm\half}(\nu\ z_0 r)\\
K_{\nu\pm\half}(q r) \ra K_{\nu\pm\half}(\nu\ z_0 r)
\label{q_to_nu_z0} 
\eea

which are listed explicit in  Math.App \ref{besselasym} (\ref{KI_asymp}-\ref{S_asymp_I}). Furthermore, also the expansion
\be
p=\sqrt{q^2+m_0^2}=\sqrt{(z_0 \nu)^2-m_0^2} \ \sim z_0 \nu -\fr{m_0^2 r}{2
  z_0} \fr{1}{\nu}\\
\label{p_of_k_to_p_of_ik} 
\ee
is to substitute. The procedure has been executed with help of {\it
Mathematica} algebraic package.
The final separation of powers $\nu$ provides the following expressions
for the first order of $\nu$

\bea
&&\ln f^{as}(ik)\sim \fr{1}{\nu} \int \fr{dr}{r} \left\{
\fr{t_0(t_0^2-1)}{2 r^2}v(r)^2-m_0 \tilde{\mu}(r)t_0 r^2 -
\fr{\tilde{\mu}(r)^2t_0^2}{2}r^2+ i\fr{t_0^2\tilde{\mu}(r)r\sqrt{1-t_0^2}}{2}
  \right\}\nn\\
&&\hspace*{6cm}+O(\fr{1}{\nu^2})
\label{nu-1_order}
\eea

Note that this expression must depend on the $k$- variable, the momentum
"outside of the vortex core", related to $r\ra \infty$, but it looks not
explicit. To restore the dependence on $k$ we use the corresponding
expressions for

\bea
z_0=z_0(z)=q/\nu =\sqrt{z^2+\fr{m_0^2-m_e^2}{\nu^2}}\\
t_0=1/\sqrt{1+z_0^2r^2}
\label{z0_to_z}
\eea
we can express it in terms of $t=1/\sqrt{1+z^2 r^2}$ and expand for small
$1/\nu$ , so we obtain the contribution to the next non zero order of $1/\nu$

\be
t_0 \sim t-\fr{t^3(m_0^2-m_e^2)r^2}{2\nu^2}+O(\fr{1}{\nu^4})
\label{t0_to_t}
\ee

Furthermore we omit the imaginary term in \ref{nu-1_order} since it does not
contribute to the vacuum energy (because of the symmetry properties of the
\jf mentioned in Sec.\ref{jostfunc}). The terms proportional to odd powers of
potentials $v(r)$ and $\mu(r)$ are not present because of symmetry properties
of exact \jf mentioned above Sec.\ref{jostfunc}.
 Finally we have for the order $1/\nu$:

\be
\ln f^{as}(ik)= -\fr{1}{\nu} \int \fr{dr}{r} \left\{
\fr{t(t^2-1)}{2}v(r)^2 + \fr{r \mu(r)\mu'(r) }{z^2 t} \right\}
+O(\fr{1}{\nu^3}),
\label{nu-1_order_fin}
\ee
where $\mu$ is redefined as $\mu(r)= \tilde{\mu}(r)-m_e$.

The analogous procedure of collecting all powers of $1/\nu^3$ in $\ln
f_{as}^{(1)}(ik)\dots f_{as}^{(4)}(ik$ with the help of {\it Mathematica } algebraic packages one
arrives at the contribution of $1/\nu^3$ order in $\ln f^{as}(ik)$:
{\small
\bea
&&\fr{1}{\nu^3} \int \fr{dr}{r} \left\{
t^3(1-t^2) \left[\fr{1}{4} \left( 1-\fr{35}{4} t^2 (1-t^2) \right)v(r)^2 -\fr{1}{8} r^2v'(r)^2-\fr{1}{8}(1-5t^2)v(r)^4
\right]-\rl
\rl
\fr{1}{4}r^2t^3(1-3t^2)v(r)^2
(\mu(r)^2-m_e^2)-\fr{1}{16}r^2t^5(3-5t^2)(\mu(r)^2-m_e^2)-\fr{1}{8}
r^4t^3\mu'(r)^2-\rl\rl \fr{1}{8} r^4t^3(\mu(r)^2-m_e^2)^2
 \right\}-\fr{(m_e^2-m_0^2)^2}{4z^4}
+O(\fr{1}{\nu^5}),
\label{nu-3_order_fin}
\eea}
what leads to the representation in terms of $X_{i,j}$ in (\ref{repres_xij}).


 \subsection{Performing of the spectral sum for Bag boundary conditions}
\label{bag_perform}

\ \ \ Consider the expression (\ref{B+B-}) for regularized \gse Sec.\ref{spectralsum}.
 We proceed like the case of scalar field and represent the \f ${\cal B}^{(+E)}_{\nu}(k)$
in the form:

\bea
{\cal B}^{(+E)}_{\nu}(k)=
f^+_{\nu}(k)\left(\sqrt{p_0+m} H^{(2)}_{\nu+\half-\delta}(k R)+\sqrt{p_0-m}
H^{(2)}_{\nu-\half-\delta}(k R)\right)\times\nn\\
\left\{
  1+\fr{ \bar{f}^+_{\nu}(k)
    \left(
      \sqrt{p_0+m} H^{(1)}_{\nu+\half-\delta}(k R)+\sqrt{p_0-m}
      H^{(1)}_{\nu-\half-\delta} (k R)
    \right) }
    { f^+_{\nu}(k)
      \left(
        \sqrt{p_0+m} H^{(2)}_{\nu+\half-\delta}(k R)+\sqrt{p_0-m}
 H^{(2)}_{\nu-\half-\delta}(k R)
\right)}
\right\}
\eea

 Then by applying of the $\ln$ on ${\cal B}^{(+E)}_{\nu}(k)$ and keeping in mind
 the asymptotics of Hankel \fs (\ref{H_asymp_infty}), we obtain in the limit
 of $R\ra \infty$

 \bea
 \ln {\cal B}^{(+E)}_{\nu}(k) =
 \ln f^+_{\nu}(k)+\half \ln \fr{2}{\pi k R} + i\left(kR -\fr{\pi}{4}\right)+i
 \fr{\pi}{2}(\nu-\half-\Omega)+\nn\\ \ln [i\sqrt{p_0+m}+\sqrt{p_0-m}]+
 \ln [1+exponential\  suppressed\  term]
 \label{B_expanded}
 \eea
 The last term tends to zero in the limit $R\ra \infty$, under consideration,
 that $k$ lies on the upper branch ($k=\Re k+i\epsilon$). 
Further, by applying of the derivation with respect to $k$, we get the
 following terms survived:
 
\bea
\fr{\pd}{\pd k} \ln {\cal B}^{(+E)}_{\nu}(k)=
\fr{\pd}{\pd k} \ln f^+_{\nu}(k)+ \fr{\pi k R^2}{4}+iR + \fr{-i}{\sqrt{k^2+m^2}}e^{-2i\theta}
\eea
where the phase $\theta$ is defined by $\theta=\arctg
\sqrt{\fr{p_0+m}{p_0-m}}$.

The second term is obviously proportional to the empty two-dimensional
volume. The contribution of this term in $E^{reg}$ is now easy to calculate
as
\be \fr{\mu^{2s}}{\sqrt{\pi}} \fr{\Gamma(s-1)}{\Gamma(s-\half)} \fr{\pi
  R^2}{8} \fr{(k^2+m^2)^{2-s}}{s-2}
\ee
that in the large mass limit $k\ra 0$ is proportional to $m^4$ and
corresponds to the contribution of the coefficient $a_0$ in \hke (\ref{large_mass_expansion}).

The third term is a pure imaginary value  and can be omitted since the energy
is defined to be real. Consider the last term in the large mass limit. Then
the real part can be integrated immediately and it turns out to be
proportional to $m^3$ and represents therefore the non-local contribution
from the boundary, corresponding to the coefficient $a_{1/2}$. In the
considered case it cancels with the corresponding term arising from
${\cal B}^{(-E)}_{\nu}(k)$.

The crucial fact for this procedure is the vanishing dependence on the
background through the relation $\fr{\pd}{\pd k}i\fr{\pi}{2}(\nu-\half-\Omega)=0$.

The performing of  ${\cal B}^{(-E)}_{\nu}(k)$ is analogous, and the final
expression results in (\ref{e0_of_lnf}).


 \subsection{Calculation of the asymptotic part of \gse}
\label{calcEAS}
\subsubsection{Useful relations}

{\bf The Abel-Plana formulae}

 The sum over half-integers $\nu$ has been transformed to integrals
 according to the Abel-Plana formula as follows:
\begin{equation}
 \sum\limits_{l=0}^{\infty}(l+\half)\e\sum\limits_{\nu=\half,\fr{3}{2}}=\int\limits_{0}^{\infty}d\nu
f(\nu)+\int\limits_{0}^{\infty}\fr{d\nu}{1+e^{2\pi\nu}}
 \fr{f(i\nu)-f(i\nu)}{i}
 \label{abelplan}
\end{equation}

The summation over integers $l$ can be performed as
\be
\sum\limits_{l=1}^{\infty}F(l) = \int\limits_{0}^{\infty}F(l)dl-\half
 F(0)+
 \int\limits_{0}^{\infty} \frac{dl}{1-e^{2\pi l}}
 \frac{F(il)-F(-il)}{i}
\label{abel-plan-integers}
\ee

{\bf Integral relations}\\

The integrations over $\nu$ and $k$ can be done by using the identities:
\be
\int\limits_{0}^{\infty} d\nu \int\limits_{m}^{\infty}
dk(k^2-m^2)^{1-s}\fr{\pd}{\pd k}\fr{t^j}{\nu^n}= -\fr{m^{2-2s}}{2}
\fr{\Gamma(2-s)\Gamma(\fr{1+j-n}{2})\Gamma(s+\fr{n-3}{2})}{(rm)^{n-1}\Gamma(\fr{j}{2})}
\label{ident1}
\ee
and
\be
 \int\limits_{m}^{\infty}
dk(k^2-m^2)^{1-s}\fr{\pd}{\pd k}t^j= -m^{2-2s}
\fr{\Gamma(2-s)\Gamma(s+\fr{j}{2}-1)}{\Gamma(\fr{j}{2})}\fr{(\fr{\nu}{rm})^j
 }{(1+(\fr{\nu}{rm})^2)^{s+\fr{j}{2}-1}}
\label{ident2}
\ee

\subsubsection{Scalar field with scalar background}
\label{easSCALAR}

{\bf Scalar field}
 
The sum $\sum\limits_{l=1}^\infty$ over $l$ in ${\cal E}^{as}_\pm$  (\ref{e_pm}) can be
replaced by the integration over $dl$ according to the Abel-Plana formula
(\ref{abel-plan-integers}). Thus the calculation of ${\cal
  E}^{as}_\pm$ is reduced to the calculation of the following three parts:

\bea
&&{\cal E}^{as}_1=\\&&-\fr{1}{2\sqrt{\pi^3}} \fr{\Gamma(s-1)\sin \pi
  s}{\Gamma(s-\half)}\int\limits_m^\infty dk (k^2-m^2)^{1-s}
\fr{V_0}{2}\int\limits_{R_1}^{R_2}r\ dr\int\limits_0^\infty dl\fr{\pd}{\pd k}
\sum\limits_{n=1}^3\sum\limits_{j=1}^7 X_{nj}\fr{t^j}{l^n}\nn\\
&&{\cal E}^{as}_2=\\ &&-\fr{1}{2\sqrt{\pi^3}} \fr{\Gamma(s-1)\sin \pi
  s}{\Gamma(s-\half)}\int\limits_m^\infty dk (k^2-m^2)^{1-s}
\fr{-V_0}{4}\int\limits_{R_1}^{R_2}r\ dr\fr{\pd}{\pd k}
\sum\limits_{n=1}^3\sum\limits_{j=1}^7 X_{nj}\fr{t^j}{l^n} \left.\right|_{l=0}\nn\\
&&{\cal E}^{as}_3=\nn\\&&-\fr{1}{2\sqrt{\pi^3}} \fr{\Gamma(s-1)\sin \pi
  s}{\Gamma(s-\half)}\int\limits_m^\infty dk (k^2-m^2)^{1-s}
\fr{V_0}{2}\int\limits_{R_1}^{R_2}r\ dr\int\limits_0^\infty \fr{dl}{1-e^{2\pi
    l}}\fr{-i\pd}{\pd k}\times\nn\\
&&\sum\limits_{n=1}^3\sum\limits_{j=1}^7 X_{nj} \left[ \fr{t^j(il)}{(il)^n}- \fr{t^j(-il)}{(-il)^n}\right]
\label{Eas1-Eas3}
\eea

 To calculate the ${\cal E}^{as}_1$, ${\cal E}^{as}_3$ for each of $n,j$ we
 use the identity (\ref{ident1}), 
and for the expression of type ${\cal E}^{as}_2$ the (\ref{ident2}) respectively
\be{\cal E}^{as}_1=\fr{V_0 m^2}{32\pi}\left( \fr{1}{s}+\ln \fr{4\mu^2}{m^2}-1
 \right)(R_1^2-R_2^2)+\fr{V_0^2}{64\pi} \left( \fr{1}{s}+\ln \fr{4\mu^2}{m^2}-2
 \right)(R_1^2-R_2^2)\label{Eas1scalar}
\ee

\be{\cal E}^{as}_2=\fr{V_0}{\pi}\left\{ \fr{1}{64m}\left(
    \fr{1}{R_2}-\fr{1}{R_1}\right)+ \fr{m}{8}(R_2-R_1) \right\}+
    \fr{V_0^2}{32\pi m}(R_2-R_1)\label{Eas2scalar}
\ee
The results for $E^{as}_3$ contain expressions of kind
\be
\sin (\pi s+\fr{j\pi}{2})\int\limits_0^\infty\fr{l^n}{1-e^{2\pi l}} [l^2-(mr)^2]^{s+j/2},
\ee
 which are not convergent in the limit $s\ra 0$ for several $n,j$ and must be
further performed. The partial integration executed several times over $l$ allows to
abrogate divergent powers of $[l^2-(mr)^2]$. The similar procedure is
described in details also in this section below, for the problem of spinor
field inn NO-vortex background.

All the remaining surface terms turn out to be zero. Finally the ${\cal E}^{as}_3$ results in
\bea
&&{\cal E}^{as}_3(V_0, R_1, R_2) =
\frac{V_0}{2\pi}\left[ f_1-\fr{1}{8}f_2-\fr{1}{4}f_3-\fr{1}{24}f_4\right]+
\frac{V_0^2}{24\pi m^2}f_5,\nn\label{Eas3scalar}
\eea
where

\bea
&&f_1\e \int\limits_{mR_1}^{\infty}
F_1(l)G_1(l,R_1)dl-\int\limits_{mR_2}^{\infty} F_1(l)G_1(l,R_2)dl\nn\\
&&f_2\e  \int\limits_{mR_1}^{\infty}F_2(l)G_1(l,R_1)dl-\int\limits_{mR_2}^{\infty} F_2(l)G_1(l,R_2)dl\nn\\
&&f_3 \e \int\limits_{mR_1}^{\infty}F_3(l)G_1(l,R_1)dl-\int\limits_{mR_2}^{\infty} F_3(l)G_1(l,R_2)dl\nn\\
&&f_4\e \int\limits_{mR_1}^{\infty}F_4(l)G_1(l,R_1)dl-\int\limits_{mR_2}^{\infty} F_4(l)G_1(l,R_2)dl \nn\\
&&f_5\e \int\limits_{mR_1}^{\infty}F_2(l)G_2(l,R_1)\frac{dl}{l}-\int\limits_{mR_2}^{\infty}F_2(l)G_2(l,R_2)\frac{dl}{l},
\eea
and 

\bea
&F_1(l)=\frac{l}{1-e^{2\pi l}}\ ,& F_2(l)=l( \frac{1}{l(1-e^{2\pi
    l})} )'\\
&F_3(l)=l(\frac{1}{l}(  \frac{l}{l(1-e^{2\pi l})})' )', 
&F_4(l)=l(\frac{1}{l}(\frac{1}{l} (\frac{l^3}{l(1-e^{2\pi l})})' )')'\nn\\
\nn\\
&G_1(l,x)=\ln \left[ \frac{l}{mr}+\sqrt{ \left(\frac{l}{mr}\right)^2-1}\right]-
\sqrt{1-\left(\frac{mr}{l}\right)^2} & G_2(l)=(l^2-m^2r^2)^{3/2}\nn
\eea

Thus the finite part of $E^{as}$ is represented in terms of integrals not
complete calculable analytically, but well convergent numerically.\\

\subsubsection{Spinor field with magnetic background}

\ \ \ The procedure of calculation of $E^{as}$ for spinor field is described
completely in \cite{borkir1}, and the similar problem is considered for
the spinor field in NO-vortex background, so we omit all details for this problem and
give the final result only: 

\bea
&&E^{as}=\nn\\&& \fr{-4}{\pi}\int\limits_{0}^{\infty}\fr{dr}{r^3}[\Omega^2 a(r)^2
g_1(rm)-\Omega^2 r^2 a'(r)^2 g_2(rm)+\Omega^4 a(r)^4 g_3(rm)]\nn\\&&
=\Omega^2 e_1(R_1,R_2)+\Omega^4 e_2(R_1,R_2)
\label{Eas_spinor}
\eea

here $g_i$ are

\begin{eqnarray}
g_1(x)&=& \half f_{1, 1}(x)-\half f_{1, 3}(x)+\fr{1}{4}f_{3,
  3}-\fr{39}{16}f_{3, 5}(x)+\fr{35}{8}f_{3, 7}(x)-\fr{35}{16} f_{3,
  9}(x)\nn\\&-&
\half x \pd_x( -\fr{1}{4} f_{3, 3}(x)+\fr{7}{8}f_{3, 5}(x)-\fr{5}{8}f_{3, 7}(x))\nn\\
&+&\half x \pd_x^2(\fr{x}{8}f_{3, 3}(x)- \fr{x}{8}f_{3, 5}(x) ), 
\nn\\
 g_2(x)&=& \fr{1}{8}(f_{3, 3}(x) - f_{3, 5}(x)), \nn\\
 g_3(x)&=& -\fr{1}{8}(f_{3, 3}(x) - 6 f_{3, 5}(x)+5 f_{3, 7}(x) ),
 \label{g_i_of_x}
\end{eqnarray}
with $f_{i, j}$ are

\begin{eqnarray}
f_{1, 1}(x)&=& - \fr{1}{1+e^{2\pi x}}\nn\\
f_{1, 3}(x)&=& - (\fr{}{1+e^{2\pi x}})' \nn\\
f_{3, 3}(x)&=& (\fr{1}{x} \fr{1}{1+e^{2\pi x}} )'\nn\\
f_{3, 5}(x)&=& \fr{1}{3}( \fr{1}{x}( \fr{x}{1+e^{2\pi x}})')'\nn\\
f_{3, 7}(x)&=& \fr{1}{15}( \fr{1}{x}( \fr{1}{x}( \fr{x^3}{1+e^{2\pi x}} )')')'\nn\\
f_{3, 9}(x)&=& \fr{1}{105}( \fr{1}{x}( \fr{1}{x}( \fr{1}{x}(
\fr{x^5}{1+e^{2\pi x}})')')')'
\label{f_i_of_v}
\end{eqnarray}
 
The derivation of this result can be viewed as a particular case of more
general problem, which is given in details below. 

\subsubsection{Spinor field with NO vortex background}

\ \ During the calculation of the $ E^{as}$ one has to evaluate the following
types of integrals over $\nu$:
\be
 {\cal E}^{as}_\infty=2 C_s \sum\limits_{\nu=\half,\fr{3}{2},\dots}   \int_m^\infty dk\
(k^2-m^2)^{1-s}\fr{\pd}{\pd k}\int\limits_0^\infty  \fr{dr}{r}
\sum\limits_{n=1}^3 \sum\limits_{j=1}^{3n} X_{nj} \fr{t^j}{\nu^n}= 2 C_s
\sum\limits_{\nu=\half,\fr{3}{2},\dots} {\cal F(\nu)}
\label{F_nu}
\ee
Treated by means of the Abel-Plana formula (\ref{abelplan}) it can be
written as
\be  {\cal E}^{as}_\infty= {\cal E}^{as}_1+ {\cal E}^{as}_2 \label{eas_intfy}\ee
where
\bea
&&{\cal E}^{as}_1=2 C_s \int\limits_0^\infty d\nu {\cal F}(\nu)\nn\\ 
&&{\cal E}^{as}_2=2 C_s \int\limits_0^\infty \fr{d\nu}{i(1+e^{2\pi\nu})}\left[ {\cal F}(i\nu)- {\cal F}(-i\nu)\right].
\label{eas1_eas2}
\eea

It is known from the experience of previous calculations that the ${\cal E}^{as}_2$
contributes to the finite and the ${\cal E}^{as}_1$ to the divergent part of ${\cal E}^{as}_\infty$.
 Proceeding from the explicit form of $ {\cal F(\nu)}$ (\ref{F_nu}) one left
with the evaluation of the integrals of kind

\be\label{contrib_eas1}
\I_{ij}\e 2 C_s \int\limits_0^\infty d\nu  \int_m^\infty dk\ (k^2-m^2)^{1-s}\fr{\pd}{\pd k} \fr{t^j}{\nu^n}
\ee
in ${\cal E}^{as}_1$ and
\be
\J_{ij}\e 2 C_s \int\limits_0^\infty  \fr{d\nu}{i(1+e^{2\pi\nu})} \int_m^\infty dk\
(k^2-m^2)^{1-s}\fr{\pd}{\pd k}\left[
\fr{t^j}{(i\nu)^n}-\fr{t^j}{(-i\nu)^n}\right]
\label{contrib_eas2}
\ee
in ${\cal E}^{as}_2$ respectively. It comes about by using of the indentities
(\ref{ident1},\ref{ident2}).
The results for (\ref{contrib_eas1},\ref{contrib_eas2}) corresponding to $\fr{t^j}{\nu^n}$ for each of $n=1,3$
and $j=1,3,5,7,9$ are:
\bea 
&&\I_{11}=\I_{13}= -C_sm^2m^{-2s}\Gamma(2-s) \Gamma(s-1) \nn\\
&&\I_{33}=\I_{35}=\I_{37}=\I_{39}=-\fr{2}{(j-2)r^2}C_sm^{-6s}\Gamma(2-s) \Gamma(s-1),
\label{terms_eas1}
\eea
for ${\cal E}^{as}_1$ and

\bea
&&\J_{11}= 4C_s \cos (\pi s)\fr{r^{2s-2}}{\sqrt{\pi}}\Gamma(2-s)\Gamma(s-\half) \int \fr{d\nu}{(\nu^2-[mr]^2)^{s-1/2}} f_{1,1}(\nu)\nn\\
&&\J_{13}=-8C_s \cos (\pi s)\fr{r^{2s-2}}{\sqrt{\pi}(1-2s)}\Gamma(2-s)\Gamma(s+\half) \int \fr{d\nu}{(\nu^2-[mr]^2)^{s-1/2}} f_{1,3}(\nu)\nn
\eea\bea
&& \J_{33}= 8C_s \cos (\pi s)\fr{r^{2s-2}}{\sqrt{\pi}(1-2s)}\Gamma(2-s)\Gamma(s+\half) \int \fr{d\nu}{(\nu^2-[mr]^2)^{s-1/2}} f_{3,3}(\nu)\nn\\
&& \J_{35}=16C_s \cos (\pi s)\fr{r^{2s-2}\Gamma(2-s)\Gamma(s+3/2) }{3\sqrt{\pi}(2s+1)(1-2s)} \int \fr{d\nu}{(\nu^2-[mr]^2)^{s-1/2}} f_{3,5}(\nu)\nn\\
&& \J_{37}=\nn\\&& 32C_s \cos (\pi
s)\fr{r^{2s-2}\Gamma(2-s)\Gamma(s+5/2)}{15\sqrt{\pi}(3+2s)(2s+1)(1-2s)} \int \fr{d\nu}{(\nu^2-[mr]^2)^{s-1/2}} f_{3,7}(\nu)\nn\\
&& \J_{39}=\nn\\&&64C_s \cos (\pi
s)\fr{r^{2s-2}\Gamma(2-s)\Gamma(s+7/2)
}{105\sqrt{\pi}(5+2s)(3+2s)(2s+1)(1-2s)} \int
\fr{d\nu}{(\nu^2-[mr]^2)^{s-1/2}} f_{3,9}(\nu)\nn\\
\label{terms_eas2}
\eea
for  ${\cal E}^{as}_2$ respectively.\\

 Here the relation 
\be\left.[(mr)^2+x^2]^{-(s+j/2)}\right|_{x=-i\nu}^{x=i\nu} =-2 (mr)^{2s+j}[v^2-(mr)^2]^{-(s+j/2)}\sin[\pi s+j\pi/2]\ee
has been applied, and the partial integration several times has been executed
to abrogate the singular denominator $[(mr)^2+x^2]^{-(s+j/2)}$ at $s\ra 0$.\\

The ${\cal E}^{as}_1$ cancels against the $E^{div}$ (Sec.\ref{efineas}), and we arrive finally at the finite part of the asymptotic energy: 
\bea
&&E^{as}={\cal E}^{as}_2=-\fr{2}{\pi} \int\limits_0^\infty\fr{dr}{r^3}\int\limits_x^\infty d\nu
\times\nn\\
&&\left\{
v(r)^2\left[
f_{11}(\nu)-f_{13}(\nu)-\half f_{33}(\nu)+\fr{3}{18}f_{35}(\nu)-\fr{7}{12}f_{37}(\nu)+\fr{1}{24}f_{39}(\nu)
  \right]+\right. \nn\\&&v(r)^4\left[
  \fr{1}{4}f_{33}(\nu)-\half f_{35}(\nu)+\fr{1}{12}f_{37}(\nu)
  \right]+v(r)^2 r^2(\mu(r)^2-m_e^2) \left[ \half f_{33}(\nu) -\half f_{35}(\nu)  \right]+\nn\\
&& v'(r)^2 r^2 \left[ \fr{1}{4} f_{33}(\nu) -\fr{1}{12} f_{35}(\nu)  \right]+
  r^2(\mu(r)^2-m_e^2)\left[f_{11}(\nu)+\fr{1}{8} f_{35}(\nu)- \fr{1}{24}
    f_{37}(\nu) \right]+\nn\\
 &&\left.\fr{1}{4}r^4[(\mu(r)^2-m_e^2)^2+\mu'(r)^2] f_{33}(\nu)
  \right\}
  \label{NO_Eas}
\eea
 
where $x$ means $mr$ and functions $f_{ij}(\nu)$ are as usual

\bea
&f_{11}(\nu)=\fr{1}{1+e^{2\pi\nu}} &f_{13}(\nu)= \left[\fr{1}{1+e^{2\pi\nu}}\right]'\nn\\
&f_{33}(\nu)= \left[\fr{1}{\nu(1+e^{2\pi\nu})}\right]'&f_{35}(\nu)=\left[\fr{1}{\nu}\left[\fr{1}{1+e^{2\pi\nu}}\right]'\right]'\nn\\
&f_{37}(\nu)=\left[\fr{1}{\nu}\left[\fr{1}{\nu}\left[\fr{1}{1+e^{2\pi\nu}}\right]'\right]'\right]'&f_{39}(\nu)=\left[\fr{1}{\nu}\left[\fr{1}{\nu}\left[\fr{1}{\nu}\left[\fr{1}{1+e^{2\pi\nu}}\right]'\right]'\right]'\right]'\nn
\label{NO_fij_Eas}
\eea\\


 \subsection{The \rn of the NO-vortex on-shell}
\label{NOrenorm}
\ \ \ We start with the classical energy density of the NO-vortex per unit
length for the particular case $n=1$, Sec.\ref{hamiltonians}, (\ref{class_energy})
 
 \be
 {\cal E}=\pi\int r\ dr\left\{\fr{1}{q^2r^2}v'(r)^2+\eta^2
 \left[f'(r)^2+\fr{f(r)^2}{r^2}(1-v(r))^2 \right] + \fr{\la\eta^4}{2}(f(r)^2-1)^2 \right\}
 \label{NOclass_energy}
 \ee

Here the expression of ${\cal E}$ is given in terms of $r$-variable,(but not $\rho=q\eta).$

  There are various possibilities to rewrite the classical energy "on shell" using the equations of motion 
  \bea
f''(r)+\fr{f'(r)}{r}-n^2\fr{f(r)}{r^2}[1-v(r)]^2+\lambda\eta^2(1-f(r)^2)f(r)=0,\nn\\
v''(r)-\fr{v'(r)}{r}+q^2\eta^2 f(r)^2[1-v(r)]=0.
\label{no_vortex_equ}
\eea


Let us transform the mixed term in the (\ref{NOclass_energy}):
 \be
    \ \ \fr{1}{r^2}(1-v(r))^2 f(r)^2;
 \ee
 from the second \eqq of (\ref{no_vortex_equ}) we have:
\be
 \fr{f(r)^2}{r^2} (1 -v(r))^2 = \fr{1}{q^2 r^2 \eta^2}\left(
   v''(r)-\fr{v'(r)}{r} \right)(v(r)-1).
 \label{rew_2_eq}
 \ee

Now we rewrite the vortex equations in the form
\bea
f(r)\left( f''(r)+\fr{f'(r)}{r}
\right)+\la\eta^2(1-f(r)^2)f(r)^2=\fr{f(r)^2}{r^2} (1 -v(r))^2\nn\\
\fr{1}{q^2\eta^2}\left(v"(r)-\fr{v'(r)}{r}
\right)\fr{v-1}{r^2}=\fr{f(r)^2}{r^2} (1 -v(r))^2,
\label{rewr_no}
\eea
so that right-hand sides of the both equations are equal and therefore the
left-hand sides as well.
 Further, if we rewrite the (\ref{NOclass_energy}) using (\ref{rew_2_eq})
 
 \bea
 {\cal E}=&\pi&\int r\
 dr\left\{
   \fr{1}{q^2r^2}\left[v'(r)^2+(v(r)-1)(v''(r)-v'(r)/r)  \right]
   +\eta^2 f'(r)^2+ \rl \fr{\la\eta^4}{2}(f(r)^4-1)-\la\eta^4(f(r)^2-1)
 \right\}
 \eea
and use a dummy trick (represent the coefficient at $\eta^2 f'(r)^2$ as
$1+\xi-\xi$), we obtain:

\bea
 {\cal E}=&\pi&\int r\
 dr\left\{\fr{1}{q^2r^2}\left[v'(r)^2+(v(r)-1)(v''(r)-v'(r)/r)  \right]
   +\eta^2 \xi f'(r)^2+\rl  \fr{\la\eta^4}{2}(f(r)^4-1) -\la\eta^4(f(r)^2-1)
 +\eta^2(1-\xi)f'(r)^2
 \right\}.
 \eea

Note, that the integral of full derivative
 \be
 \int_0^\infty dr\half [r(f(r)^2)']'
  \ee
 is zero, and $\fr{1}{r}(f(r)f'(r))'=f'(r)^2+f(r)\left( f''(r)+\fr{f'(r)}{r}
\right)$. Then it follows from (\ref{rewr_no}), that
\bea
&&f'(r)^2=-f(r)\left( f''(r)+\fr{f'(r)}{r}\right)=\\
&&\hspace*{4cm}-\fr{1}{q^2 \eta^2}\left(v''(r)-\fr{v'(r)}{r}
\right)\fr{v-1}{r^2}+\la\eta^2(1-f(r)^2)f(r)^2.\nn
\eea

So the classical energy can be rewritten in the form:
\bea
{\cal E}=\pi\int r\
 dr\left\{
\xi\eta^2f'(r)^2+(2\xi-1)\fr{\la\eta^4}{2}(f(r)^4-1)+\xi\la\eta^4(1-f(r)^2)
+\right.\nn\\
\left.\fr{1}{q^2r^2}\left[v'(r)^2+\xi(v(r)-1)(v''(r)-v'(r)/r)  \right]
\right\}.
\eea

We remark, that the integral of the full derivative
\be
\int\limits_0^\infty dr \left\{ \fr{v(r)v'(r)}{r}-\fr{v'(r)}{r}\right\}' 
\ee
provides a finite contribution $-2a$ under assumption that
$v(r)\ra a r^2+b r+...$ at $r\ra 0$, and we find that
\be
\pi\int r\
 dr\left\{\fr{1}{q^2r^2}\xi(v(r)-1)(v''(r)-v'(r)/r)
 \right\}=
 -2a \pi\fr{\xi}{q^2}- \pi\int r dr\left\{ \fr{\xi v'(r)^2}{q^2 r^2}
  \right\}. 
\ee

Finally we arrive at the expression:
\bea
&&{\cal E}=-2a \pi\fr{\xi}{q^2}+\\&&\pi\int r\
 dr\left\{
\xi\eta^2f'(r)^2+(2\xi-1)\fr{\la\eta^4}{2}(f(r)^4-1)+\xi\la\eta^4(1-f(r)^2)
+\fr{1-\xi}{q^2r^2}v'(r)^2
\right\}.\nn
\eea

Now we have a number of possibilities to define $\xi$. We require for example:
\be
\xi=\fr{\la}{2}(2\xi-1), \ \ \ \xi=\fr{\la}{2(\la-1)}.
\label{example_1}
\ee

Then
\bea
{\cal E}=-2a \pi\fr{\xi}{q^2}+&\pi&\int r\
 dr\left\{
\fr{\la}{2(\la-1)}[\eta^2
f'(r)^2+{\eta^4}(f(r)^4-1)]-\rl \fr{\la^2}{2(\la-1)}\eta^4(f(r)^2-1)+\fr{1}{q^2 r^2}\fr{\la-2}{2(\la-1)}v'(r)^2
\right\}
\label{energy_example_1}
\eea

Finally, we can give an interpretation of counter terms entering in ${\cal
  E}^{cl}$. 
To this end it is suitable to represent the divergent energy $E^{div}$
(\ref{NO_ediv}), Sec.\ref{efineas} as

\bea
&&E^{div}=-\fr{1}{4\pi}
\left\{
   \int\limits_0^\infty r\ dr \fr{f_e^4 \eta^4}{4} (f(r)^2-1)+\right.\\&&\left.
\left( \fr{1}{s}+ \ln \fr{4\mu^2}{m^2}-2  \right)
  \int\limits_0^\infty r\ dr \left[\fr{f_e^4 \eta^4}{8} (f(r)^2-1)^2+\fr{f_e^2
\eta^2}{4} f'(r)^2+ \fr{1}{3}\fr{v'(r)^2}{r^2}\right]
\right\}.\nn
\eea
 Now the subtraction ${\cal  E}^{cl}-E^{div}$ provides the ordering of
coefficients as follows:

\bea
&& {\cal  E}^{cl}-E^{div}=\nn\\&&\fr{1}{4\pi}
\int\limits_0^\infty r\ dr
 \left\{
(f(r)^2 -1) \left( \fr{f_e^4 \eta^4}{4} - \fr{4\pi^2\eta^4\la^2 }{2(\la-1)} \right)+\rl
\fr{v'(r)^2}{r^2}\left[ \fr{1}{3}\left( \fr{1}{s}+\ln \fr{4\mu^2}{m^2}-2
 \right)+\fr{4\pi^2}{q^2} \fr{\la-2}{2\la-1}  \right]+\rl
(f(r)^4 -1) \left[ \left( \fr{1}{s}+\ln \fr{4\mu^2}{m^2}-2  \right) \fr{f_e^4
 \eta^4}{8}+\fr{4\pi^2\eta^4\la }{2(\la-1) }   \right]+\rl
f'(r)^2\left[ \left( \fr{1}{s}+\ln \fr{4\mu^2}{m^2}-2  \right) \fr{f_e^2
 \eta^2}{4}+\fr{4\pi^2\eta^2\la }{2(\la-1) }   \right]
\right\}.\nn
\eea

 Thus, the possible treatment of the \rn scheme can be established as
 follows: If we state, that the first term, proportional to $(f(r)^2 -1)$
 corresponds to the finite redefinition of the coupling constant $\la$,
and the second term with $v'(r)^2$  implies the finite redefinition of the
 U(1)-coupling (electric charge), then only one free parameter remains,
 namely the Higgs condensat $\eta$, which must be fixed by two remaining
 terms. This fixing condition is then easy to obtain:
\be \eta = \fr{2\sqrt{2}}{f_e} \exp\left\{
 \fr{\pi^2\la}{\la-1}\left(\fr{4}{f_e^2}-1 \right)-1\right\}.\ee
 Here we recalled, that \be m=m_e\e \fr{f_e \eta}{\sqrt{2}}. \ee
 (as introduced in Sec.\ref{hamiltonians}).

This example shows, that one of the free parameter of this model must be
fixed anyway, since we have four relations for three parameters at the given
structure of the counter terms.


\end{document}